Development and Validation of a Scale for Measuring Instructors' Attitudes toward
Concept-Based or Reform-Oriented Teaching of Introductory Statistics in the Health
and Behavioral Sciences

A Dissertation

Presented to the Faculty of the College of Health Sciences

of Touro University International

in Partial Fulfillment of the Requirements

for the Degree of Doctor of Philosophy  in Health Sciences
(Researcher-Educator Concentration)

By

Rossi A. Hassad

August, 2007

**TOURO UNIVERSITY INTERNATIONAL**
**COLLEGE OF HEALTH SCIENCES**
**CYPRESS, CALIFORNIA 90630**

Development and Validation of a Scale for Measuring Instructors' Attitudes toward
Concept-Based or Reform-Oriented Teaching of Introductory Statistics in the Health
and Behavioral Sciences

*This dissertation, written by*

Rossi A. Hassad

*Submitted to the Faculty of Touro University International in partial fulfillment of the
requirements for the degree of*

**DOCTOR OF PHILOSOPHY**
**IN HEALTH SCIENCES**

*Approved by:*  _______________________________

        *Dr. Edith Neumann, Vice President for Academic affairs*

        _______________________________

        *Dr. Mihaela Tanasescu, Doctoral Program Director*

*We, the undersigned, certify that we have read this dissertation and approve it as
adequate in scope and quality for the Doctor of Philosophy.*

*Dissertation Committee:*

_______________________________       _______________________________

*Dr. Anthony Coxon, Chair*             *Dr. Edith Neumann, Member*

_______________________________

*Dr. Frank Gomez, Member*

# ACKNOWLEDGEMENTS

I am profoundly grateful to the members of my dissertation committee, Dr. Anthony Coxon, Dr. Edith Neumann, and Dr. Frank Gomez for the expert guidance and support they provided, in particular, their emphasis on scholarship. Added gratitude is in order for Dr. Coxon, for serving as my dissertation Chair, exposing me to the beauty of multidimensional scaling techniques, and always being positive, encouraging, and supportive. I will always be indebted to my mother, brothers, and sisters for their love, understanding, and support, which helped to shape me into the person I am, and facilitated my achievements.

Special thanks to the administration of Mercy College, as well as Hunter College (Department of Psychology) for the opportunity to develop and advance my pedagogical knowledge and skills. And to all my friends who embraced me on this long and winding journey, you have my heartfelt gratitude. I must recognize and thank the many statistics educators, researchers, and practitioners who served as content and measurement specialists, and hence contributed significantly to this study. In this regard, special mention must be given to the ASA/MAA Joint Committee on Undergraduate Statistics (2004).

# TABLE OF CONTENTS

# LIST OF TABLES

# LIST OF FIGURES

# ABSTRACT

Development and Validation of a Scale for Measuring Instructors' Attitudes toward
Concept-Based or Reform-Oriented Teaching of Introductory Statistics in the Health and
Behavioral Sciences

By

**Rossi Alim Hassad**

Doctor of Philosophy in Health Sciences

Touro University International, Cypress, California

**Dr. Anthony P.M. Coxon, Dissertation Committee Chair**

The emergence of evidence-based practice has highlighted the training of
healthcare professionals, particularly with regard to evaluating and using research data to
facilitate effective decision-making and optimum patient care. Consequently, statistics is
increasingly becoming a core requirement for college majors in the health and behavioral
sciences. Associated with this development, is ongoing reform in statistics education, as
in general, traditional teaching strategies have failed to promote statistical literacy. But in
spite of a decade of reform efforts, desired learning outcomes are lacking. Moreover,
there is strong theoretical and empirical evidence that instructors' attitudes impact the
quality of teaching and learning.

The objective of this study was to develop and validate a scale for measuring
instructors' attitudes toward concept-based (reform-oriented or constructivist) teaching of
introductory statistics in the health and behavioral sciences, at the tertiary level. Data
were obtained from a cross-sectional study of 227 instructors (USA and international),
and analyzed using primarily factor analysis and multidimensional scaling. The scale will
be referred to as **FATS** (**F**aculty **A**ttitudes **T**oward **S**tatistics), and it consists of five subscales **(perceived usefulness, personal teaching efficacy, perceived difficulty, avoidance-approach**, and **intention).** These represent cognition, affect, and intention (consistent with the tripartite attitude model) with a total of 25 items, and an overall alpha of .89. These 5 subscales (alphas between .65 and .85) explained 51% of the total variance, and all of the common variance underlying the items.

Construct validity of the attitude scale was established. Specifically, the overall scale, and subscales plausibly, and significantly (except perceived difficulty) differentiated between low-reform and high-reform practice instructors, and explained 28% of the variance in teaching practice. The teaching practice scale consists of two dimensions (behaviorist and constructivist). Statistically significant differences in overall attitude and some subscales were observed with respect to age and teaching area. This scale can be considered a reliable and valid measure of instructors' attitudes toward concept-based teaching of introductory statistics, and can facilitate the use of reform-oriented strategies. Faculty development programs should address perceived usefulness, the primary determinant of intention, which was the strongest predictor of practice. Additional studies are required to confirm these structural and psychometric properties.

# CHAPTER 1

## INTRODUCTION

Statistical thinking will one day be as necessary for efficient citizenship as the ability to read and write. (Herbert George Wells, 1952)

## Background

There is widespread emphasis on reform in the teaching of introductory statistics at the college level (Batanero et al., 1994; Chance, delMas, & Garfield, 2004; Delmas, Garfield, Ooms, & Chance, 2006; Garfield et al., 2000; Hodgson, 1996; Mills, 2002; Saldanha & Thompson, 2001). Underpinning this reform is a consensus among educators and practitioners that traditional curricular materials and pedagogical strategies have not been effective in promoting statistical literacy[1] (Cobb, 1992; Mills, 2002; Moore, 1997), a skill that is becoming increasingly necessary for advancement and effective decision-making in many aspects of everyday life (Batanero et al., 1994; Wild et al., 1997; Ottaviani, 1998).

Statistical literacy is especially relevant given the current global information explosion (Fisher & Lee, 2005; Pakenham-Walsh, 2002), and the emphasis on evidence-based practice and decision-making (Belar, 2003). The fundamental benefit of statistics, "is that it deals with variability and uncertainty which is everywhere" (Bartholomew, 1995, p.5), and therefore, it is as an important tool for analyzing the "uncertainties and complexities of life and society" (Mosteller, 1989, p.ix), and a "catalyst for investigation and discovery" (George Box quoted in Roberts, 1992, p. 109). Nowhere is there a greater need to measure and explain variability than in the health and behavioral sciences, which issues directly influence and inform quality of life, a top priority in any society.

---

[1] Statistical literacy is used interchangeably (and most times synonymously) with statistical thinking and quantitative reasoning, however, there is some debate as to the correctness of this. Nonetheless, logically, statistical literacy can be viewed as a product of statistical thinking and quantitative reasoning.

**A Brief Historical Perspective of Statistics Education**

The practice of statistics dates back to around the mid eighteen century, when it was regarded as "political arithmetic", given its initial sole focus on population and economic data. The practice expanded, and evolved into a scientific and independent discipline, the formalization of which can be attributed to the Royal Statistical Society (RSS, founded in 1834), and subsequently the American Statistical Association (ASA, founded in 1839). Another organization, the International Statistical Institute (ISI) was established in 1885, and it was the founding of the Committee on Statistical Education within the ISI in 1948 that initiated serious and focused dialogue on the training needs of the discipline, and research in statistics education (Ottaviani, 1999; Vere-Jones, 1995). The ISI-Committee on Statistical Education (and its successor, the International Association of Statistical Education – IASE, established in 1991) emerged as the leader in this regard, with a broad international focus, whereas the ASA took the charge in the USA.

For over a decade, the academic community, primarily in the USA, has witnessed reform in undergraduate statistics education, coordinated mainly by the ASA, the Mathematical Association of America (MAA), and the National Science Foundation (NSF). Specifically, there is a well-defined movement focused on reform in the teaching and learning of introductory statistics at the college level (Cobb, 1993; Franklin & Garfield, 2006; Garfield et al., 2002). In particular, the ASA sections on Teaching Statistics in the Health Sciences, and Statistical Education are directly involved in reform efforts and programs. In 2003, The American Statistical Association (ASA) funded the Guidelines for Assessment and Instruction in Statistics Education (GAISE) project, one component of which was the introductory college statistics course. In 2005, the ASA Board of Directors endorsed the final report (and recommendations), which now serves as the blueprint for reform-oriented teaching of introductory statistics (Franklin & Garfield, 2006)

**Targeting Introductory Statistics**

> In response to the question: "If you were to start up a statistics department at a new university, what advice would you give to the new department head", Sir David Cox (the distinguished statistician) said, "first the importance of aiming to make the first course in statistics that students receive of especially high quality and relevance." (Cox, 2004, p. 13)

There is a long-held consensus among statistics educators that introductory statistics should be a general education requirement, and an integral part of the post-secondary curriculum (Cobb, 1992; Hogg, 1992; Moore, 1993). Toward this end, and amidst mounting empirical evidence of students manifesting mathematics anxiety, fear, lack of interest, frustration, and emerging from the introductory course deficient in basic statistical knowledge, and with "no useful skills" (Dallal, 1990, p.266; Gal & Ginsburg, 1994; Garfield & Ahlgren, 1988; Perney & Ravid, 1991), the statistics reform movement was formalized (Moore, 1997). Over the years, the reform movement has recommended a shift in the teaching of introductory statistics from the predominantly mathematical and theoretical approach to a more concept-based pedagogy, aimed at fostering statistical literacy, thinking and reasoning (Chance, 2002; Garfield, 2002; Moore, 1993).

Introductory courses in any discipline are intended to provide students with exposure to the fundamentals of the field, and serve as a basis for pursuing advanced courses in that or related fields. Such introductory courses can influence students' beliefs about, and attitudes toward the discipline, and hence, to a large extent, determine whether they choose to pursue the field or go beyond the first course (Dallal, 1990; Moore, 2007, 2005). Moreover, for the majority of college students, the introductory statistics course will be their only formal exposure to statistics (Moore, 1998). Therefore, according to Macnaughton (1996), "rather than being a worst course, and possibly irrelevant, the introductory statistics course ought to be a friendly introduction to the simplicity, beauty, and truth of the scientific method". Indeed, this recommendation captures the spirit of reform-based teaching of introductory statistics, which is intended to facilitate statistical literacy, thinking and quantitative reasoning through active learning strategies, by emphasizing concepts and their applications rather than calculations, procedures and formulae. Toward this end, David Moore (1997, p.136), a former president of the American Statistical Association, wrote:

> I feel strongly, for example, that statistics is not a subfield of mathematics, and that in consequence, beginning instruction that is primarily mathematical, or even structured according to an underlying mathematical theory, is misguided.

**Reform-Oriented Pedagogy and Statistical Literacy**

In general, the primary objective of reform-oriented teaching and learning of introductory statistics is to facilitate students to become informed consumers of statistical information (Chance, 2002) by addressing conceptual issues about data such as distribution, center, spread, association, uncertainty, randomness and sampling (Garfield, 2002). The reform-oriented (concept-based or constructivist) approach (Erickson, 2001; Rumsey, 2002) to teaching introductory statistics is generally defined and operationalized as a set of related strategies intended to promote statistical literacy by emphasizing concepts and their applications rather than calculations, procedures and formulas. It involves active learning strategies such as projects, group discussions, data collection, hands-on computer data analysis, critiquing of research articles, report writing, oral presentations, and the use of real-life data. Statistical literacy is the desired outcome of reform-oriented teaching and learning strategies, and refers to the ability to understand, critically evaluate, and use statistical information and data-based arguments (Gal, 2000; Garfield, 1999).

Moore (1990) posited that the key to achieving statistical literacy, thinking and reasoning is facilitating students to recognize and appreciate the omnipresence of variation, and understand how such variation is quantified and explained. Additionally, Chance (2002) noted that importance must be given to the context from which the data emerged, and to which the findings will be applied. These guidelines are germane to reform-based pedagogy, which situates introductory statistics in an applied and research context, and according to Iversen (1992, p.37):

> The goal of applied statistics is to help students to form, and think critically about, arguments involving statistics. This construction places statistics further from mathematics and nearer the philosophy of science, critical thinking, practical reasoning and applied epistemology.

Consistent with this approach, Hogg (1991) suggested that statistics at the introductory level should be promoted as a tool of research by addressing the formulation of appropriate questions, effective data collection, interpretation, summarization, and presentation of data with attention to the limitations of statistical inference. Hogg (1991) further observed that "good statistics is not equated with mathematical rigor or purity, but is more closely associated with careful thinking". In this regard, the focus of reform-oriented teaching is on creating active learning environments (Garfield, 1993) which address real world problems with real world data, facilitating learning which is deep and meaningful rather than rote and mechanical. In other words, these strategies allow students to experience the material, and construct meaning (Steinhorst & Keeler, 1995), an approach that is premised on the theory of constructivism (Cobb, 1994; delMas et al., 1998; Von Glasersfeld, 1987).

Reform-based teaching and learning strategies for introductory statistics are relatively innovative, representing a shift from the traditional, predominantly mathematical focus, to an applied, research-oriented, and concept-based approach. Consequently, the implementation of reform-oriented pedagogy may be associated with cognitive and affective conflict on the part of instructors. And consistent with theories of adoption and diffusion of innovations (Hall & Hord, 1987; Rogers, 1995) varying predispositions or attitudes toward reform should be expected among instructors (Ajzen & Madden, 1986; Fishbein & Ajzen, 1975; Jaccard & Blanton, 2004; Zimbardo & Leippe, 1991). Such attitudes may serve as facilitators of or barriers to successful reform (Liljedahl, 2005; Tobin, Tippins, & Gallard; 1994).

**Statistics as a Core Course in the Health and Behavioral Sciences**

Statistics is increasingly becoming a core requirement for most college majors, and especially in the evidence-based disciplines such as health and psychology, it is regarded by some academics as "the single most important course in terms of admittance into graduate schools" (Alder & Vollick, 2000). According to Cobb and Moore (1997, p.801): "Statistics is a methodological discipline. It exists not for itself but rather to offer to other fields of study a coherent set of ideas and tools for dealing with data." In almost every discipline, the ability to critically evaluate research findings (often expressed in statistical jargon) is recognized as an essential core skill (Giesbrecht, 1996), especially for college students interested in becoming practitioners (Buche & Glover, 1988). Consequently, undergraduate students in the health and behavioral sciences are generally required to take an introductory statistics course as a core requirement of their degree program (Albert, 2000). This course may be titled statistics, applied statistics, biostatistics, data analysis, or taught as a component of an epidemiology or quantitative research methods course. The importance and relevance of statistics to the health sciences, in this regard, is reinforced in the following (MCPHS, 2004, p.92):

> Statistics is a core course because it provides tools needed to accurately assess statistical analyses that are reported in both the mass media and scholarly publications. The ability to effectively interpret numerical and graphical statistics is necessary for advanced study in the health professions and it is essential that healthcare professionals demonstrate knowledge of the statistical terminology and methodologies found in the biomedical and professional literature. The formal study of statistics complements the sciences because it also requires that students learn to formulate and test hypotheses and draw appropriate conclusions.

The health and to a lesser extent the behavioral sciences undergraduate programs are unique compared to most other disciplines, in that they produce graduates who serve as licensed or certified practitioners (e.g., nursing, radiologic technology, pharmacy, health education, physical therapy, laboratory technology, counseling, dietetics, nutrition, respiratory therapy, emergency medical technology, occupational therapy, and medical ultrasound). Also, it is not uncommon for health sciences divisions to include programs in medicine, veterinary medicine and dentistry, and which students usually take a similar statistics course as the undergraduate health and behavioral sciences students. Graduates of these health and behavioral sciences undergraduate, and first professional programs are much more likely than their counterparts in other disciplines to be required to use, produce, and communicate statistical information (Buche & Glover, 1988), toward evidence-based decision-making and practice (Belar, 2003). Furthermore, for most of these students, the introductory course may be their only formal exposure to statistics, a realization that has resulted in a call for greater importance to be given to instructional methodology, as this will affect the quality of knowledge and skills acquired (Garfield et al., 2002; Moore, 1988).

This implication for teaching and learning was underscored in the Boyer commission's report on educating undergraduates (Boyer Commission, 1998), which emphasized that our classrooms are in crisis, and what we need, are educators in every discipline, not just subject matter experts. Toward this end, Aukstakalnis and Mott (1996, p.14) noted that: "The great challenge faced by educators in every discipline is to present foreign concepts to students in forms which achieve the greatest measure of clarity and understanding." Additionally, Batanero et al. (1994, p.1) cautioned that "many teachers need to increase their knowledge of both the subject matter of statistics and appropriate ways to teach the subject". This is particularly necessary given that "statistics has its own substance, its own distinctive concepts and modes of reasoning" (Moore, 1997). Furthermore, several researchers and educators have noted that statistical knowledge appears to necessitate rules and ideas that to many are counterintuitive, and therefore difficult to understand (Broers, 2001; Garfield, 1995; Tversky & Kahneman, 1974; Wild & Pfannkuch, 1999).

**Statistics and Evidence-Based Practice**

The growing importance of statistics to the health and behavioral sciences is largely linked to the emergence of evidence-based practice (EBP) which is defined as "the conscientious, explicit, and judicious use of current best evidence in making decisions about the care of individual patients" (Sackett et al., 1996, p.71). EBP requires that practitioners are able to identify, access, and critically evaluate relevant research evidence for reliability, validity, applicability, and overall quality, toward optimum patient care. Such appraisal and use of data necessitate statistical competence (Cox, 1997). Underlying and facilitating EBP is the availability of, and greater accessibility to increasingly large amounts of research data.

The proliferation of research data in the health and behavioral sciences can be attributed to (among other factors) the shift in the epidemiological trend, from primarily infectious diseases to predominantly chronic non-communicable diseases, such as diabetes, heart disease, and mental health illness (De Flora et al., 2005; McQueen, 2007). Whereas the infectious diseases have defined and established causal pathways involving specific biological agents, the chronic non-communicable diseases present the challenge of multiple causality, as reflected in the biopsychosocial model of disease (Engel, 1977). That is, there is generally no definite causal agent, but implication of multiple risk factors (including physical, biological, psychological, cultural, behavioral, and environmental) requiring scientific research and data analysis of different designs, and varying levels of complexity. Another stimulus for increased research activity and data production is the expanded definition of health as "a state of complete physical, mental and social well-being and not merely the absence of disease or infirmity" (WHO, 1978, 1986).

Specifically, these developments created a greater need for, and gave more importance to biomedical, behavioral and bio-behavioral research (NIH, 2003), which resulted in a broad spectrum of data of varying quality and complexity, and a challenge to decision-making, in the context of patient care.

In addition to the changing epidemiological trend, and the expanded definition of health, the importance of statistical literacy was heightened by the paradigm shift in patient care associated with the primary healthcare approach (WHO, 1978). Primary healthcare is intended to achieve a basic level of healthcare for all, and emphasizes multidisciplinary team-work rather than the traditional physician-centered model. This approach places responsibility and accountability for patient care on all healthcare practitioners, and not only the physician. This new model of patient care demanded change in the health and behavioral sciences curricula, including facilitating students' understanding of statistics and research toward participative and effective decision making, and evidence-based practice. The primary implication of the team-work patient care paradigm is that health and behavioral sciences practitioners must now be equipped with statistical knowledge and skills to enable them to critically evaluate research evidence for reliability, validity, and applicability toward effective decision-making, in the context of patient care (Sackett et al., 1996).

## Statement of the Problem

In spite of more than a decade of active reform in the teaching of introductory statistics, focused on content, pedagogy, assessment, and integration of technology (Moore, 1993, 2005; Garfield et al., 2000; Garfield et al., 2002), there is a growing body of evidence indicating that the achievement of desired learning outcomes is still lacking (Butler, 1998; Chance, delMas, & Garfield, 2004; Delmas, Garfield, Ooms, & Chance, 2006; Garfield, 2003; Garfield, delMas, & Chance, 1999; Hodgson, 1996; O'Connell, 2002; Saldanha & Thompson, 2001; Schafer & Ramsey, 2003; Schield, 2000; Verkoeijen et al., 2002). In particular, research reports indicate that students are experiencing difficulties with introductory statistics, and emerge with a lack of understanding of core concepts, such as statistical variation (delMas & Liu, 2005), chance (Garfield, 2003), graphical representations of distributions (Bakker & Gravemeijer, 2004), sampling variation (Reading & Shaughnessy, 2004), and sampling distributions (Saldanha & Thompson, 2001).

Specific barriers to effective learning have also been documented. Onwuegbuzi and Leech (2003) observed that many students experience high levels of anxiety about statistics, which can pose "a major threat to the attainment of their degrees". And according to Gal and Ginsberg (1994) a considerable proportion of students do not regard statistics as a relevant or important component of their degree programs. Furthermore, and paradoxically, Delmas, Garfield, Ooms and Chance (2006, p.3), have reported that "many students can still have conceptual difficulties even after the use of innovative instructional approaches and software". Indeed, the most recent comprehensive report on the impact of reform, noted a high level of reported use of active learning strategies by instructors (Garfield, 2000). This discord raises questions about instructors' understanding, acceptance and effective use of these pedagogical strategies, as well as their actual effectiveness.

In general, these negative outcomes have been attributed to passive teaching and learning strategies with a mathematical focus (Garfield, et al., 2002), geared toward rote memorization and surface learning instead of deep and conceptual understanding. In the words of Moore (2005, p.205), there is "an overemphasis on the mathematical aspects of the subject at the expense of experience with data". Specifically, leading educators and researchers have been consistent in linking these difficulties primarily to the preparation and training of instructors, in particular, mathematics (versus statistics) background, and the lack of data and research experience (Cobb & Moore, 1997; Garfield et al., 2002; Moore, 1993). In this regard, and with reference to psychology, Haller and Krauss (2002, p.17) concluded that:

> Teaching statistics to psychology students should not only consist of teaching calculations, procedures and formulas, but should focus much more on statistical thinking and understanding of the methods. Although it can be assumed that this is an implicit aim of lectures on statistics, our survey suggests that these efforts are not sufficient. Since *thinking* as well as *understanding* are genuine matters of scientific psychology, it is astounding that these issues have been largely neglected in the methodology instruction of psychology students.

A major consequence of ineffective teaching and learning strategies is underscored by Hammersley (2004, p.4) who in referring to nurses, stated that their "lack of basic skills in statistics can lead to a situation he refers to as "the mystery of the invisible mid-section". He notes that this common occurrence describes the process of reading the introduction to an article, passing over the method and analysis section with glazed eyes, and then reading the conclusion." This spells great concern for scientific progress (Glencross & Binyavanga, 1996) and patient care, and can exacerbate the existing dearth of healthcare researchers (NIH, 1991; U.S. Senate, 2004).

In general, reform programs are lacking, albeit there are extensive empirical guidelines on curricular materials and strategies for effective teaching. Amidst this situation, evidence is emerging (primarily from other, but related disciplines) that instructors' beliefs about, and attitudes toward reform-oriented teaching and learning strategies are pivotal to the adoption and effective application of this approach (Czerniak & Lumpe, 1995, 1996; Czerniak, Lumpe, & Haney, 1999; Garfield, 1995; Garfield, 2002; Henderson, 2002; Lumpe, Haney, & Czerniak, 2000). More recently, Liljedahl (2005, p.1) noted that:

> Too often, however, the emphasis within teacher education programs is placed on the infusion of content knowledge, pedagogy, and pedagogical content knowledge, with only a cursory treatment of the beliefs that, for better or for worse, will govern the eventual application of what has been acquired within these programs.

Fishbein and Ajzen (1975; Ajzen & Fishbein 1980) have outlined a pathway showing that the combined evaluation of multiple beliefs about an object (in this case, reform-oriented teaching and learning) results in an attitude toward that object. Each belief is associated with an attribute or dimension of the object, and can be evaluated as negative, positive or neutral. Beliefs and concerns about *self*, *task* and *impact* have been implicated in shaping attitudes toward reform (Hall & Hord, 1987) and innovation. According to Allport (1935) and subsequent research, attitude is a mental state of readiness and hence a disposition or intent which influences behavior. The importance of examining instructors' beliefs in this context is further underscored by Tobin, Tippins and Gallard (1994, p.64) in their major review of science education research:

> Future research should seek to enhance our understanding of the relationships between teacher beliefs and science education reform. Many of the reform attempts of the past have ignored the role of teacher beliefs in sustaining the status quo. The studies reviewed in this section suggest that teacher beliefs are a critical ingredient in the factors that determine what happens in classrooms.

Specifically, instructors' beliefs about, and attitudes toward technology, teaching, and learning have been reported as significant in shaping and predicting teaching behavior and learning outcomes in the disciplines of mathematics, science and physics, amidst reform (Henderson, 2002; Janiak, 1997; Liljedahl, 2005; Mcnamara et al., 2002; Tiberghien, Jossem, & Barojas, 1997,1998; Tobin, Tippins, & Gallard 1994). The accessible research literature on statistics education reflects a paucity of empirical data, in this regard. Based on case study data from fourteen (14) instructors of introductory statistics in the USA, Garfield (2000, p.7) reported that "the nature and extent of the implementation" of reform recommendations were "often due to the instructor's experience and beliefs about teaching". A related observation was noted by Schau, Dauphinee and Vecchio (1992) in a small-scale informal qualitative study which identified teacher disposition as a general theme in students' explanations of their feelings regarding mathematics and statistics. Besides these two basic qualitative research reports, instructors' beliefs about, and attitude toward teaching and learning remain virtually unexplored among college instructors of introductory statistics.

## Study Objective

The objective of this study was to develop and preliminarily validate a scale for measuring instructors' attitude (cognitive, affective and behavioral intent components) toward reform-oriented or concept-based teaching and learning of introductory statistics in the health and behavioral sciences at the undergraduate level. The structure and dimensionality of the scale were examined, and its psychometric properties were assessed with regard to reliability (internal consistency) and construct validity (with reference to teaching practice). Selected personal and sociodemographic variables (age, gender, ethnicity, geographic location, teaching experience, highest academic degree, degree concentration, teaching area, membership status in professional organizations, and employment status) were examined as possible correlates and predictors of attitude.

## Significance of the Study (Rationale)

There is strong theoretical and empirical evidence that instructors' beliefs about, and attitudes toward teaching, directly influence the quality of learning (Czerniak & Lumpe, 1995; Henderson, 2002; Liljedahl, 2005; Mcnamara et al., 2002; Tobin, Tippins, & Gallard, 1994). Specifically, unfavorable beliefs and attitudes, in this context, can inhibit the use of effective teaching strategies, and result in negative learning outcomes, which are widespread among students of introductory statistics (Chance, delMas, & Garfield, 2004; Delmas, Garfield, Ooms, & Chance, 2006; O'Connell, 2002; Saldanha & Thompson, 2001; Schafer & Ramsey, 2003; Schield, 2000; Verkoeijen et al., 2002;). However, there is a notable absence of empirical data on statistics instructors' beliefs and attitudes, in this regard. The belief and attitudinal data referenced herein are primarily from the disciplines of physics, mathematics and science. In order to address this gap, this study developed and validated a multidimensional scale to measure instructors' attitudes toward reform-oriented teaching of introductory statistics in the health and behavioral sciences.

This multidimensional scale can be used as a tool for characterizing instructors with regard to specific components of attitude (cognitive, affective and behavioral intent), and for explaining and predicting their teaching practice. In particular, this scale can serve to identify (at the individual level), specific dimensions of attitude which can be targeted for modification or reinforcement (in relation to reform-oriented teaching) through continuing education, and other faculty development programs. Additionally, profiles of subgroups of instructors with reference to attitude and practices, can be established, based on personal and sociodemographic factors, rendering them easily identifiable, and targetable for appropriate professional development programs.

## Conceptual Framework and Theoretical Assumptions

A primary focus of statistics education reform is to facilitate instructors to adopt and maintain effective use of active learning strategies, a relatively innovative pedagogical approach. In this regard, instructors' attitude toward teaching and learning was explored as a proximal antecedent of teaching behavior consistent with the Theory of Reasoned Action (Ajzen & Fishbein, 1980; Fishbein & Ajzen 1975), the Theory of Planned Behavior (Ajzen, 1991), and with attention to change and innovation. However, while attitude in these models is operationalized as purely cognitive (that is, formed solely by beliefs), and separate from intention, this study adopted the tripartite (or three-component) conceptualization (detailed below). Identification of issues and concerns which are likely to arise during the change process, and shape beliefs and attitudes was guided by the "Stages of Concern" component of the Concerns Based Adoption Model (Hall & Hord, 1987). Additionally, the recommendations of the ASA/MAA (American Statistical Association/Mathematical Association of America) joint committee on the teaching of undergraduate statistics (Cobb, 1992) as well as the GAISE (Guidelines for Assessment and Instruction in Statistics Education) report (Franklin & Garfield, 2006; Garfield, 2004) were used to inform the development of a teaching practice scale used for characterizing teaching approach.

**The Theory of Reasoned Action (TRA)**

The Theory of Reasoned Action (TRA) is an expectancy-value model, and the most widely used premise for conceptualizing and operationalizing the attitude-behavior relationship (Ajzen & Fishbein, 1980; Fishbein & Ajzen 1975). This theory posits that attitude is a major determinant of a person's intention to perform the behavior in question. The model assumes rational behavior and details a systematic and largely linear process assuming a "causal" pathway linking beliefs to attitudes, attitudes to intentions, and intentions to behaviors. This assumed "linear progression" from beliefs to behavior has been recognized as a limitation (Kippax & Crawford, 1993) of the TRA especially with regard to behaviors that are not fully under volitional control, and for which, intention to act may not necessarily result in action.

**The Theory of Planned Behavior (TPB)**

The Theory of Planned Behavior (TPB) is also an attitude-intention-behavior model (Ajzen, 1991), which emerged as an extension of the TRA to compensate for its lack of predictability of some behaviors from intention.  In order to address this potential gap, the TPB includes perceived behavioral control (PBC) as an antecedent of intention and behavior. PBC is akin to the concept of perceived self-efficacy (Bandura, 1977), and refers to people's perception of their ability to perform a given behavior (Ajzen, 1991). Specifically, PBC is generally conceptualized as beliefs about one's control over the decision to act. Behavioral control is viewed as a continuum that extends from complete control (no constraints or barriers) to a total lack of control (implementation may require resources or skills which may be lacking), and can be influenced by both internal and external factors. Internal factors include skills, abilities, knowledge, and emotions, and external factors comprise situational and environmental characteristics. Control beliefs may facilitate or impede performance of a behavior (Ajzen, 1991)**.**

Consistent with the expectancy-value model the TPB posits that attitude results from one's beliefs (expectations) that a behavior will result in a particular outcome (its subjective probability) and one's evaluation of that outcome (its subjective utility). Like perceived behavioral control (PBC), subjective norms are considered a critical input in behavioral intentions, and refer to perceived social pressure to engage or not to engage in a behavior. According to Ajzen (2001) each belief associates the behavior with a certain attribute, and it is the subjective summation of these attributes (interacting with the strength of the associations) which results in an attitude. He further argued that a person may possess multiple different beliefs about a behavior, however, it is assumed that only salient or readily accessible beliefs influence attitudes at any given moment. It is now generally accepted that attitude represents a "summary evaluation of a psychological object captured in such attribute dimensions as good-bad, harmful-beneficial, pleasant-unpleasant, and likeable-dislikeable" Ajzen (2001, p.28).

For this study, attitude was conceptualized as an evaluative disposition toward some object, based upon cognitions, affective reactions, and behavioral intentions (Jaccard & Blanton, 2004; Rosenberg & Hovland, 1960; Zimbardo & Leippe, 1991). In other words, attitude is an informed predisposition to respond, and is comprised of three components, namely beliefs, feelings, and a readiness or intent for action (Jaccard & Blanton, 2004; Zimbardo & Leippe, 1991). This attitude model is referred to as the tripartite or three-component theory, and these three facets have consistently been reported as separable components of attitude (Fabrigar et al., 1999).

**The Concerns-Based Adoption Model (CBAM)**

The Concerns-based Adoption Model (CBAM) (Hall & Hord, 1987) allows for a multi-faceted and systematic examination of the change process, especially with regard to teachers. In particular, it focuses on how individuals respond to, and implement innovations, such as curriculum and instructional practices. The CBAM is premised on the "concerns model" a concept put forward by Fuller (1969) who explored pre-service teacher concerns toward innovations and observed a developmental sequence, moving from concerns about *self* to concerns about *task* followed by concerns about *impact*. This sequence of concerns was reproduced by Hall, Wallace and Dosset (1973) among in-service teachers faced with innovations. The original research of Fuller (1969) revealed patterns in teacher concerns that correlated with maturity and teaching experience. These concern domains (self, task and impact) have been expanded and constitute one component (Stages of Concern) of the CBAM (Table 1).

Hall, George and Rutherford (1977, p.5) stated that:

> The composite representation of the feelings, preoccupation, thought, and consideration given to an issue or task is called concern. To be concerned means to be in a mentally aroused state about something. The intensity of the arousal will depend on the person's past experiences and associations with the subject of the arousal.

The authors noted that it is perception that stimulates concern, not the reality. According to Hall (1976), an individual's concerns directly affect performance; and since concern levels are associated with levels of performance, lower level concerns must be removed before higher-level concerns can emerge. "If these early concerns remain intense, then the user is apt to modify the innovation or their use of the innovation, or perhaps discontinue use, in order to reduce the intensity of these concerns" (Hall, George, & Rutherford, 1978, p.13). The CBAM is a conceptual framework that describes, explains, and predicts probable teacher behaviors in the change process. The assumptions underlying the CBAM (Hall & Hord, 1987) are:

- Change is a process, not an event;
- Systemic change requires that individuals change, hence the individual must be the focus if change is to be facilitated;
- Change is a very personal experience;
- Change is developmental, the individual progresses through various stages with specific emotions, beliefs and attitudes relating to the innovation;
- the focus of facilitation should be on individuals, innovations, and the context for change; and
- Change is best understood in operational terms (allowing for a systematic and adaptive process with constant monitoring).

Table 1: The Stages of Concern Component of the CBAM

| Dimension | Stage | Expression of Concern |
|-----------|-------|-----------------------|
| **Impact** | 6. Refocusing | I have some ideas about something that would work even better. |
| | 5. Collaboration | How can I relate what I am doing to what others are doing? |
| | 4. Consequence | How is my use affecting learners? How can I refine it to have more impact? |
| **Task** | 3. Management | I seem to be spending all my time getting resources ready. |
| **Self** | 2. Personal | How will using it affect me? |
| | 1. Informational | I would like to know more about it. |
| | 0. Awareness | I am not concerned about it. |

Reproduced from *Taking Charge of Change*, Shirley M. Hord, William L. Rutherford, Leslie Huling-Austin, and Gene E. Hall. Alexandria, VA: Association for Supervision and Curriculum Development and Southwest Educational Development Laboratory, 1987.

Unlike the popular Diffusion of Innovations Theory (Rogers, 1983, 1995) which considers characteristics of the innovation as the primary determinants of adoption, the CBAM posits that the change process is more successful if the concerns of the teacher are considered. The CBAM comprises of three diagnostic components (subscales) designed to measure "Stages of Concern", "Levels of Use" and "Innovation Configurations". This study will utilize the "Stages of Concern" subscale only, which is widely used independent of the other two subscales (Anderson, 1997; Bailey & Palsha, 1992).

The "Stages of Concern" subscale (Table 1) is a measure of seven different reactions that educators are likely to experience when they are implementing a new program (curriculum and/or instructional practices), and which are characterized as concerns associated with **self**, **task** and **impact**. Anderson (1997) rationalizes Stages of Concern as a framework that describes the feelings and motivations a teacher might have about change at different points in the implementation. At Stage 0 (AWARENESS), the teacher has minimal knowledge about or interest in the innovation and change. At Stage 1 (INFORMATIONAL), the teacher demonstrates interest in learning about the innovation and the implications of its implementation. Concerns at Stage 2 (PERSONAL), manifest as strong feelings and emotions about the teacher's ability to implement change, the relevance of the change, and the personal cost of getting involved. In other words, the teacher performs a cost-benefit analysis. Stage 3 (MANAGEMENT), is reached when the teacher begins to experiment with implementation; at this point, teacher concerns are primarily related to the logistics (including new behaviors) of putting the change into practice. At Stage 4 (CONSEQUENCE), teacher concerns predominantly reflect the impact of the change on students, other evaluation of the innovation, and the teacher's use of it. At Stage 5 (COLLABORATION), teacher concerns are centered around their interest in working with their peers in the school toward learning more about the application of the innovation, as well as promoting the benefits of change implementation. During the change process, teachers may reach Stage 6 (REFOCUSSING). At this stage, their concerns are focused on comprehensive and structured evaluation of the innovation so as to improve it or replace it with a better approach.

Rakes and Casey (2002) caution that teachers' response to innovations does not always conform to this structured and seemingly linear progression. They note that typically, teachers can have concerns at more than one of the Stages of Concern at any given time. While there is much empirical evidence to support the developmental sequence of concerns in the teaching context, some studies have suggested a re-thinking of the seven stages of concern toward obtaining a more parsimonious measure. Redundancy has been observed among the seven stages, specifically suggesting similarity between "awareness and informational concerns" and "refocusing and consequence concerns" (Bailey & Palsha, 1992). With regard to the reliability and validity of the Stages of Concern measure; Hall, George and Rutherford (1986) reported acceptable estimates of internal consistency (Cronbach's alpha ranging from .64 to .83) with six of the seven coefficients being above .70. They note that multiple validity studies were conducted, all of which supported that the Stages of Concern Questionnaire measures the hypothesized Stages of Concern. Their studies were conducted among teachers. While reliability and construct validity have been established in different settings for the individual components of the CBAM, Anderson (1997) questions the applicability of the model to current innovations which he classifies as organizationally focused initiatives (e.g. "outcomes-based education" and "effective schools") versus the contemporary strategies which led to the development of the model, which he classifies as discrete innovations in curriculum and instruction, innovations that were intended to address selected curriculum areas and classrooms.

# CHAPTER 2

## LITERATURE REVIEW

> The concept of attitudes is probably the most distinctive and indispensable concept in contemporary American social psychology. (Allport, 1935, p. 798)

So wrote Gordon Allport (a Harvard University Psychologist) in 1935. Undoubtedly, his research and commentaries on this phenomenon, are among the seminal works on attitude theory. Allport (1935, p. 810) defined attitude as "a mental state of readiness which exerts a directing influence upon the individual's response to all objects and situations with which it is related". Attitude measurement continues to be a core focus of social psychology, and is emerging as an important dimension of teaching and learning. The scientific research on this construct, centers around its structure, function, measurement, and the attitude-behavior relationship (Ajzen, 2001; Dawes & Smith, 1985).

### Attitude Functions

According to attitude functions theory, attitudes are formed, maintained, and changed in order to satisfy personal needs, and achieve psychological benefits. Accordingly, we possess attitudes for different reasons (O'Keefe, 2002). Katz's (1960) typology of attitude functions which expands on an earlier taxonomy put forward by Smith, et al. (1956) is the generally accepted framework for understanding the motivational underpinnings and functions of attitudes (Abelson & Prentice, 1989; Julka & Marsh, 2000). Such knowledge is essential for elucidating the attitude-behavior relationship, and facilitating change. Katz (1960) posited that a given attitude held by any individual will serve one or more of the following four personality functions (see also Daugherty et al., 2005).

1. Utilitarian (Adjustment): Where the attitude serves to achieve personal gains and avoid negative consequences. In particular, this function represents attitudes based on self-interest.
2. Ego-defensive: Where the attitude is motivated by the need to protect people from internal insecurities or external threats. The attitude object may be viewed as an extension and expression of self (ego). Under such motivation, individuals would embrace the attitude object in order to minimize their own self-doubts, and experience a sense of belonging or strength.
3. Value-expressive: Where the attitude functions as a means of expressing or projecting strongly held personal values. Individuals motivated by this function may embrace the attitude object to achieve status, recognition, and visibility through membership. Increased self-esteem and social support, or a reduction in anxiety, are common psychological benefits.
4. Knowledge (object-appraisal): Where the attitude is motivated by need to gain and organize information in order to better understand, adapt, orient to the environment, and make it more predictable.

Evidently, these functions are neither mutually exclusive nor dichotomous (either present or absent), but may exist on a continuum, and interact, providing the motivational basis of attitude expression. Also, and particularly with reference to the ego-defensive function, an individual's expressed attitude may not reflect his or her true disposition.

**Defining and Measuring Attitude**

Attitude represents a summary evaluation of a psychological object captured in such attribute dimensions as good-bad, harmful-beneficial, pleasant-unpleasant, and likeable-dislikeable. (Ajzen, 2001, p.28)

The abovementioned is a commonly used definition of attitude, which is generally interpreted as a disposition to behave or act in a certain way (Jaccard & Blanton, 2004). Nonetheless, consensus is lacking among researchers and practitioners on to how to define, operationalize and measure this construct (attitude), a phenomenon which is not directly observable, but is rationalized based on manifest behavior (Ajzen, 2001; Dawes & Smith, 1985). Ajzen (1991, 2001) posited that attitude results from one's beliefs (expectations) that a behavior will produce a particular outcome (its subjective probability) and one's evaluation of that outcome (its subjective utility). While this conceptualization emphasizes belief structures as fundamental to attitude, Thompson (1992) argued for a more inclusive underpinning, that is, conception.

In his pioneering research among mathematics teachers, Thompson (1992) defined conception as "a general mental structure, encompassing beliefs, meanings, concepts, propositions, rules, mental images, preferences, and the like" (p. 130). Accordingly, conceptions (rather than mere beliefs) can be viewed as a more meaningful and comprehensive representation of the set of perceptual positions that an instructor may possess about teaching and learning (Contreras, 1998). Such conceptions are considered central to attitude formation, and hence constitute a major determinant of teaching practice (Brown & Cooney, 1982). Numerous studies guided by the Concerns-Based Adoption Model (Hall & Hord, 1987) have established that in the context of an innovation (such as reform-oriented pedagogy), teachers are likely to possess concerns (or conceptions) about *self*, *task* and *impact*, which can influence the formation of beliefs and attitudes.

**Selected Methodological Considerations of Attitude Research**

Empirical reports of a weak statistical relationship between attitude and behavior continue to raise questions about the explanatory and predictive value of attitude in relation to behavior, and have resulted in greater attention to the methodological underpinnings of attitude-behavior measures (Eagley, 1992; Ogden, 2003). In particular, context (Brown & Cooney 1982) and attitude strength (Ajzen, 2001) have emerged as key factors in establishing reliable and valid attitude scales. McConnell et al. (1997) reported that individuals have been shown to possess different evaluations of the same object based on the context in which it is situated or observed. And strong attitudes have been observed to be relatively stable over time, resistant to persuasion, and to predict manifest behavior (Ajzen, 2001). According to Visser and Krosnick (1998), a major determinant of attitude strength is the accessibility of underlying beliefs, which is considered a function of their salience or importance (to the individual), their valence (negative or positive), and the context. The authors further noted that attitude strength can vary by age, education, gender and race. Moreover, based on a methodological framework put forward by Crano and Messe (1982), attitude can more reliably and meaningfully explain and predict behavior if attention is given to the following during research:

1. **Specificity**: the attitude measure must be specific to the behavior with reference to target (who), action (what), context (where), and time (when). This is also known as the principle of correspondence or compatibility (Ajzen & Fishbein, 1977).
2. **Intentionality**: include some measure of the respondent's intention to adopt the behavior of interest.
3. **Ambiguity**: refers to respondent's personal experience and knowledge of the attitude object or elements thereof. Personal knowledge and experience, in this regard, reduces ambiguity and increases the likelihood that the expression of attitude will be consistent with behavior.
4. **Level**: refers to the centrality or salience of beliefs. Attitudes based on salient rather than peripheral beliefs are more likely to predict behavior.
5. **Vested interest**: the extent to which the respondent perceives the attitude to impact his her well-being.
6. **Self-monitoring**: refers to the extent to which respondent values feedback from the social context in evaluating the acceptability and appropriateness of his/her actions.

Incongruence with regard to instructors' pedagogical knowledge, beliefs, attitudes, and practices (Prosser & Trigwell, 1999) has been attributed primarily to the nature of these constructs. Specifically, beliefs and attitudes are viewed as dynamic and permeable mental structures, which influence, and can be influenced by practice (Muijs & Reynolds, 2001; Thompson et al., 1992). This two-way relationship may lead to temporal ambiguity, that is, lack of clarity as to whether the belief or attitude preceded the practice or vice versa. Additionally, given the varying functions of attitudes, counterintuitive and inconsistent findings can be expected, especially if cognitive dissonance occurs (Bender et al., 1991; Festinger, 1957), and the individual modifies his or her belief(s) or action(s) in order to rationalize the experience.

**Attitude Structure**

Another, and indeed critical aspect of attitude measurement is the structure of this construct, and this has been configured differently over the years. The various conceptualizations of attitude are; **one-component**, that is, purely affective (feelings) (Thurstone, 1931; Allport, 1935) or purely cognitive (beliefs) (Fishbein & Ajzen, 1975; Ajzen & Fishbein, 1980), **two-component**, detailed as cognitive and affective (Crites, Fabrigar, & Petty, 1994; Trafimow & Sheeran, 1998), and the **tripartite or three-component** framework. According to the tripartite model (Breckler, 1984; Rosenberg & Hovland, 1960; Smith, 1947), attitudes have affective, cognitive, and behavioral intent (intention to act) dimensions, and these have repeatedly been reported as separable components of attitude structure (Fabrigar et al., 1999). Consistent with the tripartite model, a practical and operational definition of attitude is given as "relatively lasting clusters of feelings, beliefs, and behavioral tendencies directed towards specific persons, ideas, objects or groups" (Baron & Byrne, 1984, p. 126). This is depicted in the figure below.

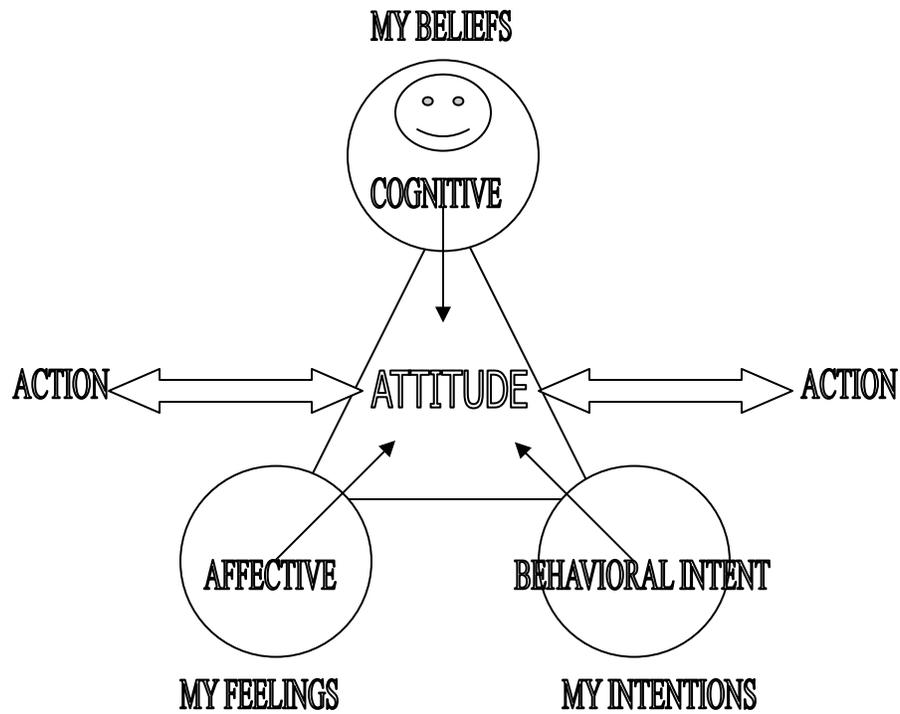

**The Tripartite Attitude Model**

Whereas cognition and affect have been reported as separate components of attitude, there exists a body of literature suggesting that these two components are not mutually exclusive (Ajzen, 2001; Dickerson, 1993). As Piaget noted (cited in Clark & Fiske, 1982, p. 130), "at no level, at no state, even in the adult, can we find a behavior or a state which is purely cognitive without affect nor a purely affective state without a cognitive element involved". Apparent support for this position is generally evidenced by a moderate to strong statistical relationship between cognition and affect (Estrada & Batanero, 2004) or a cross-clustering of items from scales intended to measure affect and cognition separately (Cho, Kim, & Choi, 2003). Adding to this debate, is the conjecture that both affect and cognition may be hierarchical structures (Krathwohl, 1964; Mcleod, 1991).

Other attitude research focused on affect and cognition as separate components of attitude, have shown that "evaluations of emotions and feelings that are associated with an attitude object are more accessible in memory [requiring a shorter response time] than evaluations of thoughts about the object" (Verplanken, 1998, p.32). The authors noted that this differential accessibility may be explained by the primary and less complex nature of affect (feelings) compared to cognition-based evaluations (beliefs), which are more involved. This finding is akin to that of Edwards and von Hippel (1995), who reported that affect-based attitudes tended to be expressed with greater confidence than cognition-based attitudes, an observation that can have implications for the relative weighting of affect and cognition in contributing to attitude expression.

**Approaches to Collecting and Analyzing Data for Attitude Measurement**

In general, two different but complementary approaches have been used for obtaining data in order to test the **one**, **two** and **three-component** conceptualizations of attitude. The more common method utilizes magnitude techniques (primarily rating scales), whereby the attitudinal dimensions and items are detailed (a priori), and presented to the subjects for rating (Dawes & Smith, 1985; Rosenthal & Rosnow, 1991). This approach, albeit popular, has been criticized for presupposing specific dimensions about the subject's perception of the psychological object. While the respondent is likely to rate these as instructed, they may have little or no importance or salience to the respondent (Dawes & Smith, 1985), and therefore threaten the reliability and validity of the resulting measure or scale. In contrast, the other approach involves proximity techniques, whereby the subject is required to assess the psychological proximity (similarity) of pairs of stimuli. The perceptual dimensions used by the subject are then inferred from a map of distances (which represent the judgment of the subject about similarity). As these dimensions are not predetermined, they may be more meaningful (salient, reliable and predictive) than those resulting from the use of magnitude techniques (Dawes & Smith, 1985).

The data obtained from both approaches are usually analyzed using selected data reduction or classification techniques, in order to identify underlying concepts or dimensions. This allows for explaining the phenomenon of interest, and assessing the psychometric properties of the measure or scale of interest (Rosenthal & Rosnow, 1991). For magnitude techniques (rating scales), the commonly used statistical methods include principal components analysis (PCA), factor analysis (FA), and cluster analysis. By far, PCA and FA are more often used, and both assume a linear relationship among the items (Fabrigar, 1999). More recently, researchers have been augmenting these data analytic techniques with multidimensional scaling (MDS) so as to explore both the metric (linear) and non-metric (ordinal) properties of the data. This combined analytic approach allows for a more formative, comprehensive, and pragmatic assessment, and produces spatial maps which make for easier identification, and interpretation of the dimensions underlying the data.

For similarity-dissimilarity measures (relationship among objects or stimuli), the primary statistical techniques are referred to as multidimensional scaling (MDS) (Coombs, Dawes, & Tversky, 1970; Torgerson, 1958; Rosenthal & Rosnow, 1991; Young & Hamer, 1987). MDS (metric and non-metric models) results in both numerical and graphical representations of the data. Specifically, unlike PCA and FA, the assumption of linearity need not be imposed on the data, and attributes and dimensions of similarity do not have to be stated a priori. MDS has the potential to facilitate the identification of truly salient dimensions of judgments about similarity-dissimilarity (Kruskal & Wish, 1978; Coxon, 1982; Young, 1987)

There is much evidence that PCA is being used inappropriately to develop and validate measurement scales. Whereas both PCA and FA are data reduction techniques (Hair et al., 2006), with PCA, all sources of variance (unique, shared, and error) are analyzed for each measured item. In other words, total variance is analyzed. With FA techniques, only shared or common variance is analyzed. That is, covariance is examined. Widaman (1993) noted that "principal component analysis should not be used if a researcher wishes to obtain parameters reflecting latent constructs or factors" which is generally the goal when a new scale is being developed or an existing scale is being validated. Rather, FA techniques should be used. Given the same data, an "attractive" outcome of PCA over FA is that for a given set of components or factors, PCA is inclined to explain a greater proportion of variance, and hence give more "favorable" results (albeit misleading if the objective is to identify latent constructs or the common factor structure).

**Beliefs and Attitudes as Predictors of Teaching Effectiveness**

Reform initiatives, particularly in the teaching of statistics, mathematics and science have given new meaning, and greater importance to understanding the relationship between teaching practice and student performance (Garfield et al., 2002; Le et al., 2004; Muijs & Reynolds, 2001). Over the years, numerous variables and constructs have been explored and proposed as important in explaining and accounting for student performance. Initially, the research focus was on teacher personality, which yielded no meaningful and consistent predictive value for student performance. (Borich, 1996). Subsequent research was guided by the systems view of teaching and learning (Biggs, 1989; Le et al., 2004; Ramsden 1992), with attention to ***presage, process,*** and ***product factors***, as well as their interaction.

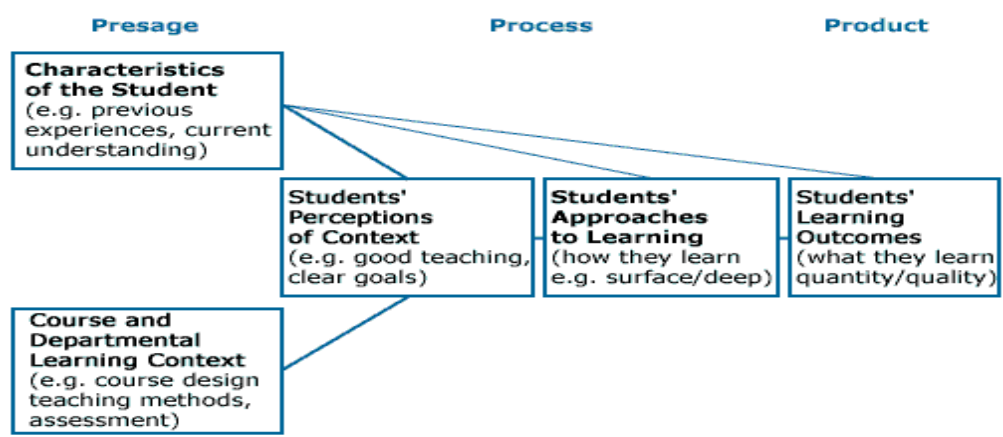

**Presage, Process, and Product Factors (Biggs, 1989)**

This model has been operationalized with a bias toward student characteristics (cognitive and affective). And its failure to reliably explain and predict student learning outcomes has been attributed to the narrow focus on teacher characteristics, that is teaching methods only. According to Muijs and Reynolds (2001), there has been much criticism leveled at researchers for their earlier preoccupation with teachers' behaviors as predictors of effective teaching and student performance. They noted that "the deeper structures [beliefs and attitudes] are more important to teaching quality than immediately observable behaviors" (p.4).

Instructors' beliefs about, and attitudes toward teaching and learning have been researched in many disciplines (primarily physics, mathematics, statistics and science) and reported as significant factors in explaining teaching behavior, amidst reform (Cho, Kim, & Choi, 2003; Henderson, 2002; Janiak, 1997; Koballa & Crawley, 1985). Teachers' attitudes are not "simply an affective factor" that contributes to teaching, but a critical element in determining the quality of pedagogy (Cho, Kim, & Choi, 2003; Tobin, Tippins, & Gallard, 1994). Above all, it has been reported that affective, belief, and attitudinal traits manifested by prospective teachers may better explain instructional potential than conventional academic variables (Bascom, Rush, & Michel, 1994; Schmitz & Lucas, 1990).

Accordingly, Askew et al. (1997; Prosser & Trigwell, 1999), have shown that belief systems can reliably and plausibly differentiate between behaviorist teachers (focused on transmission of knowledge), and those who conform to the constructivist approach (consistent with reform-oriented or concept-based teaching). A similar dichotomous teaching classification is supported by Handal (2003). Based on factor analytical modeling, Handal (2003) suggested two clusters of beliefs, which characterized the teaching of mathematics as thematic (concept-based, constructivist or reform-oriented) versus behaviorist (focused on knowledge transmission).

This constructivist-behaviorist (high-low reform) teaching practice classification is well-established. Specifically, according to Trigwell and Prosser (2004), the "Approaches to Teaching Inventory" scale can discriminate between teaching practices, as either information transfer (teacher-focused, traditional or behaviorist), or conceptual change (student-focused, constructivist or reform-oriented). However, the authors emphasized that teacher behavior is context dependent, that is, teachers who adopt one approach in one context may not adopt the same one in a different context. Additional support for this classification is evidenced by Le et al. (2004), who, using multidimensional scaling techniques, grouped mathematics and science teachers as either high-reform (emphasizes conceptual understanding, inquiry, application, and communication of mathematical or scientific ideas), or low-reform (focuses on acquisition of discrete skills and factual knowledge).

The validity of this high-low reform teaching classification is also empirically supported. In particular, instructors' level and frequency of use of core reform strategies, such as discussion, group-work (Le et al., 2004), and critical thinking (Haas & Keeley, 1998; Keyser and Broadbear, 1999) are known to be strongly, and positively correlated with student achievement (specifically deep and conceptual learning). In this regard, Le et al. (2004) cautioned that weak associations may be attributed to measurement quality, specifically the set of indicators of reform instruction used in the study. The use of cross-sectional studies (surveys) has also been implicated as a threat to the validity of this teaching classification. However based on comparative analyses of survey data, direct observations and logs, Burstein et al. (1995; Meyer, 1999) concluded that surveys can reliably differentiate between "low-reform and high-reform teachers", but noted that surveys may be less sensitive to the finer gradations in frequency of reform practices, and implementation.

**Selected Measures of Faculty Attitudes (Related to Reform)**

A multitude of studies (in the teaching and learning context), aimed at developing and validating attitude scales have been conducted over the past decade. The target groups have been primarily students (Gal, Ginsburg, & Schau, 1997; Schau, 2003), pre-service and elementary school teachers (Bascom, Rush, & Machell, 1994; Maney,

Monthley, & Carner, 2000; Christensen, 2002; Cho, Kim, & Choi, 2003; Koszalka, Prichavudhi, & Grabowski, 2000; Troutman, 1991), and to a much lesser extent, college and university faculty. At the college level, researchers have focused primarily on faculty attitudes toward computers, internet use, and other aspects of information technology (Christensen & Knezek, 1998; Gilmore, 1998). Undoubtedly, computer use is a critical component of reform-based teaching of statistics, and teachers' attitudes have consistently been reported as significant predictors of computer use (Yuen et al., 1999). In particular, Mathew (2001) noted that faculty attitudes toward computer-based instruction were significantly different between faculty adopters (positive) and laggards (negative) of computer-based instruction.

Other studies have examined barriers to the teaching of critical thinking skills (another important element of reform-based teaching of statistics). While faculty deficiency in this area has been attributed to lack of exposure to pedagogical styles and assessment methods, large class sizes, and lack of support and incentives (Haas & Keeley, 1998; Hannah & English, 1999), major studies have concluded that teachers' beliefs and attitudes are more important, in this regard. For example, Trigwell (1995; Kember, 1997) found a correlation between teachers' conception of teaching, and classroom practice. They observed that teachers who view instruction primarily as the transmission of knowledge tend to lecture (behaviorist practice), whereas those who aim to facilitate students to think, pursue and embrace techniques that promote deeper learning (reform-oriented or constructivist practice). Kember (1997, p.273) summarized this as follows:

> An understanding of teaching conceptions … becomes important if measures to enhance the quality of teaching are to have any impact. If teaching approaches are strongly influenced by the underlying beliefs of the teacher, quality assurance measures should take into account conceptions rather than concentrate exclusively upon approaches.

**Approaches to Measuring Instructors' Attitudes**

The recent literature on scales for measuring faculty attitudes, reflects a bias toward use of the two-component conceptualization (affect and cognition), and each component or domain is usually operationalized as a multi-item scale. Commonly used measures of affect (feelings) in the teaching and learning context, include concern, anxiety, comfort, confidence, enthusiasm, enjoyment, interest, fear, acceptance, aversion, avoidance, liking, and reluctance (Anastasiadou, 2002; Christensen, 1998, 2002; Christensen & Knezek, 1999; Estrada & Batanero, 2004; Fabrigar et al., 1999; Gilmore, 1998; Jaccard & Blanton, 2004; Perez, Luquis, & Allison, 2004; Race, 2001; Winstead, 1996). On the other hand, cognition in this context, is generally represented by beliefs about difficulty, usefulness, benefits, value, utility, relevance, understanding, competence, self-efficacy, behavioral control, importance, effort, and preparedness (Anastasiadou, 2002; Bandura, 1977; Cho, Kim, & Choi, 2003; Estrada & Batanero, 2004; Fabrigar et al., 1999; Jaccard & Blanton, 2004; Winstead, 1996). However, these beliefs are sometimes not given the cognitive label by researchers, a practice which contributes to research ambiguity. Additionally, there is now increasing interest in the effect of teachers' epistemological beliefs on instruction (Hofer, 2002).

In addition to cognition and affect, there is an emerging consensus that intentionality should be an integral component of the attitude construct, as it has been shown to increase predictability and consistency of attitude measures in relation to behavior (Estrada & Batanero, 2004; Maney, Monthley, & Carner, 2000; Perez, Luquis,

& Allison, 2004). Intentionality has been operationalized both as multiple items in a subscale (Koszalka, 2001; Yuen & Ma, 2001**)**, and as a single item (Cho, Kim, & Choi, 2003). Each item usually contains a stem connoting "I plan to…", or "I will…". Some researchers have also included perceived behavioral control (PBC) (often linked to perceived self-efficacy) as a component of attitude (Yuen & Ma, 2001), albeit, theory supports PBC more as a moderator or mediator of the attitude-behavior relationship (Ajzen, 2001). Inclusion of PBC can be rationalized based on its established meaningful relationship with perceived difficulty. However, major empirical research on this issue (Trafimow et al., 2002) supports a distinction between perceived behavioral control and perceived difficulty, and suggests that perceived difficulty is a better predictor of most behavioral intentions and behaviors than is perceived behavioral control. Notably, perceived behavioral control, perceived self-efficacy, and perceived difficulty are specific beliefs, and hence cognitive in nature.

**Personal and Sociodemographic Factors as Determinants of Faculty Attitude**

In general, faculty attitude appears to operate independent of selected personal and sociodemographic characteristics, especially with regard to certain aspects of reform-oriented pedagogy. For example, according to Mathew (2001), academic discipline, rank, gender and age, were non-significant with respect to attitudes of faculty adopters and laggards of computer use in instruction. In another study which examined allied health instructor's attitude toward faculty development issues, no significant difference was found based on academic preparation (White, 1979). Furthermore, Masse and Popovich (1998) noted that faculty attitudes toward the teaching of writing, differentiated between two groups of teachers, "product" (rigid, traditional, and behaviorist) versus "process" (flexible, creative, and constructivist-oriented), however, age and length of experience were unrelated to attitude. It is noteworthy, however, that according to Visser and Krosnick (1998), attitude strength (commonly defined as the degree to which an attitude is stable, accessible, and resistant to change) can vary by age, education, gender and race.

**Theoretical Foundations of Reform-Oriented (or Concept-Based) Teaching**

There is nothing more practical than a good theory. (Lewin, 1951, p.169)

The conceptual underpinning of reform-based teaching of introductory statistics is generally attributed to constructivism (Cobb, 1994; delMas et al., 1998; Von Glasersfeld, 1987), which can be considered a family of concepts and principles (about the construction of knowledge and meaning, and hence teaching and learning). In general, this philosophy defines knowledge as being "temporary, developmental, socially and culturally mediated, and thus, non-objective" (Brooks & Brooks, 1993, p. vii). In other words, constructivism provides an ontological and epistemological model for exploring and understanding the teaching-learning process. There are two recognized forms of constructivism; cognitive or Piagetian constructivism (Piaget, 1921, 1929, 1932), and social or Vygotskian constructivism (Vygotsky, 1962, 1978). Piagetian constructivism emphasizes the mind of the individual, and views learning as simply the assimilation and accommodation of new knowledge by learners, that is, merely a process of adjusting our mental models to accommodate new experiences. On the other hand, social constructivism focuses on learning as a function of social interaction, but is considered a variant of cognitive constructivism. It is social constructivism that is more relevant to reform-based teaching.

Social or Vygotskian constructivism is aimed at social transformation and emphasizes the socio-cultural context in which the individual or student is situated. It posits that individual meaning and understanding result from social or group interactions (primarily through collaboration and negotiation). Individuals construct knowledge through a mutually beneficial relationship with the environment. Vygotsky posited that all learning takes place in the 'zone of proximal development', which he defined as the difference between what a learner can do alone, and what he or she can do with assistance.

Social constructivism is sometimes labeled as activity theory (Leont'ev, 1972) and embodies situated or anchored learning (Open Learning Technology Corporation, 1996). These approaches emphasize cognitive apprenticeship (Singer & Willett, 1993) through authentic activities, encompassing projects and other tasks, which model discipline-specific real-world activities, including practitioner behaviors. The underlying goal is to facilitate discovery-based learning (Bruner, 1973, 1983, 1990)

Instructional design based on constructivism is generally contrasted with instruction based on behaviorism, which is described as a rigid procedural approach intended to use fixed stimuli and selected reinforcements to promote a fixed world of objective knowledge measured primarily in terms of observable behavior (Caprio, 1994; Skinner, 1974). It focuses on discrete and compartmentalized knowledge and skills rather than integration of knowledge and conceptual understanding. The key difference between these two approaches is that behaviorism is centered around mere transmission of knowledge from the instructor to the student (passive student and a top-down approach) whereas constructivism is focused on the construction of knowledge by the student (active student and a bottom-up approach). According to Askew et al. (1997) highly effective teachers possess constructivist or connectionist beliefs rather than a transmission orientation (behaviorist beliefs).

In the constructivist context, the instructor serves as a coach to facilitate students to explore information, discover concepts and construct knowledge and meaning. This allows for the development of deep and conceptual understanding, that is, the ability to know "what to do and why" (Skemp, 1987, p.9) rather than surface knowledge (from rote learning associated with behaviorist pedagogy). The instructor should utilize strategies such as asking probing and open-ended questions during discussion, as well as authentic activities (Barrows & Tamblyn, 1980; Boyle, 1999) toward promoting cognitive challenge as a stimulus for thinking and reasoning, and hence meaningful learning.

# CHAPTER 3

## METHODOLOGY

**Study Design**

This cross-sectional study was aimed at developing and preliminarily validating a scale for measuring instructors' attitudes toward the teaching of introductory statistics in the health and behavioral sciences, at the undergraduate level, using the concept-based (reform-oriented or constructivist) pedagogy. Qualitative methods (indepth interviews and focus group discussions) were also employed, especially for item generation, item analysis, and in general, for establishing content validity. A teaching practice scale was concurrently developed, and used for assessing the criterion validity of the attitude scale. The data were analyzed using primarily exploratory factor analysis (FA) and multidimensional scaling (MDS) toward elucidating the structural and psychometric properties of both scales. In general, the research conceptualization and procedures were guided by the principles of measurement of data put forward by Coombs (1964), and discussed in Dawes and Smith (1985, p. 513).

1. First, decide what to observe.

2. Second, decide which empirical relations among the things or phenomena we observe are worth trying to measure.

3. Third, decide which formal relational system to use to represent these relations.

A cross-sectional research design involves the collection of data at one point in time (Polit et al., 2001), and allows for examining the association between different measures, as information can be obtained concurrently. This approach has been documented as the most efficient way of testing and validating a measurement tool (CEBMH, 2002). While factor analysis is the conventional data analytic framework for scale development and theory testing (Fabrigar et al., 1999; Floyd & Widaman, 1995; McKinley et al., 1997), it can be limited and limiting, and it is for this reason that multidimensional scaling (MDS) was also performed (Coxon, 1982; Kruskal & Wish, 1978; Young, 1987). Specifically, factor analysis requires assumptions of linearity and interval or ratio-level data, the absence of which may lead to misspecified models. Furthermore, given the complex nature of attitude (Jaccard & Blanton, 2004; Oppenheim, 1982; Rosenberg & Hovland, 1960; Smith, 1947; Zimbardo & Leippe, 1991), as well as the use of Likert-type scales (ordinal data), such assumptions can be tenuous. However, these assumptions are relaxed for MDS, and non-metric MDS is well-established for use with ordinal data (Coxon, 1982; Carroll & Wish, 1976).

Additionally, MDS allows for the spatial representation of psychological stimuli, and such graphical representations (maps) can facilitate a formative understanding of the mental process underlying reported attitudes and perceptions (Carroll & Wish, 1976). The use of both FA and MDS can therefore result in a more plausible and parsimonious latent structure or model that can explain the patterns of relationship and similarity among the measured variables. The primary analysis was conducted with a view to identifying possible separable latent factors and dimensions underlying the attitudinal items. The derived factors constitute the subscales of the instrument. In general, the study approach was exploratory because the investigator had no **firm** a priori expectations based on theory or prior research about the composition of the attitude scale (Floyd & Widaman, 1995).

**Relevant Operational Definitions**

Throughout this study, the **concept-based (constructivist or reform-oriented)** approach to teaching introductory statistics is operationally defined as a set of strategies intended to promote ***statistical literacy*** by emphasizing concepts and their applications rather than calculations and formulas (Rumsey, 2002). It involves active learning strategies such as projects, group discussions, data collection, hands-on computer data analysis, critiquing research articles, report writing, oral presentations, and the use of real-life data. ***Statistical literacy*** is the ability to understand, critically evaluate, and use statistical information and data-based arguments (Gal, 2000; Garfield, 1999). These definitions and explanations were included in the questionnaire so as to facilitate a common understanding of the concept-based pedagogy (the attitude object in this context). Additionally, **Attitude** was conceptualized and defined as an evaluative disposition toward some object based upon cognitions, affective reactions, and behavioral intentions. In other words, attitude is an informed predisposition to respond, and is comprised of three components: beliefs, feelings, and a readiness or intent for action (Jaccard & Blanton, 2004; Oppenheim, 1982; Rosenberg & Hovland, 1960; Smith, 1947; Zimbardo & Leippe, 1991). This three-component **attitude theory** is referred to throughout this study as the tripartite attitude model.

**Definition of Subjects and Inclusion Criteria**

The subjects of this study were volunteer instructors of introductory statistics courses in the health and behavioral sciences at four-year regionally accredited degree-granting institutions in the USA (and the equivalent in foreign countries). Both full-time and adjunct (part-time) instructors who had full responsibility for an introductory statistics course were eligible to participate.

**Sample Methodology (and Size Determination)**

A purposive (maximum variation) sample (Patton, 1990) was used to reflect the range of instructors that the final measure is intended to be used on (Sackett et al., 2000), and to allow for meaningful statistical analysis. Purposive sampling has been widely used in major studies to explore teachers' beliefs, attitudes and practices in school reform situations (Goertz, Floden, & O'Day, 1995; Ravitz, Becker, & Wong, 2000; Tschannen-Moran et al., 2000). Specifically, this sampling approach helps to guard against a restricted range in measurement (likely from a largely homogeneous sample) which can result in attenuated correlations among items and lead to falsely low estimates of factor loadings and correlations among factors (Comrey & Lee, 1992; Gorsuch, 1983; Tucker & MacCallum, 1997). Furthermore, it must be recognized that this was an initial exploratory study, and therefore, purposive sampling was desirable in order to "maximize discovery of the heterogeneous patterns and problems that occur in the particular context under study" (Erlandson et al., 1993, p.82). Moreover, according to Viswanathan (2005, p.70): "convenience sampling is suited for these studies rather than probabilistic sampling because the aim is not to establish population estimates, but rather to use correlational analysis to examine relationships between items and measures".

The sample size was determined primarily by guidelines for best practices in factor analysis, specifically, the standard of a "participant to variable" ratio of 5 to 1, and a minimum of 100 subjects (Gorsuch, 1983) along with a "coefficient of overdetermination" (item to factor ratio) of between 3 and 7 (MacCallum et al., 1999). MacCallum et al. (1999) argued that large samples are not necessary when factors are highly overdetermined, and that in such a situation, relatively small samples may produce quite meaningful and stable results. Such results can be rendered more reliable by ensuring that a battery of salient items is used (Floyd & Widaman, 1995).

Based on an anticipated maximum of 5 factors or subscales, a conservative coefficient of overdetermination of 5 (Tabachnick & Fidell, 2001), and a participant to variable ratio of 6, a minimum sample size of 150 (**5 x 5 x 6**) was deemed necessary. In order to be more conservative, it was decided to aim for a sample of at least 200 (Comrey & Lee,1992). The final effective sample size was 227 (Table 2). A response rate is not reported, as the number of eligible recipients of the questionnaire is not known. The proportions of statistics instructors who reported teaching either in health sciences or behavioral sciences were 94 (41%) and 102 (45%) respectively, whereas 31(14%) reported teaching courses in both health and behavioral sciences.

| Table 2: Determination of the Effective Sample Size | |
| --- | --- |
| Number of questionnaires received | 258 |
| Number of duplicates | 4 |
| Number of unique questionnaires | 254 |
| Number of questionnaires disqualified* | 27 |
| Effective sample size (n) | 227 |

*Respondents were disqualified because of (1) no teaching experience in the health and behavioral sciences, (2) they taught graduate courses only, or (3) a grossly incomplete questionnaire.

**Development of the Attitude Instrument**

In general, this process addressed content validity, that is, ensuring that the array of items adequately and plausibly reflected the theoretical and empirical dimensions of attitude in relation to reform-based teaching. **Attitude** was conceptualized and defined as an evaluative disposition toward some object based upon cognitions, affective reactions, and behavioral intentions. In other words, attitude is an informed predisposition to respond, and is comprised of three components: beliefs, feelings, and a readiness or intent for action (Jaccard & Blanton, 2004; Oppenheim, 1982; Rosenberg & Hovland, 1960; Smith, 1947; Zimbardo & Leippe, 1991). This three-component **attitude theory** is referred to throughout this study as the tripartite attitude model.

The first draft of the questionnaire contained 64 items intended to represent the tripartite conceptualization of attitude as follows: (1) **cognition** (perceived difficulty, usefulness, behavioral control, and beliefs about preparation in relation to the concept-based approach to teaching introductory statistics), (2) **affect** (feelings toward the concept-based pedagogy), and (3) **intention** (perceived likelihood of teaching according to the concept-based pedagogy). Items relating to general and epistemological beliefs about the concept-based approach and introductory statistics were also included. The abovementioned domains and dimensions were derived from a review of the extant literature, followed by a mini-survey of pioneer and "expert" statistics educators, researchers, and practitioners from the health sciences, psychology, education, and statistics (including psychometrics). These measurement and content specialists rated the relevance and salience of each attitude domain and dimension to the study context (Appendix A)

The initial set of items for each attitude dimension was culled from related studies, as well as formulated by the researcher. Item analysis (in-person and email) was performed by college instructors from the disciplines of education, health sciences, psychology, and statistics, as well as language and communications. In general, and based on consensus, items were added, rephrased, or removed. All items were assessed for face validity, salience, clarity (including double-barrel items), readability, theoretical relevance, and redundancy. In general, each a priori defined subscale contained items reflecting concerns and issues about self, process, and impact, in relation to use of the concept-based pedagogy. This structure is consistent with the Concerns-Based Adoption Model (Hall & Horde, 1987).The specific items reflected course content, pedagogy, assessment, and integration of technology.  Item content and wording were also guided by the principle of correspondence or compatibility (attitude-behavior) with reference to the action (using the concept-based approach), target (introductory statistics instructors), context (health and behavioral sciences at 4-year colleges), and time (when you teach statistics) (Ajzen, 1988; Ajzen & Fishbein, 1977, 1980; Fishbein & Ajzen, 1975).

A five-point Likert-type scale ranging from 1 (strongly disagree) to 5 (strongly agree) with a neutral midpoint, was used for each attitudinal item in the questionnaire. Higher scores indicate a more favorable disposition. While other approaches such as Thurstone scaling (Thurstone, 1929), Guttman scaling (Guttman, 1971) and semantic differential scaling (Osgood, 1957) can be used for measuring attitudes, the Likert format (Likert, 1932) is preferred, given its ease of use, general familiarity, and the resulting data being amenable to MDS and FA for scale development. Furthermore, other formats such as continuous and graphic scales are inclined to give rise to spurious exactness and uninterpretable results, especially with regard to complex constructs such as attitude.  In particular, a five- point scale is recommended, as data from scales beyond 7 to 9 points may decrease in reliability and stability (Nunnally 1978; Miller, 1956)**.** It will also be less cognitively burdensome, and may increase completion and response rates while rendering a higher quality of data over other approaches. Furthermore, this format will allow for meaningful comparison of outcome models to research from other disciplines (mathematics, science and physics) in which this measurement approach has been widely used.

There were both negatively and positively-worded items, and reverse scoring was performed for the negatively worded items for the purpose of statistical analysis. Attention was given to the order of the items so as to reduce the likelihood of acquiescent response set bias. Also, for each set of items intended to represent a specific dimension, at least one marker item (focal point of the subscale) was included (Froman, 2001; Tabachnick & Fidell, 2001). In the final analysis, if the marker item(s) correlated strongly with that factor or subscale, this served as a guide in naming the factor.  The final version of the instrument included 45 attitude items (see Appendix B).

**Development of the Teaching Practice Instrument**

In order to assess the criterion validity (among other dimensions of validity) of the attitude scale, a teaching practice scale was developed. This was used to characterize instructors as either high reform (concept-based, constructivist, or reform-oriented) or low-reform (traditional, mathematical, or behaviorist), and then an analysis was conducted to determine if there is a significant difference between these two groups with respect to overall (composite) attitude and subscale scores. In other words, can attitude (scores) explain or predict type of teaching practice as is theoretically and empirically supported? A similar approach was used by Riel and Becker (2000) in their study of teacher professionalism, pedagogy, and computer use, part of a national survey funded by the National Science Foundation. They formulated a "Pedagogy Index" with teachers in the lowest quartile classified as having a "direct instruction" pedagogy (low-reform or behaviorist) and those in the highest quartile as having a "knowledge construction" pedagogy (high-reform or constructivist).

In general, behaviorist-oriented teachers tend to be more preoccupied with subject content, and the transmission of information, and are more likely to engender surface learning, whereas constructivist-oriented teachers are more student-centered, concept-based, and seek to facilitate deep and conceptual learning (Askew et al., 1997; Prosser & Trigwell, 1999; Trigwell & Prosser, 2004). The practice scale contains 10 items (5 behaviorist and 5 constructivist) (see Appendix B). Development of the scale content was guided by the seminal recommendations of the **ASA/MAA** Joint Committee on Undergraduate Statistics (Cobb, 1992) and the GAISE (Guidelines for Assessment and Instruction in Statistics Education) preliminary report on introductory statistics (Garfield, 2004) which represents an expansion and updated operationalization of the Cobb's report. Active learning strategies with reference to course content, pedagogy, assessment and integration of technology are emphasized in these reports (see summary of ASA/MAA report below). These recommendations were agreed upon by a multidisciplinary team of subject specialists (statistics, mathematics, and education) as effective for facilitating statistical literacy.

1. "Emphasize the elements of statistical thinking:

    a.   the need for data,

    b.   the importance of data production,

    c.   the omnipresence of variability,

    d.   the measuring and modeling of variability.

2. Incorporate more data and concepts, fewer recipes and derivations and wherever possible, automate computations and graphics. An introductory course should

    a.   rely heavily on real (not merely realistic) data,

    b.   emphasize statistical concepts such as causation vs. association, experimental vs. observational studies, and longitudinal vs. cross-sectional studies,

    c.   rely on computers rather than computational recipes,

    d.   treat formal derivations as secondary in importance.

3. Foster active learning, through the following alternatives to lecturing:

    a.   group problem solving and discussion,

    b.   laboratory exercises,

    c.   demonstrations based on class-generated data,

    d.   written and oral presentations, projects, either group or individual."

Practice items were formulated following an extensive literature review. Specifically, the teaching practice checklist (Handal, 2003) used to classify teachers of mathematics as constructivist (thematic/concept-based) versus behaviorist (non-thematic/formula-based) was used as a guide. Toward content validity, and in keeping with a best practice in this regard (Haynes, Richard, & Kubany, 1995; Nunnally & Bernstein, 1994), the initial list of teaching practice items was presented to multiple pioneer and "expert" introductory statistics educators (Appendix A), for their rating of relevance, salience, and frequency of use. Two item analysis sessions, simultaneous with the attitudinal items were conducted. The responses were obtained on a frequency of use scale of 1 (never) through 5 (always), and where applicable, items were reverse-coded so that higher values reflect a more favorable teaching approach (reform-oriented, constructivist or concept-based).

**Pilot Testing of the Questionnaire**

Both scales (attitude and teaching practice) were merged into a single instrument (including questions ascertaining personal and sociodemographic information). A pilot[2] test was then conducted via email (n = 30). Open-ended feedback also provided valuable information. This exercise allowed for refinement of the instrument with respect to salience, variance, phraseology, ordering, and ambiguity of items, as well as possible subject burden. Item responses were evaluated for variability, and discriminant value (in relation to teaching practice items). Items were removed, replaced, rewritten or added. The revised version of the questionnaire was reviewed by a group of introductory statistics educators before it was finalized. The final version contained a total of 64 items, 45 attitudinal, 10 practice, and 9 personal and sociodemographic (Appendix B)

---

[2] These participants were not in the final sample.

**Recruitment of Subjects**

This activity sought to maximize similarities and differences with respect to attitudes and practices. It  involved targeting instructors of introductory statistics in the health and behavioral sciences at  four-year colleges where  pioneer educators active in the  reform  movement were  employed.  This approach  was  assumed  to   increase  the likelihood of including in the sample, instructors who have adopted or are moving toward reform  recommendations  (the  concept-based  approach).  Such  sampling  strategy  for identifying  institutions  was  used  by  Riel  and  Becker  (2000)  in  their  study  titled:  "The Beliefs,  Practices,  and  Computer  Use  of  Teacher  Leaders".  Pioneer  educators  were identified from the membership database (accessible online to members) of the American Statistical Association (Sections on Statistics Education, and Teaching of Statistics in the Health Sciences). The ASA Directory of Minority Statisticians was also consulted.

 Instructors were also targeted based on their publications, reseach interests, and a review  of   their  course  outlines. Faculty  who  could  be  charcaterized  as  having  a mathematical/traditional/beahviorist  focus  were  also  targeted.  Additional  contact information was obtained from the following sources:

1. Articles in the Journal of Statistics Education (online)
2. The  first  USCOTS  (United  States  Conference  On  Teaching  Statistics) resource notebook
3. JSM (Joint Statistical Meetings, American Statistical Association) conference proceedings
4. Online institutional faculty lists
5. ICOTS  (International  Conference  On  Teaching  Statistics)  conference proceedings
6. ASA/MAA  (American  Statistical  Association/Mathematical  Association  of America) Joint Committee on Undergraduate Statistics

7. The Stanford University online directory of statistics departments (USA and international)

8. The ASA (American Statistical Association) list of schools offering degrees in staistics

9. CAUSE - **C**onsortium for the **A**dvancement of **U**ndergraduate **S**tatistics **E**ducation

10. The RSS (Royal Statistical Society, UK) – Center for Statistics Education

11. IASE (The International Association for Statistical Education)

The Chairs of Departments were also contacted and asked to cirulate the questionnaire link to relevant faculty. To supplement the sample, participation was solicited from instructors of introductory statistics who participated in a preliminary mini email survey (Hassad, 2003), and self-characterized their teaching as either concept-based or calculation-based. Participation was also solicited via statistics and research online discussion forums, and listservs such as:

1. ALLSTAT@JISCMAIL.AC.UK - A UK-based worldwide email broadcast system for the statistical community.

2. TEACHING-STATISTICS@JISCMAIL.AC.UK - A UK-based worldwide email broadcast system, concerned with the initial learning and teaching of statistics.

3. SRMSNET@LISTSERV.UMD.EDU - A mailing list of the Survey Research Methods Section of the ASA (American Statistical Association).

4. EDSTAT-L@LISTS.PSU.EDU - An email forum devoted to discussion of topics related to the teaching and learning of statistics at the college level.

Deparments contacted include, statistics, mathematics, health sciences, biostatistics, public health, epidemiology, psychology, behavioral sciences, social sciences, sociology, and mathematics. An incentive for participation was offered. All participants who completed the instruments were given a chance to win one of three $100 (one hundred dollar) awards toward conference registration, journal subscription, continuing education courses or other professional development activities. The final sample of 227 comprised of respondents from the USA (74%), as well as twenty-three other countries (see Appendix C). In general, these respondents represented 133 academic institutions (see Appendix D).

**Data Collection**

This study was approved by Institutional Review Board of Touro University International (Appendix E). Two instruments (attitude and teaching practice) merged into a single questionnaire were administered online. Personal and sociodemographic data were also ascertained. The questionnaire was programmed in Hyper Text Markup Language (HTML) using the radio-button response format, and permitting one selection only per item or question. The script was validated before use, and the instrument was programmed so that each respondent's data sheet was emailed as a text file to the researcher, for ease of data entry. There was a total of 64 closed-ended items, and an open-ended option for general comments. The link to the questionnaire was provided in an email invitation to participate, and respondents had to first view (and read) the online informed consent (Appendix E), and agree to participate, before the questionnaire could be viewed. A total of three emails were sent (one week apart) to each target, and all contained the online link to the questionnaire. The first was the survey notification and invitation to participate, the second, a reminder to participate, and the third, a last call to participate. Data collection took place between August and October of 2005. Names and

email contacts were provided by the participants, and this facilitated follow-up for missing data (mostly one item). The data were screened for possible redundant or duplicate sumissions by way of name and/or email address.

**Data Analysis**

The data analysis centered around the assessment of the structural (underlying factors or subscales) and psychometric (reliability and validity) properties of the both the attitude and practice scales. The conventional methods were performed using exploratory factor analysis. The data were then analyzed using multidimensional scaling (MDS), a more general framework which involves less rigorous demands on the data. Specifically, the assumptions of interval or ratio-scaled data, and linear relationships among the items (required for FA) were relaxed allowing for both metric and non-metric MDS applications. In other words, the factor model was augmented by using a distance model (MDS). Whereas the FA model is driven by product moment correlation coefficients, MDS accommodates other measures (including Spearman's rho and Kendall's tau), and hence a more comprehensive exploration of the data was achieved (Coxon, 1982). The factor analysis procedures are detailed below followed by the MDS approach.

**Data Screening**

Data entry was performed using SPSS version 14.0 (Statistical Package for the Social Sciences). The dataset was then checked against a random sample of questionnaires for data entry accuracy. All negatively worded items were reverse-coded (or reflected) in order to obtain meaningful scores, and interpretable factor loadings and MDS dimensions. Univariate distributions (item level) were examined for possible data entry error, as well as response bias toward extreme or middle categories.The data were screened for the basic assumptions underlying exploratory factor analysis, specifically factorability of the data. A data set is factorable if there is a substantial number of meaningful relationships among the items (usually Pearson's correlation coefficient of .3 or greater). This was based on a visual inspection of the full correlation matrix, as well as the Bartlett's test of sphericity and the Kaiser-Meyer-Olkin measure of sampling adequacy (Tabachnick & Fidell, 1996). Attention was also given to the recommended thresholds for questioning the assumption of multivariate normality (skew > 2, kurtosis > 7) (West et al., 1995), albeit is is recognized that this is not a required assumption for principal axis factor analysis (PAF) , but it is for maximum likelihood extraction methods and confirmatory factor analysis (not performed in this study).

**Model Fitting (Factor Extraction Procedure)**

Exploratory factor analysis (EFA) (Cudeck & O'Dell, 1994; MacCallum et al., 1999), specifically Principal axis factor analysis (PAF) was used to explore the relationships among the items (attitudinal and practice) and to identify underlying factors and dimensions. PFA analyzes common variance only, which is required for scale development, as the objective is to identify common latent constructs. Additionally, it has been established that PAF with squared multiple correlations (SMC) as the initial estimate of communality provides more accurate results in terms of the population factor loadings over principal components analysis (Widaman, 1993; Russell, 2002). According to Fabrigar, et al. (1999, p4):

> The primary purpose of EFA is to arrive at a more parsimonious conceptual understanding of a set of measured variables by determining the number and nature of common factors needed to account for the pattern of correlations among measured variables. That is, EFA is used when a researcher wishes to identify a set of latent constructs underlying a battery of measured variables.

**Communalities**

The variance in an item consists of common, unique and error variance. Common variance is the variance that an item shares with the other items. Unique variance is specific to a single item, and error variance is random variance associated with a particular item (Costello & Osborne; 2005; Kline, 2005; Tinsley & Tinsley, 1987). As the objective of factor analysis is to identify latent constructs underlying the common variance, it is necessary to approximate what proportion of the variance in the matrix is common variance. This is estimated prior to the analysis (estimated communality). The variance that an item shares with the factors is referred to as its communality. In other words, communality is the proportion of each item's variance that can be explained by the factors.

These communality values are used in the correlation matrix for the factor analysis. Unlike principal component analysis (PCA), which assumes that all of the variance in an item can be explained by the components extracted, factor analysis attempts to eliminate unique and error variance from factors, hence a reduced correlation matrix or covariance matrix is used with communality estimates (rather than 1) in the diagonal of the correlation matrix (Russell, 2002). This initial estimate is the squared multiple correlation (SMC) obtained when each item is regressed on the other items. And this is initially taken as a reasonable estimate of each item's variance that can be accounted for by the extracted factors. The SMC represents a lower-bound estimate of the communality, and its use ensures that the analysis will not capitalize on error variance. (Tinsley & Tinsley, 1987). With reference to a particular item, a low communality (e.g. .2 or less than 20%) (Fullagar, 1986) suggests that the item has little in common with the other items, and hence is not important to the analysis, and can be removed barring strong conceptual or theoretical relevance.

**Determining the Number of Factors to Retain**

The following criteria, with attention to parsimony (a model with the fewest possible factors) and plausibility (a model that makes theoretical and empirical sense) were used to determine the optimal number of factors to retain in the factor analysis model (Fabrigar, et al., 1999). This approach facilitated the identification of major and salient common factors underlying the battery of items.

1. The Kaiser criterion of selecting factors with eigenvalues greater than one (Gorsuch, 1983).

2. The Cattell's scree test (Cattell, 1966; Cattell & Jaspers, 1967), which is generally considered to be more conservative than the Kaiser's criterion. Conventionally, this is a plot of the eigenvalues (and corresponding factors) of the full correlation matrix, and the maximum number of factors is indicated by the point before the plot levels off, that is, the point preceding the elbow. Beyond this point, factors contribute negligible variance. The plot of the eigenvalues of the reduced correlation matrix was also examined, as the focus of this study in on common variance (Fabrigar et al., 1999).

**Factor Rotation**

Rotation is a method used to reorient the factor loadings so that the factors are more interpretable. Mathematically, exploratory factor analysis (EFA) models with more than one factor do not have a unique solution, as there could be an infinite number of alternative orientations of the factors in multidimensional space that will explain the data equally well (Fabrigar et al., 1999). The challenge is therefore to select one solution from amongst these. This aspect of the analysis was guided by the principle of simple structure (Thurstone, 1947). Thurstone noted that in this context, the solution with the best "simple structure" would generally be the most easily interpretable, psychologically meaningful, and replicable. Thurstone recommended that factors be rotated in multidimensional space to obtain such a solution (Fabrigar et al., 1999).

Thurstone's concept of simple structure refers to solutions in which each factor is characterized by a cluster of measured items that have large loadings relative to the other items (i.e., high with-in-factor variability in loadings) and in which each measured item loads highly on only a subset of the common factors (i.e., low factorial complexity in defining items) (Fabrigar et al., 1999). To achieve this, the retained factors were rotated to simple structure using the oblique rotation algorithm, promax with kappa = 4 (Fabrigar et al., 1999; Russell, 2002). Promax first conducts an orthogonal varimax rotation and then allows correlations between the factors toward achieving simple structure. Therefore, if the factors are orthogonal (uncorrelated) with one another, that will be revealed with promax rotation. Moreover, it is reasonable to expect that psychological constructs will be correlated (Fabrigar et al., 1999). Any relationship among the factors is critical to the interpretation and meaning of the factors. In this regard, oblique rotation is advantageous as it produces a correlation matrix of factors (unlike orthogonal rotation) which also allows for checking for possible higher order factors (strong correlation among factors) thereby gaining a more realistic conceptual understanding of the factors, and model in general (Gorsuch, 1983). The percentage of variance explained by each rotated factor was calculated using Cattell's formula (Cattell, 1978; Barrett & Kline, 1980). That is, the variance explained by each rotated factor is the product of the sum of the structure loadings and pattern coefficients for all variables.

**Interpretation and Naming of the Factors (Subscales)**

Following oblique rotation, each cluster of variables was carefully examined to determine the underlying construct and its substantive meaning, in other words, what the items (in each cluster) have in common (Kim & Meuller, 1978). Specifically, interpretation and naming of the factors were based on the factor pattern matrix coefficients (Field, 2000; Hair, et al., 1998; Russell, 2002; Stevens, 1992), previously identified marker items, and the general standard of at least three variables per factor (Anderson & Rubin, 1956; Comrey, 1988). The pattern matrix coefficients are standardized regression coefficients (weights) which reflect the relative and independent contribution of each factor to the variance of the item on which it loads (Russell, 2002).

A pattern coefficient ("loading") of .4 and higher (that is, a factor explaining at least 16% of an item's variance) was considered salient for a sample size of about 200, so as not to capitalize on chance (Field, 2000; Hair, et al., 1998; Stevens, 1992). Nonetheless, both high and low pattern coefficients were noted as they tell us what the factor most likely is, and is not (Rummel, 1970). Notably, while the structure matrix loading is a measure of the association (Pearson's correlation coefficient) between each item and the factor on which it loads, when the factors are correlated (as in this study) there is overlap among the loadings, which makes the structure matrix loadings biased estimates of the independent relationship between the item and the factor (Russell, 2002). It is for this reason that interpretation of the factors was based on the pattern matrix coefficients (rather than the structure matrix loadings).

**Factor (Subscale) Score**

A composite factor score was calculated for each subject based on the score of items (with salient loadings) which constitute each factor (Cattell, 1957; Gorsuch, 1974; Kline, 2005). Factor scores were used for correlation, regression, t-test, and ANOVA among other statistical analyses. All items were equally weighted, and the mean of the item scores was obtained in order to retain the original scale properties (Hogue, Liddle, Singer, & Leckrone, 2005; Russel 2002). In particular, the factor scores were not automatically generated based on the factor loadings (weights), as this approach uses all the items that loaded (high and low) on a factor to generate the score. Factor loadings are sample-specific, and therefore, such scores may not be replicated in future studies (as items may not have the same loadings). The approach of summing the scores of items with salient loadings on each factor is generally recommended (Cattell, 1957; Gorsuch, 1974; Gorsuch, 1983; Kline, 2005), especially for exploratory factor analysis in scale development, as repeated studies for reliability and validation checks will be necessary.

**Reliability Analysis**

Following rotation, as well as preliminary interpretation and naming of the factors (subscales), Cronbach's alpha (Cronbach, 1951; Nunnally, 1967) was calculated for each factor (subscale) as well as the overall scale. Cronbach's alpha quantifies the degree of internal consistency (reliability), that is, the extent to which a set of items measures a single unidimensional latent construct or dimension of a construct (van Os, 2002). In other words, it reflects the degree of homogeneity or coherence of the scale or each subscale (Freeman & Tyrer, 1995). The factors (subscales) were also examined for possible multidimensionality. In general, a Cronbach's alpha of at least .7 is the criterion used to establish an acceptable level of internal consistency (reliability) (Ghiselli, 1981; Nunnally, 1967). However, the recommended minimum Cronbach's alpha for exploratory studies is .6 (Nunnally, 1974; Robinson, Shaver, & Wrightsman, 1991). Also, in order to establish the contribution of individual items to each subscale, the change in Cronbach's alpha was noted following the deletion of each item. Furthermore, according to Nunnally and Bernstein (1994), item-total correlations should be at least .3 in order to be considered a meaningful contribution to the scale, and this criterion was followed. Test-retest reliability was not performed for this initial exploratory study.

**Validity Analysis**

Reliability is necessary but not sufficient to establish validity, which refers to whether the scale measures the construct (attitude or practice) as theorized and operationally defined (Nunnally & Berstein, 1994). Validity is a multidimensional concept (Messick, 1995) more appropriately referred to as "construct validity", which encompasses ***content validity***, ***criterion*** (concurrent or predictive) ***validity***, **convergent** and ***discriminant validity*** (Cronbach & Meehl, 1955; Viswanathan, 1993). Muldoon et al. (1998) noted that "while validity can be examined in several ways, comparison with the best indicator available (criterion validity) is the preferred method." Specifically, "criterion validity assesses the measure's ability to distinguish between groups that it should theoretically be able to distinguish between" (Maclnnes, 2003). Psychological constructs (for example attitude) are not directly observable, and in order to determine whether such hypothetical constructs are being measured, we must show that a measure of a given construct relates to a measure of another construct (the criterion) in a theoretically predictable way (Cronbach & Meehl, 1955). In accordance with the conceptual framework of this study, attitude scores should meaningfully differentiate between high and low reform practice instructors.

The developed teaching practice scale (which was established to be theoretically and empirically plausible) was used to achieve the dichotomous high-low reform practice classification, and the independent samples t-test, ANOVA (and their non-parametric equivalents) were used to compare the subgroups based on the mean total and subscale scores. Additionally, multiple regression analysis of teaching practice score on attitude subscale scores was performed to determine the extent to which attitude can explain and predict teaching practice in this context (Asher, 1997; Meehl, 1954). The coefficient of determination (R-squared, the proportion of variance explained or accounted for by the model), and standardized regression coefficients (beta) were reported. Multiple regression analysis of intention on the other attitude subscales was also conducted. An alpha level of .05 was used for all tests of significance.

**Content Validity**

Content validity was achieved primarily during the early stages of instrument development, and refers to the extent to which the items in the scale capture or reflect the theoretically and empirically supported facets of the construct being measured (Nunnally, 1978). This was facilitated by a thorough review of the related literature, consultation with experts, and a multidisciplinary team approach to item generation and item analysis. In keeping with a best practice in this regard (Haynes, Richard, & Kubany, 1995; Nunnally & Bernstein, 1994), the initial list of facets of attitude and teaching practice, were presented in a grid format to multiple experts, for their evaluation of relevance and salience in this context. There was also an open-ended section for general comments. To ensure adequate coverage of the underlying construct of each subscale, the criterion of a minimum of three variables (to establish a subscale) was adhered to (Anderson & Rubin, 1956; Comrey, 1988).

**Convergent Validity**

For convergent validity, a modified version of the multi-trait multi-method (MTMM) approach (Campbell & Fiske, 1959) used by Bagozzi et al. (1979) to examine the construct validity of the tripartite classification of attitude was applied. According to the principle of convergent validity, measures of theoretically similar constructs should be substantially intercorrelated. Each of the five subscales (of the final solution) was considered a "method" for measuring the trait (attitude), given the conceptual relatedness of the factors. As Campbell and Fiske (1959, p. 82) notes, in order to establish convergent validity, the relevant correlations "should be significantly different from zero and sufficiently large". Correlation coefficients, corrected for attenuation due to measurement error were also obtained and reported (Lord & Novick, 1968).

**Discriminant Validity**

According to the principle of discriminant validity, measures of theoretically different but related constructs should not correlate highly with each other. Toward this end, the inter-factor correlations (observed and corrected) were examined as well as the extent of "simple structure". Also, a more rigorous test of discriminant validity based on the average variance extracted (AVE) for each construct, was applied. Fornell and Larcker (1981) recommended that in order to demonstrate discriminant validity, the AVE for each construct (within construct variance) should be greater than the squared correlation (variance) between that construct and another.

# Multidimensional Scaling (MDS)

MDS is a set of multivariate analytical techniques aimed at reducing and organizing data so as to elucidate how and why items are related. Both the attitudinal and teaching practice items were subjected to MDS. Integral to the use of MDS is the assumption that the items in the dataset are related through some underlying psychological concept (Fitzgerald & Hubert, 1987). MDS seeks to achieve a spatial representation (geometric map or configuration usually in 2-dimensions) of the latent or hidden structure that underlies and explains the relationships among the items which constitute the map (Coxon, 1982; Fitzgerald & Hubert, 1987; Kruskal & Wish, 1978). By using MDS, we can therefore discern the dimensions of the perceptual space of subjects used to evaluate the set of items (Coxon, 1982; Fitzgerald & Hubert, 1987; Kruskal & Wish, 1978; Pinkley, Gelfand, & Duan, 2005). Unlike factor analytic techniques which require the assumption of interval- or ratio-scaled (metric) data, and linear relationship among the items, all MDS models do not impose such restrictions on the data, as there are both metric (linear transformation) and non-metric (ordinal transformation) variants of MDS (Coxon, 1982; Kruskal & Wish, 1978; Schiffman, Reynolds, & Young, 1981; Young & Lewyckyj, 1979) Moreover, compared to factor analysis, MDS can result in more parsimonious and interpretable solutions (Fitzgerald & Hubert, 1987). Indeed, as these attitudinal data (measured using Likert-type scales) are truly non-metric, MDS is quite suitable for this study (Pinkley, Gelfand, & Duan, 2005).

The input information for MDS is a numerical measure of distance indicating how similar (or dissimilar) each item is to every other item. Both metric (MRSCAL[3]) and non-metric (MINISSA[4]) MDS were carried out, and for each approach both Pearson's correlation coefficient (based on the interval properties of the data) and Kendall's tau (based on the rank order of the data) were used as measures of similarity of the items.

---

[3] MRSCAL refers to **M**et**R**ic **SCAL**ing (Roskam, 1972).

INDSCAL[5] (INDividual SCALing) was not considered necessary for this initial exploratory study as there were no significant or meaningful subgroup differences (Coxon, 1982) in attitude with respect to gender, ethnicity, age group, employment status, membership status in professional organizations, highest academic degree, degree concentration, and duration of teaching. In other words, there was no compelling evidence to suggest that the different subgroups of instructors (based on the abovementioned variables) would attach different weights or levels of salience to the overall underlying dimensions of the attitude construct.

**Interpretation of the MDS Maps**

Interpretation of the MDS maps or configuration involved identifying and assigning meanings to patterns, neighborhoods or regions with related characteristics, and which plausibly differed from other regions, and for this, "a two-dimensional configuration is far more useful than one involving three or more dimensions" (Kruskal & Wish, 1978, p.58). This is called an internal analysis, as only the original data (measures of similarity in this study) are used in the interpretation (Coxon, 1982). Notably, it has been established that the neighborhood or regional interpretation approach can give rise to different and more informative maps or solutions, compared to dimensional interpretation (Coxon 1982, Kruskal & Wish, 1978), as its focus is primarily on the small distances (large similarities), whereas the dimensional approach attends mostly to the large distances (Guttman, 1965; Kruskal & Wish, 1978). The neighborhood or regional approach is commonly used as the primary technique for interpreting MDS maps (Bilsky & Jehn, 2002; Cooper & Donald, 2003; Fitzgerald & Hubert, 1987; Guttman, 1965; Papanastasiou, 2005; Pfleiderer, 2003). According to Guttman (1965) this approach is more advantageous, and should be preferred over dimensional interpretation, which is based solely on mathematical properties of the data (Coxon, 1982, Kruskal & Wish, 1978).

---

[4] MINISSA refers to **S**mallest **S**pace **A**nalysis (see Roskam and Lingoes, 1970).
[5] INDSCAL (**IND**vidual **SCAL**ing analysis), developed by Carroll and Chang (1970).

In order "to guard against the human tendency to find patterns whether or not they exist" (Kruskal & Wish, 1978, p.36), hierarchical cluster analysis was used to guide the identification of structures within the spatial configurations or maps (Coxon, 1982). Also, to achieve interpretability, all the configurations were rotated to "simple structure" (Kruskal & Wish, 1978). Rotation allows for identifying meaningful clusters if these in fact exist. Regions or clusters with conceptually and theoretically related items are delineated in the maps to show potentially separable structures underlying the construct. For both the metric and non-metric MDS approaches, measures of goodness of fit (between the Euclidean distances and the input proximities) were generated and used in conjunction with evidence of interpretability, parsimony, stability, plausibility, and construct validity to evaluate the adequacy of the MDS solutions.

**Adequacy of the MDS solutions**

The adequacy of the MDS solutions is evidenced by the stress[6] and the coefficient of determination (R-squared) values[7]. Both stress and R-squared (RSQ) are measures of fit, indicating how well the solution or model (MDS map or configuration) fits the data, in terms of the percentage of variance in the proximity data that is explained or remains unaccounted for by the MDS model (Coxon, 1982; Kruskal & Wish, 1978). Smaller stress values (approximate amount of variance not accounted for by the MDS model) indicate better fit, whereas larger RSQ values (the amount of variance explained) reflect better fit. Additionally, the stability of the solutions was assessed using the empirical guideline of at least $4k + 1$ objects (items) for a $k$-dimensional solution (Kruskal & Wish, 1978, p. 34), as well as consistency across all the MDS maps or configurations (Kruskal & Wish, 1978).

---

[6] Stress 1 = Residual sum of squares from monotonic regression.
[7] Stress and R-squared values range from 0 to 1.

**Personal and Sociodemographic Factors as Determinants of Attitude and Practice**

Secondary analyses were conducted using one-way ANOVA, and the t-test, as well as linear correlation and regression to check for variation in total and subscale scores (attitude and practice) based on gender, ethnicity, age group, geographic location, employment status, membership status in professional organizations, highest academic degree, degree concentration, teaching area, and duration of teaching.

**Significance Testing**

Throughout this study, statistical significance was determined based on an alpha level of .05, and Bonferroni adjustment for multiple comparison testing (post-hoc analysis) was performed where applicable. For subgroup comparisons, the assumption of homogeneity of variances was examined using the Levene's test, and for most analyses, particularly, where this assumption was not met, the non-parametric equivalent of the independent samples ANOVA and the t-test (Kruskal-Wallis and Mann-Whitney U tests respectively) were also performed. Ninety-five percent (95%) confidence intervals of means were also generated.

# CHAPTER 4

## RESULTS & DISCUSSION

Because of the large amount of data, a combined results and discussion chapter is presented instead of separate chapters, a format that is highly recommended for studies of this nature (Medawar, 1979). This chapter presents, explains, and discusses the outcomes of the various statistical analyses used to explore the relationships among the attitudinal items, toward developing, and validating a scale for measuring instructors' attitudes toward the teaching of introductory statistics using the concept-based (reform-oriented or constructivist) approach. Specifically, the structural and psychometric properties of the latent factors which account for this shared variance (covariance) are detailed. The teaching practice items are similarly detailed and discussed. The results will also establish how these latent factors (hereinafter referred to as components, subscales, facets, dimensions, or domains of attitude) relate to teaching practice, in assessing the criterion validity of the attitude scale.

All data were obtained from a purposive (maximum variation) sample of 227 tertiary level instructors of introductory statistics in the health and behavioral sciences. According to Viswanathan (2005, p.70), this type of "convenience sampling is suited for these studies rather than probabilistic sampling because the aim is not to establish population estimates, but rather to use correlational analysis to examine relationships between items and measures". Obtaining maximum variation in the measures of interest (in this case, attitude and practice) through targeted and purposeful sampling, helps to guard against attenuation of correlation coefficients associated with a restricted range in measurement (Comrey & Lee, 1992; Fabrigar et al.,1999; Gorsuch, 1983; Tucker & MacCallum, 1997).

As this is an initial exploratory study, multiple analytical methods were used to explore the data from different perspectives, and under different assumptions. In particular, linear and non-linear statistical techniques (univariate, bivariate, and multivariate) were applied. This data analytical strategy allowed for a more formative, objective and comprehensive approach to identifying and interpreting the dimensions of instructors' attitude and practice.  In general, missing data (item non-response) was minimal (less than 4% for almost all analyses), and no imputation was performed, therefore, **n** varies slightly for some analyses.

**Respondents' Background Characteristics**

There was a total of 227 participants (Table 3), and of the 222 (98%) who supplied country identifying information, 165 (74%) were from the USA, and the remainder, 57 (26%) from international locations (primarily the UK, Netherlands, Canada, and Australia).  In general, the respondents represented 24 countries (Appendix C), and 133 academic institutions (Appendix D). All were active faculty at the time of participation, and taught either introductory statistics in the health or behavioral sciences, or both at regionally accredited 4-year institutions in the USA, or the equivalent in foreign countries. One hundred and seventy instructors (76%) reported full-time employment status. Of the 165 instructors from the USA, 135 (82%) identified as Caucasian, and the remainder ethnic minorities. Ethnic classification was not obtained for the international participants.

In general, the majority of the participants were male, 139 (61%). The modal age category was 31 to 40 years, 65 (30%), and the median 41 to 50 years. The distribution of duration of teaching was positively skewed with mean, median and standard deviation of, 14years, 10 years, and 11 years respectively. Similar distributions were observed for all teaching categories (health sciences, behavioral sciences, and both). One hundred and seventy-nine (79%) respondents were academically prepared at the doctoral level, and the remainder reported primarily master's degrees. Additionally, almost equal proportions of health sciences, 41 (40%) and behavioral sciences, 43 (42%) instructors held membership in professional organizations concerned with the teaching of statistics.

The proportions of statistics instructors who reported teaching either in health sciences or behavioral sciences were 94 (41%) and 102 (45%) respectively, whereas 31(14%) reported teaching courses in both health and behavioral sciences (Table 4). Overall, the modal category for academic degree concentration was statistics, 92 (41%), followed by psychology/social/behavioral sciences, 71(31%). The academic specialization least reported was mathematics/engineering, 17 (8%). The majority, 60 (59%) of instructors in psychology/behavioral sciences reported their highest academic degree in psychology/behavioral sciences. For instructors in health sciences only, as well as those who taught in both the health and behavioral sciences, slim majorities, 52 (55%), and 16 (52%) respectively reported statistics as their academic degree concentration.

**Table 3: Background Characteristics of Participants (N = 227[8])**

| CHARACTERISTICS | Total | TEACHING AREA | | |
|---|---|---|---|---|
| | | Health Sciences | Psychology/Behavioral Sciences | Health & Behavioral Sciences |
| **Gender** | | **n (percent)** | **n (percent)** | **N (percent)** |
| Male | 139 | 52 (37%) | 69 (50%) | 18 (13%) |
| Female | 88 | 42 (48%) | 33 (38%) | 13 (15%) |
| **Ethnicity** | | | | |
| Caucasian **(USA)** | 135 | 47 (35%) | 67 (50%) | 21 (16%) |
| Ethnic Minority **(USA)** | 30 | 11 (37%) | 11 (37%) | 8 (27%) |
| [9]International | 57 | 34 (60%) | 22 (39%) | 1 (02%) |
| **Age Group** | | | | |
| 26 – 30[10] | 19 | 10 (53%) | 8 (42%) | 1 (5%) |
| 31 – 40 | 65 | 28 (43%) | 27 (42%) | 10 (15%) |
| 41 – 50 | 62 | 30 (49%) | 22 (36%) | 10 (16%) |
| 51 – 60 | 50 | 17 (34%) | 28 (56%) | 5 (10%) |
| 60 + | 23 | 7 (30%) | 11 (48%) | 5 (22%) |
| **Employment Status** | | | | |
| Full-Time | 170 | 71 (42%) | 75 (44%) | 24 (14%) |
| Part-Time | 54 | 22 (41%) | 26 (48%) | 6 (11%) |
| **Highest Academic Degree** | | | | |
| Doctoral[11] | 179 | 67 (37%) | 84 (47%) | 28 (16%) |
| Masters | 48 | 27 (56%) | 18 (38%) | 3 (6%) |
| **Professional Membership** | | | | |
| Yes | 103 | 41 (40%) | 43 (42%) | 19 (18%) |
| No | 121 | 51 (42%) | 58 (48%) | 12 (10%) |
| **Duration of Teaching** | | | | |
| N | 223 | 91 | 101 | 31 |
| Mean (years) | | 12 | 15 | 15 |
| Median | | 8 | 11 | 10 |
| Standard Deviation | | 10 | 12 | 11 |

---

[8] **N** varies between 219 and 227 due to missing data (item non-response).
[9] Primarily from the UK, Netherlands, Canada, and Australia.
[10] Includes one subject who reported 18 –25 years.
[11] Includes three (3) respondents with ABD (all but dissertation) status.

**Table 4: Percentage Distribution of Respondents by Teaching Area and Academic Concentration**

| | | Academic Concentration (Highest Earned Degree) | | | | | |
|---|---|---|---|---|---|---|---|
| | | Statistics | Psychology/ Social/ Behavioral Sciences | Health Sciences/Public Health/ Epidemiology/ Biostatistics | Mathematics/ Engineering | Education/ Business/ Operations Research | Total |
| **Teaching area** | Health Sciences | 52 | 7 | 26 | 6 | 3 | 94 |
| | | 55.3% | 7.4% | 27.7% | 6.4% | 3.2% | 100% |
| | Psychology/Social/ Behavioral Sciences | 24 | 60 | 2 | 6 | 10 | 102 |
| | | 23.5% | 58.8% | 2.0% | 5.9% | 9.8% | 100% |
| | Health and Behavioral Sciences | 16 | 4 | 0 | 5 | 6 | 31 |
| | | 51.6% | 12.9% | .0% | 16.1% | 19.4% | 100% |
| Total | | 92 | 71 | 28 | 17 | 19 | 227 |
| | | 40.5% | 31.3% | 12.3% | 7.5% | 8.4% | 100% |

The proportions of statistics instructors who reported teaching either in health sciences or behavioral sciences were 94 (41%) and 102 (45%) respectively, whereas 31(14%) reported teaching courses in both health and behavioral sciences (Table 4). Overall, the modal category for academic degree concentration was statistics, 92 (41%), followed by psychology/social/behavioral sciences, 71(31%). The academic specialization least reported was mathematics/engineering, 17 (8%). The majority, 60 (59%) of instructors in psychology/behavioral sciences reported their highest academic degree in psychology/behavioral sciences. For instructors in health sciences only, as well as those who taught in both the health and behavioral sciences, slim majorities, 52 (55%), and 16 (52%) respectively reported statistics as their academic degree concentration.

**The Attitudinal Items**

Tables 5, 6 and 7 detail the distribution of the forty-five (45) attitudinal items contained in the questionnaire (Appendix B). Twenty-five (25) of these items were retained for the final factor analysis (FA) and multidimensional scaling (MDS) solutions (Tables 13 and 15). A mix of negatively and positively worded items were used in order to reduce acquiescent response bias (Nunnally, 1967), and responses were obtained on a five-point Likert-type scale[12]. Items were reverse-coded where necessary, so that higher values represent a more favorable disposition toward the concept-based pedagogy. There was no clear evidence of response bias toward extreme or middle categories. The reasons for removal of items are given below.

**Table 5: Descriptive Statistics of Attitudinal Items as Appeared in the Questionnaire** [a]

| Attitudinal Items | N | Mean | Std. Dev. |
|---|---|---|---|
| 1. The concept-based approach to teaching introductory statistics is just a fad that will soon be forgotten. | 227 | 1.76 | .88 |
| 2. Instructors should integrate introductory statistics with other subjects. | 227 | 3.89 | 1.00 |
| 3. Statistics in any form is mathematics. | 227 | 2.55 | 1.23 |
| 4. Statistical literacy is necessary for effective decision-making in everyday life. | 227 | 4.12 | .96 |
| 5. Statistics by nature is a difficult subject. | 227 | 2.71 | 1.16 |
| 6. The concept-based approach to teaching introductory statistics (rather than emphasizing calculations and formulas) makes students better prepared for work. | 227 | 4.03 | .89 |
| 7. It is unreasonable to expect students to achieve statistical thinking and literacy from an introductory statistics course. | 227 | 2.31 | 1.11 |
| 8. The concept-based approach to teaching introductory statistics is straightforward. | 227 | 3.16 | 1.06 |
| 9. Focusing on statistical literacy as a learning outcome of introductory statistics is a major shift from the way I was trained. | 226 | 3.41 | 1.16 |
| 10. Teaching introductory statistics with emphasis on concepts and applications rather than calculations and formulas, can be time consuming. | 226 | 3.31 | 1.19 |
| 11. I will adjust easily to teaching introductory statistics using the concept-based approach. | 226 | 3.91 | 1.00 |
| 12. The preparation required to teach introductory statistics using the concept-based approach is burdensome. | 227 | 2.66 | 1.04 |
| 13. Integrating hands-on computer analysis into the introductory statistics course is not a difficult task. | 227 | 3.46 | 1.19 |
| 14. Using active learning strategies (such as projects, group discussions, oral and written presentations) in the introductory statistics course can make classroom management difficult. | 227 | 2.83 | 1.16 |

[a]. Responses were given on a Likert-type scale (1= strongly disagree, through 3 = undecided, to 5 = strongly agree). Items were classified as follows: 1-7 (general and epistemological beliefs -cognitive), and 8-14 (perceived difficulty - cognitive). Negatively worded items were not reverse-coded for these descriptive statistics.

The original set of items was intended to represent the tripartite conceptualization of attitude as follows: (1) **cognition** (perceived difficulty, usefulness, behavioral control, and beliefs about preparation in relation to the reform-oriented or concept-based approach to teaching introductory statistics), (2) **affect** (feelings toward the concept-based approach), and (3) **intention** (perceived likelihood of teaching according to the concept-based approach). Items relating to general and epistemological beliefs about the concept-based approach and introductory statistics were also included.

---

[12] Both linear and monotonic relationships were explored.

Table 6: Descriptive Statistics of Attitudinal Items as Appeared in the Questionnaire [a]

| Attitudinal Items | N | Mean | Std. Dev. |
|---|---|---|---|
| 15. The concept-based approach to teaching introductory statistics (rather than emphasizing calulations and formulas) makes students better prepared for further studies. | 227 | 4.00 | .89 |
| 16. Emphasizing concepts and applications in the introductory statistics course (rather than calculations and formulas) is a disservice to our students. | 227 | 1.80 | .93 |
| 17. The concept-based approach to teaching introductory statistics is for low achievers only. | 227 | 1.60 | .74 |
| 18. The concept-based approach to teaching introductory statistics enables students to understand research. | 227 | 4.11 | .74 |
| 19. The concept-based approach to teaching introductory statistics is not theoretically sound. | 227 | 1.92 | .95 |
| 20. Concept-based teaching of introductory statistics may be problematic for me. | 227 | 1.96 | .94 |
| 21. The benefits of concept-based teaching of introductory statistics are clear to me. | 227 | 4.21 | .90 |
| 22. I do not understand how to organize my introductory statistics course to achieve statistical literacy. | 226 | 1.96 | .88 |
| 23. I am engaged in the teaching of introductory statistics using the concept-based approach. | 226 | 3.91 | 1.05 |
| 24. I will need training on how to integrate hands-on computer exercises into the introductory statistics course. | 226 | 2.12 | 1.10 |
| 25. I am convinced that the concept-based approach to teaching introductory statistics enhances learning. | 227 | 4.24 | .82 |
| 26. I may not use the concept-based approach to teach introductory statistics because of limited departmental resources. | 227 | 2.32 | 1.09 |
| 27. It is important for me to network with instructors who are teaching introductory statistics using the concept-based approach. | 227 | 3.41 | 1.15 |
| 28. I am hesitant to use computers in my introductory statistics class without the help of a teaching assistant. | 227 | 2.16 | 1.15 |
| 29. I am concerned that using the concept-based approach to teach introductory statistics may result in me being poorly evaluated by my students. | 227 | 1.90 | .92 |
| 30. It is mostly up to me whether or not I use the concept-based approach to teach introductory statistics. | 227 | 3.93 | 1.13 |
| 31. Using active learning strategies (such as critiquing of research articles, group discussions, and hands-on computer analysis) in my introductory statistics course, may result in students asking me questions which I cannot answer. | 224 | 2.40 | 1.18 |

a. Responses were given on a Likert-type scale (1= strongly disagree, through 3 = undecided, to 5 = strongly agree). Items were classified as follows: 15-19 (perceived usefulness -cognitive), 20-25 (beliefs about preparation-cognitive), and 26-31 (perceived behavioral control - cognitive). Negatively worded items were not reverse-coded for these descriptive statistics.

Although items 26 to 31 (Table 6) relating to perceived behavioral control (PBC) were included based on the advice and recommendation of content and measurement experts, this construct is not theoretically supported as a direct component of the attitude construct. Perceived behavioral control is defined as the degree to which an individual feels that performance or non performance of the behavior in question is under his or her volitional control, and is generally conceptualized as either a moderator or mediator of the intention-behavior relationship (Ajzen, 1991). Expectedly, therefore, most of the PBC items demonstrated low communalities or shared variance (between .1 and .2), and were removed from the item set.

**Table 7: Descriptive Statistics of Attitudinal Items as Appeared in the Questionnaire** [a]

| Attitudinal Items | N | Mean | Std. Dev. |
|---|---|---|---|
| 32. I am not confident in my ability to successfully teach introductory statistics using the concept-based approach. | 227 | 1.93 | 1.09 |
| 33. Teaching introductory statistics using the concept-based approach is likely to be a positive experience for me. | 227 | 4.10 | .92 |
| 34. I am not comfortable using computer applications to teach introductory statistics. | 227 | 1.65 | .92 |
| 35. Teaching introductory statistics with emphasis on concepts and their applications (rather than calculations and formulas) may be stressful for me. | 227 | 1.97 | 1.02 |
| 36. Using computers to teach introductory statistics makes learning fun. | 226 | 3.91 | .92 |
| 37. I am anxious about using active learning strategies (such as projects, group discussions, hands-on computer analysis, critiquing of research articles, oral and written presentations) for the teaching of introductory statistics. | 227 | 2.36 | 1.24 |
| 38. I am interested in using the concept-based approach to teach introductory statistics. | 227 | 4.27 | .83 |
| 39. I want to learn more about the concept-based approach to teaching introductory statistics. | 227 | 3.81 | .98 |
| 40. Using the concept-based approach to teach introductory statistics is not a priority for me. | 227 | 2.21 | 1.04 |
| 41. I plan on teaching introductory statistics according to the concept-based approach. | 226 | 3.96 | .96 |
| 42. I will avoid using computers in my introductory statistics course. | 226 | 1.67 | .95 |
| 43. I will incorporate active learning strategies (such as projects, hands-on data analysis, critiquing research articles, and report writing) into my introductory statistics course. | 226 | 4.14 | .86 |
| 44. I will emphasize calculations and formulas in my introductory statistics course. | 225 | 2.36 | 1.16 |
| 45. Statistics is mathematics, and that is how I intend to teach it. | 226 | 1.94 | .96 |

a. Responses were given on a Likert-type scale (1= strongly disagree, through 3 = undecided, to 5 = strongly agree). Items were classified as follows: 32-38 (affect/feeling) and 39-45 (intention). Negatively worded items were not reverse-coded for these descriptive statistics.

It is worthy of note that the marker item intended for perceived behavioral control (30 in Table 6), "*It is mostly up to me whether or not I use the concept-based approach to teach introductory statistic*" demonstrated little variability, with almost all of the subjects responding "agree" or "strongly agree". Evidently, instructors, in general, believe that they have control over the decision to use a particular teaching methodology, and this may be linked to the almost universal principle of academic freedom. Two PBC items (28 and 29 in Table 6) with communalities of approximately .3 were retained, and each loaded on a different factor (defined a posteriori). Almost all of the general and epistemological belief items (1 to 7 in Table 5) also performed poorly with regard to shared variance (low communalities), and the only item retained from this set (number 6 in Table 5), loaded highly on another of the a priori conceptualized factors. The rejection of the PBC, as well as general and epistemological belief items adds to the construct validity of this study, as theoretically, these (especially PBC) are not considered direct components of attitude.

Furthermore, most of the items initially classified as beliefs about how prepared respondents felt about using the concept-based approach (numbers 20 to 25 in Table 6), statistically redistributed into other clusters with which they proved more meaningful and conceptually relevant. For example, the item: "*I am convinced that the concept-based approach enhances learning*" which was intended to capture mental preparation, loaded on perceived usefulness (consider the phrase "enhances learning"). Indeed, "beliefs about preparation" is not a commonly used component of attitude in the teaching context, however, it was considered relevant given the relative innovative nature of the concept-based approach to teaching introductory statistics. This was also the rationale for including perceived behavioral control, as well as general and epistemological belief items.

In general, therefore, items were removed because of low communality (less than .3) that is sharing less than about 10% variance with the other items (Fullagar, 1986), in conjunction with theoretical relevance and conceptual clarity. One item: *"Statistics is mathematics, and that is how I intend to teach it"* (number 45 in table 7) proved ambivalent (and not readily interpretable), and given that it was also found to be double-barreled, it was removed from the item set. No item was removed because of cross-loading (referred to as item complexity). In fact, the three instances of cross-loading (discussed later) proved very insightful in understanding the factors and their interrelationships, especially for an initial and exploratory study.

**The Teaching Practice Items**

The questionnaire contained ten (10) items which formed the teaching practice scale (Table 8), used to characterize teaching approach on a low to high-reform continuum, consistent with the theories of behaviorism and constructivism respectively. In general, behaviorist-oriented teachers tend to be more preoccupied with subject content, and the transmission of information, and are more likely to engender surface learning, whereas constructivist-oriented teachers are more student-centered, concept-based, and seek to facilitate deep and conceptual learning (Askew et al., 1997; Prosser & Trigwell, 1999; Trigwell & Prosser, 2004). The teaching practice items were generated based on the seminal recommendations of the ASA/MAA[13] Joint Committee on Undergraduate Statistics Education (Cobb, 1992) and the ASA GAISE (Guidelines for Assessment and Instruction in Statistics Education) report on introductory statistics

---

[13] ASA (American Statistical Association)/MAA (Mathematical Association of America)

(Franklin & Garfield, 2006; Garfield, 2004). In accordance with these guidelines, the practice items reflected the following domains: content, pedagogy, assessment, and integration of technology.

Items 1, 4, 6, 8, and 10 (Table 8) reflect behaviorist teaching practices (low-reform), and items 2, 3, 5, 7, and 9 tap, concept-based (constructivist or reform-oriented teaching). There was no clear evidence of response bias toward extreme or middle categories (Table 8).

**Table 8: Descriptive Statistics of Teaching Practice Items as Appeared in the Questionnaire** [a]

| Teaching Practice Items | N | Mean | Std. Dev. |
|---|---|---|---|
| (1) I emphasize rules and formulas as a basis for subsequent learning. ** | 227 | 2.73 | .82 |
| (2) I integrate statistics with other subjects. | 226 | 3.77 | .95 |
| (3) Students use a computer program to explore and analyze data. | 226 | 4.02 | 1.08 |
| (4) I assign homework primarily from the textbook. ** | 226 | 2.87 | 1.13 |
| (5) Critiquing of research articles is a core learning activity. | 227 | 2.98 | 1.15 |
| (6) The mathematical underpinning of each statistical test is emphasized. ** | 227 | 2.78 | .95 |
| (7) I use real-life data for class demonstrations and assignments. | 226 | 4.06 | .75 |
| (8) I require that students adhere to procedures in the textbook.** | 225 | 2.77 | .93 |
| (9) Assessment includes written reports of data analysis. | 225 | 3.59 | 1.12 |
| (10) I assign drill and practice exercises (mathematical) for each topic** | 227 | 2.67 | 1.07 |

[a.] Responses were obtained on a Likert-type scale (1 = Never, to 5 = Always). Behaviorist items (**) were not reverse-coded for these descriptive statistics.

In order to identify and understand the possible structure underlying the interrelationships among the practice items, and evaluate for construct validity, both metric (linear transformation) and non-metric (monotonic transformation) MDS (multidimensional scaling) analyses were conducted. MDS plots the objects (each item) on a map, placing similar objects close to each other and dissimilar objects further apart (Coxon, 1982; (Kruskal & Wish, 1978; Young, 1987). Ordinal and interval measures of similarity (Kendall's tau and Pearson's product moment correlation coefficient respectively) were used to generate 1 to 3-dimensional MDS solutions and spatial configurations of the ten practice items. The two-dimensional solutions were most meaningful and interpretable, and are hereafter reported.

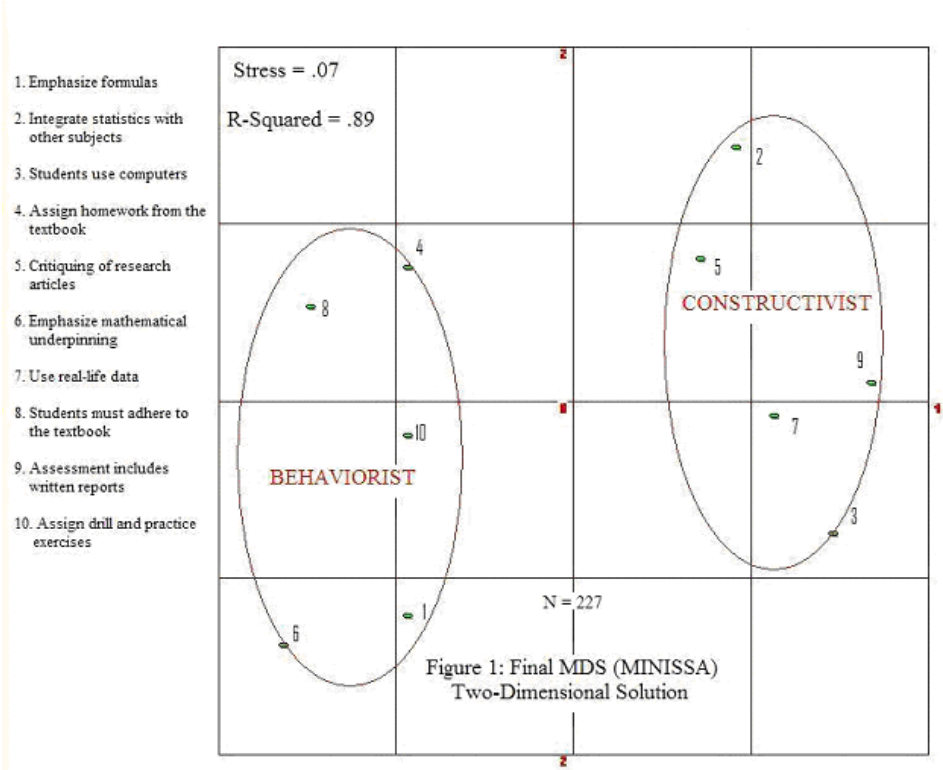

Figure 1: Final MDS (MINISSA) Two-Dimensional Solution

In general, the two metric (MRSCAL) (Roskam, 1972), and two non-metric (MINISSA) (Roskam & Lingoes, 1970) solutions and maps were similar, however, the best fit was obtained with non-metric MDS (MINISSA) using Pearson's correlation as the input measure of similarity. This method is equivalent to a linear regression simultaneous with a monotonic transformation of the original distances (represented by Pearson's correlation coefficient). This solution is therefore explained below. The spatial map (Figure 1) was minimally rotated[14] for interpretability, and reveals two distinct clusters, separating the items as theorized, and empirically supported, that is, consistent with behaviorism (low-reform practice), and constructivism (high-reform practice). A similar MDS outcome was reported by Le, et al. (2004) in their study of reform-oriented teaching practices among mathematics and science teachers.

________________________

[14] Rotation does not alter the solution in the Euclidian distance model.

Also, this solution fits the data very well, with a normalized stress value (residual sum of squares) of .07. Stress values closer to zero represent a better fit. In this case, the stress value is more than two times smaller than stress based on simulation approximation to random data (Spence, 1979). The R-squared value (coefficient of determination) of this solution is 89%, which represents the amount of variance within the data that is explained or accounted for by this two-dimensional solution, suggesting a very good fit. Additionally, the stability of the solution is in keeping with the empirical guideline of at least $4k + 1$ objects for a $k$-dimensional solution with non-metric scaling (Kruskal & Wish, 1978, p. 34). This configuration or pattern was also confirmed with hierarchical cluster analysis (nearest neighbor method with squared Euclidean distance input), as well as principal axis factor analysis (with promax rotation).

### Further Analysis of the Behaviorist Factor of the Teaching Practice Scale

It is worth considering whether or not the "behaviorist" subscale may be artifactual or a pseudofactor, that is, not a truly separate and unique dimension of teaching practice, especially given that the items constituting this factor are exclusively those which were reverse-coded. There is some empirical evidence that negatively worded (and reverse-coded) items tend to cluster together, and load on a separate factor (Cordery & Sevastos, 1993; Marsh, 1996; Schmitt & Stults, 1985). It has been posited that such a factor is a "methods" factor, that is, the items share more variance with the method of measurement (particularly the wording of the items and organization of the scale) than with the construct supposedly being measured. Level of education (a proxy for cognitive ability) has been implicated as the primary underpinning of this outcome, suggesting that less educated subjects may not understand and recognize the structure of such items (Cordery & Sevastos, 1993; Fried & Ferris, 1986; Schmitt & Stults, 1985), and hence respond in a systematically invalid manner.

However, specific elements of this study methodology and analysis establish that the behaviorist factor or subscale is not a "methods" factor, but can be considered a meaningful and separable dimension of teaching practice (Hall et al., 2002; King, n.d.). In particular, the study participants were university and college instructors, whom we can reasonably assume are educated enough to have comprehended the items correctly. Also, these behaviorist practice items are not phrased in the negative (using "not"), but are affirmatively worded reflecting behaviorist teaching practice (albeit intended to be negative in relation to reform-oriented teaching). Furthermore, these practice items were rigorously developed and pilot-tested to ensure their relevance and specificity to this subscale (behaviorist teaching practice), and were theoretically and empirically expected to cluster as emerged. Most importantly, the argument that the behaviorist scale is not a "methods" factor is reinforced by the almost absence of a correlation between the two subscales (behaviorist and constructivist) indicating that they are independent dimensions of teaching practice (Hall et al., 2002).

Table 9 shows Cronbach's alphas as measures of internal consistency (reliability) for the overall teaching practice scale, as well as each subscale. The overall scale has a Cronbach's alpha of .6 which is an acceptable (minimum) level of reliability for exploratory studies (Nunnally, 1974; Robinson, Shaver, & Wrightsman, 1991). Each of the two clusters or subscales is more internally consistent than the overall scale, supporting the MDS finding and theoretical expectation that two dimensions (behaviorist and constructivist) underlie teaching practice. In fact, as shown here (and established in the literature), when multiple latent dimensions underlie a scale, the overall Cronbach's alpha can be deflated (Yu, 2001). Deletion of any item did not appreciably improve the internal consistency, and furthermore, each item tapped a specific and necessary element of teaching practice in this context. The item-total correlations for both the behaviorist and constructivist subscales were .3 or greater, with means of .37 and .42 respectively. According to Nunnally and Bernstein (1994), item-total correlation should be at least .3 in order for the item to be considered a meaningful contribution to the scale.

**Table 9: Teaching Practice Subscales (Overall Cronbach's Alpha = .6)**

| Constructivist (alpha = .66) <br> N = 222, Mean = 18.4, SD = 3.3 | *Behaviorist (alpha = .61) <br> N = 224, Mean = 13.8. SD = 3.1 |
|---|---|
| 1. I integrate statistics with other subjects. | 1. I emphasize rules and formulas as a basis for subsequent learning. |
| 2. Students use a computer program to explore and analyze data. | 2. I assign homework primarily from the textbook. |
| 3. Critiquing of research articles is a core learning activity. | 3. The mathematical underpinning of each statistical test is emphasized. |
| 4. I use real-life data for class demonstrations and assignments. | 4. I require that students adhere to procedures in the textbook. |
| 5. Assessment includes written reports of data analysis. | 5. I assign drill and practice exercises (mathematical) for each topic. |
| Responses (1=never, 2=rarely, 3=sometimes, 4=usually, 5=always). *Items were reverse-coded for the overall teaching practice score, so that higher values reflect more favorable (reform-oriented, concept-based or constructivist) practice. Inter–subscale correlation: Pearson's r = -.06, df=217, ns. An alpha of .6 is a recommended minimum for exploratory studies (Nunnally, 1967; Robinson, Shaver, & Wrightsman, 1991). | |

## Correlation between the Teaching Practice Subscales

Individual scores were calculated for each dimension (subscale) based on the sum of the item values (not reverse-coded) in each cluster. The maximum possible score for each subscale was therefore 25 (5 items x 5) and this wider scale rather than the mean (using the original 5-point scale) was used to allow for more meaningful differentiation in teaching practice. According to Pearson's product moment correlation analysis, the two subscales are almost orthogonal (r = -.06, df = 217, ns)[15] indicating that these components are largely independent of each other. In fact, Brooks and Brooks (1993) reported that these two approaches (constructivist and behaviorist) are contrasting perspectives on teaching, consistent with the finding in this study.

Further evidence in support of this finding is noted in two major studies. Woolley, Benjamin and Woolley (2004) in their study of the construct validity of a self-report measure of teacher beliefs related to constructivist and traditional (behaviorist) approaches to teaching and learning, reported a correlation coefficient of -.011 for the relationship between the constructivist teaching and traditional (behaviorist) teaching subscales. And Handal (2002) reported a similar finding, noting that among mathematics teachers, constructivist and behaviorist beliefs and practice were independent of each other (r = -0.232). This consistent negative statistical relationship suggests the potential for an increase in the level of constructivist practice to be linked to lower levels of behaviorist practice, and vice versa.

**Selected Factors as Determinants of Teaching Practice Subscale Scores**

Subscale scores (constructivist and behaviorist) did not vary significantly with respect to gender, age, ethnicity, duration of teaching, teaching area, membership status in professional organizations, degree concentration, and employment status. However, a statistically significant and meaningful difference was noted among instructors (on the behaviorist scale) with respect to their geographic location (Table 10). Instructors from international locations, on average, reported a significantly lower level of behaviorist practice than those from the USA. A significant difference was also observed with regard to highest academic degree concentration (Table 11). Those with mathematics and engineering degrees had the highest level of behaviorist practice compared to those with health sciences degrees, who had the lowest level on this scale.

---

[15] Pearson's r = -.09 when corrected for attenuation due to measurement error.

**Table 10: Comparison of Teaching Practice Subscale Scores by Geographic Location**

| Teaching Practice Subscales | Geographic Location | N | Mean | Std. Deviation | t | Sig. |
|---|---|---|---|---|---|---|
| Constructivist | USA | 161 | 18.34 | 3.28 | .66 | .509 |
| | International | 56 | 18.68 | 3.53 | | |
| Behaviorist | USA | 164 | 14.32[b] | 2.85 | | |
| | International | 55 | 12.13[b] | 3.27 | 4.78 | .001 |

b. Items were not reverse-coded for these statistics, therefore, a higher value represents a higher level of behaviorist practice (and hence less reform-oriented and less concept-based).

**Table 11: Comparison of Teaching Practice Subscale Scores by Degree Concentration**

| Teaching Practice Subscales | Highest Academic Degree Concentration | N | Mean | Std. Dev. | F | Sig. |
|---|---|---|---|---|---|---|
| Constructivist | Statistics | 90 | 18.33 | 3.50 | .335 | .854 |
| | Psychology/Behavioral Sciences | 71 | 18.70 | 3.20 | | |
| | Health Sciences [a] | 27 | 18.33 | 3.40 | | |
| | Math & Engineering | 16 | 17.94 | 2.52 | | |
| | Education [b] | 18 | 17.89 | 3.91 | | |
| Behaviorist [e] | Statistics | 89 | 13.91 | 3.22 | 2.48 | .045[c] |
| | Psychology/Behavioral Sciences | 71 | 13.80 | 3.06 | | |
| | Health Sciences [a] | 28 | 12.39[d] | 3.22 | | |
| | Math & Engineering | 17 | 15.24[d] | 2.88 | | |
| | Education [b] | 19 | 13.89 | 1.49 | | |

a. Includes public health, epidemiology, and biostatistics

b. Includes 4 instructors from business and operations research

c. Kruskal Wallis test: p = .118

d. Post-Hoc Analysis (with Bonferroni correction): p = .026, the other pairwise comparisons are NS.

e. Items were not reverse-coded for these descriptive statistics, therefore, a higher value represents a higher level of behaviorist practice (and hence less reform-oriented, and less concept-based).

## The Overall Teaching Practice Scale (Composite Score and Practice Categorization)

The responses were obtained on as scale of 1 (Never) through 5 (Always), and a composite score was computed for each respondent by summing their values on each of the ten items, with a possible maximum score of 50 (higher scores reflect more reform-oriented practice, consistent with constructivist and concept-based teaching). For this calculation, the behaviorist items were reverse-coded so that higher values reflect more positive (and reform-oriented) practice. The distribution of practice score (Figure 2) has a mean and median of 35 and a mode of 36. The standard deviation is 5, and the lowest and highest quartiles are 32 and 38 respectively.

Respondents in the highest quartile were considered high-reform instructors (in tune with concept-based or constructivist teaching), whereas those in the lowest quartile were labeled low-reform instructors (more akin to the traditional, mathematical or behaviorist approach). This categorization was used in order to obtain a meaningful differentiation among teachers with respect to practice, and was used in exploratory factor analysis to evaluate for criterion validity of the attitude measure and subscales.

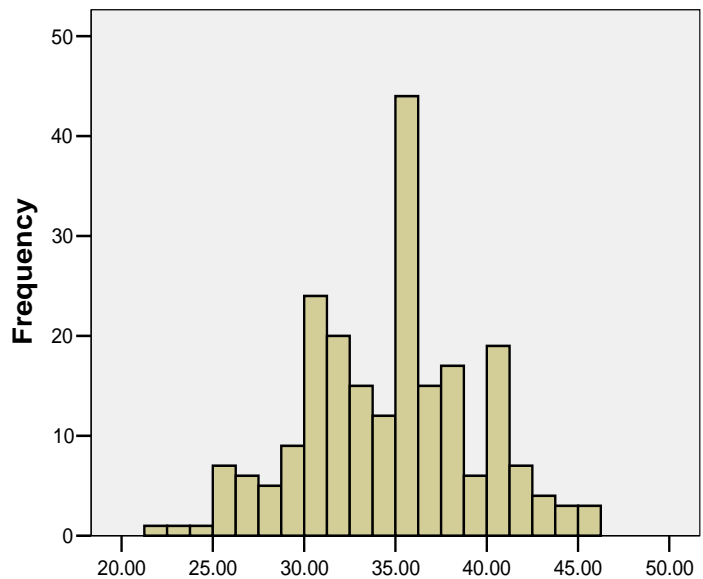

**Figure 2: Overall Teaching Practice Score**
**(N = 219)**

The validity of this classification was shown by noting that those in the lowest quartile (low-reform) had a significantly lower level of constructivist practice and higher behaviorist practice compared to instructors in the highest quartile (who showed the reverse pattern) (Table 12), as theoretically and empirically expected.  In total, the psychometric evidence establishes that these two subscales are independent measures of different dimensions of teaching practice.  Furthermore, all instructors reported some degree of each practice (behaviorist and constructivist), and hence can be considered eclectic in their teaching (pedagogical mixing) (Lenski, Wham, & Griffey, 1998; Ravitz, Becker, & Wong, 2000; Woolley, Benjamin, & Woolley, 2004). In other words, these teaching practices (behaviorist and constructivist) coexist.

**Table 12: Comparison of  Teaching Practice Subscale Scores by Overall Teaching Practice Score Quartiles**

| Teaching Practice Subscales | Overall Teaching Practice Score | N | Mean | Std. Dev. | t | Sig. |
|---|---|---|---|---|---|---|
| Constructivist | Lowest Quartile | 74 | 15.78 | 2.97 | 12.4 | .001[a] |
| | Highest Quartile | 59 | 21.29 | 2.13 | | |
| Behaviorist | Lowest Quartile | 74 | 16.26[c] | 2.21 | 13.4 | .001 |
| | Highest Quartile | 59 | 10.92[c] | 2.37 | | |

[a]. Mann-Whitney test: p <.001

[c]. Items were not reverse-coded for these statistics, therefore, a higher value represents a higher level of behaviorist practice (and hence less reform-oriented, and less concept-based).

## Exploratory Factor Analysis of the Twenty-Five (25) Retained Attitudinal Items

Principal axis factor analysis (PAF) was the factor extraction method used, and the retained factors were rotated to simple structure using the oblique rotation algorithm, promax (Fabrigar et al., 1999; Russell, 2002). In addition to analyzing common variance only, which is required when the objective is to identify underlying latent constructs, it has been established that PAF with squared multiple correlations (SMC) as the initial estimate of communality, provides more accurate results in terms of the population factor loadings over principal components analysis (Russell, 2002; Widaman, 1993). With regard to the oblique rotation, promax first conducts an orthogonal varimax rotation and then allows correlations between the factors toward achieving simple structure. Therefore, if the factors are orthogonal (uncorrelated) with one another, that will be revealed with promax rotation. Moreover, it is reasonable to expect that psychological constructs will be correlated (Fabrigar et al., 1999).

**Data Screening**

The data were screened for the basic assumptions underlying factor analysis including factorability of the data. A data set is factorable if there is a substantial number of meaningful interrelationships among the items (usually Pearson's correlation coefficient of .3 or greater). A visual inspection of the full correlation matrix showed that this standard was met, and that there was no evidence of statistical redundancy. More objectively, the level of significance of the Bartlett's test of sphericity, chi squared (df =300, N=219) = 2582, p<.001, and the high value of the Kaiser-Meyer-Olkin measure of sampling adequacy (.88) indicate that the set of items has adequate common variance, and hence acceptable factorability (Tabachnick & Fidell, 1996).

Additionally, the skew (-1.8 to .38, Mean = -.9502) and kurtosis (-1.13 to 3.75, Mean = .834) of the items are lower than the recommended upper limits for questioning multivariate normality (skew >2, kurtosis > 7) (West et al., 1995; Fabrigar et al., 1999). While the assumption of multivariate normality is necessary for factor analytic methods such as maximum likelihood, and confirmatory where significance testing of loadings and models is necessary, it is not required for PAF (or least squares factor analysis) which is a distribution-free statistical method. Nonetheless, it has been noted that "all methods of factor analysis are more likely to yield clearer, more replicable factor patterns if the data conform to multivariate normality" (Floyd & Widaman, 1995, p. 8). Also, the generally recommended conservative sample size of at least 200, and a participant to variable ratio of 5:1 were met (Floyd & Widaman, 1995; Gorsuch, 1983).

**Communalities**

Table 13 shows the initial and extracted communality values for each item. The initial estimate is the squared multiple correlation (SMC) obtained when each item is regressed on the other items. For example, if your regress items 2 through 25 on item 1 (Table 13, "initial"), the SMC would be .65 or 65%, representing the amount of variance in item 1 that can be explained by the other items, and this is initially taken as a reasonable estimate of each item's variance that can be accounted for by the extracted factors. The extracted communality is derived after the number of factors is determined, by an iterative process "until the estimates of the communalities converge (i.e., change minimally from one factor extraction to another)" (Russell, 2002).

The extracted communality indicates the proportion of each item's variance that can be explained by the retained factors. Items with high values are well represented in the common factor space. With reference to a particular item, a low communality (e.g. .2 or less than 20%) (Fullagar, 1986) suggests that the item has little in common with the other items, and hence is not important to the analysis, and could be removed barring strong conceptual or theoretical relevance. As is evident from Table 13, the final communalities (extracted) range from approximately .3 (30%) to .77 (77%) with a mean and median of .51 (51%) and .50 (50%) respectively, underscoring the relevance and meaningful contribution of these items to this analysis and solution.

**Table 13: Initial and Extracted Communalities of the Retained Attitudinal Items**

| | Initial[a] | Extracted |
|---|---|---|
| 1. The concept-based approach to teaching introductory statistics (rather than emphasizing calculations and formulas) makes students better prepared for work. | .65 | .63 |
| 2. Teaching introductory statistics with emphasis on concepts and applications rather than calculations and formulas, can be time consuming. | .37 | .44 |
| 3. I will adjust easily to teaching introductory statistics using the concept-based approach. | .50 | .50 |
| 4. The preparation required to teach introductory statistics using the concept-based approach is burdensome. | .37 | .38 |
| 5. Using active learning strategies (such as projects, group discussions, oral and written presentations) in the introductory statistics course can make classroom management difficult. | .39 | .45 |
| 6. The concept-based approach to teaching introductory statistics (rather than emphasizing calculations and formulas) makes students better prepared for further studies. | .51 | .53 |
| 7. Emphasizing concepts and applications in the introductory statistics course (rather than calculations and formulas) is a disservice to our students. | .61 | .62 |
| 8. The concept-based approach to teaching introductory statistics is for low achievers only. | .57 | .52 |
| 9. The concept-based approach to teaching introductory statistics enables students to understand research. | .49 | .51 |
| 10. Concept-based teaching of introductory statistics may be problematic for me. | .64 | .65 |
| 11. I do not understand how to organize my introductory statistics course to achieve statistical literacy. | .45 | .49 |
| 12. I am engaged in the teaching of introductory statistics using the concept-based approach. | .63 | .63 |
| 13. I am convinced that the CB approach enhances learning. | .70 | .70 |
| 14. Teaching introductory statistics using the concept-based approach is likely to be a positive experience for me. | .54 | .51 |
| 15. I am not comfortable using computer applications to teach introductory statistics. | .40 | .36 |
| 16. Teaching introductory statistics with emphasis on concepts and their applications (rather than calculations and formulas) may be stressful for me. | .37 | .36 |
| 17. Using computers to teach introductory statistics makes learning fun. | .41 | .40 |
| 18. I am interested in using the concept-based approach to teach introductory statistics. | .65 | .63 |
| 19. I want to learn more about the concept-based approach to teaching introductory statistics. | .44 | .46 |
| 20. Using the concept-based approach to teach introductory statistics is not a priority for me. | .58 | .61 |
| 21. I plan on teaching introductory statistics according to the concept-based approach. | .72 | .77 |
| 22. I will avoid using computers in my introductory statistics course. | .48 | .70 |
| 23. I will incorporate active learning strategies (such as projects, hands-on data analysis, critiquing research articles, and report writing) into my introductory statistics course. | .46 | .38 |
| 24. I am hesitant to use computers in my introductory statistics class without the help of a teaching assistant. | .31 | .29 |
| 25. I am concerned that using the concept-based approach to teach introductory statistics may result in me being poorly evaluated by my students. | .35 | .34 |

a. This initial estimate is the squared multiple correlation (SMC) obtained when each item is regressed on the other items.

## Factor Extraction

The decision regarding the number of factors to retain was based on the Kaiser's criterion (Gorsuch, 1983) and the "scree test" (Cattell, 1966; Cattell & Jaspers, 1967) in conjunction with theoretical plausibility, and the principle of parsimony (Fabrigar et al., 1999). The Kaiser's criterion recommends that only factors which account for more variance than a single variable should be extracted given that the goal of factor analysis is to account for as much variance as possible in a set of observed items, with a smaller number of latent common factors. Therefore, for this technique, factors with eigenvalues greater than 1 are analysis was performed on the correlation matrix, the variables are standardized, therefore, each variable has a variance of 1, and the total variance is equal to the number of variables used in the analysis, that is 25. Accordingly, this approach indicated 5 factors with eigenvalues ranging from 1.28 (5.1%) to 8.1 (32.55) which explained 61 % of the total variance (Table 14). Note however, that this initial extraction is based on PCA (principal components analysis).

**Table 14: Percentage of Variance Explained**

| Factor | Initial Eigenvalues[a] | | | Extraction Sums of Squared Loadings[b] | | | Promax Rotation |
|---|---|---|---|---|---|---|---|
| | Total | % of Variance | Cumulative % | Total | % of Variance | Cumulative % | Total |
| 1 | 8.12 | 32.49 | 32.49 | 7.70 | 30.79 | 30.79 | 16.36[c] |
| 2 | 2.54 | 10.17 | 42.66 | 2.00 | 7.99 | 38.78 | 13.84[c] |
| 3 | 2.00 | 7.99 | 50.65 | 1.52 | 6.07 | 44.85 | 8.76[c] |
| 4 | 1.34 | 5.37 | 56.03 | .86 | 3.42 | 48.27 | 7.08[c] |
| 5 | 1.28 | 5.13 | 61.16 | .78 | 3.11 | 51.39 | 5.36[c] |

a. Based on principal component analysis (PCA)

b. Based on principal axis factoring (PAF)

c. Following promax (oblique) rotation the percentage of variance explained was calculated using Cattell's formula: variance = the sum of the Structure loadings x Pattern coefficients for all variables (Barrett & Kline, 1980; Cattell, 1978).

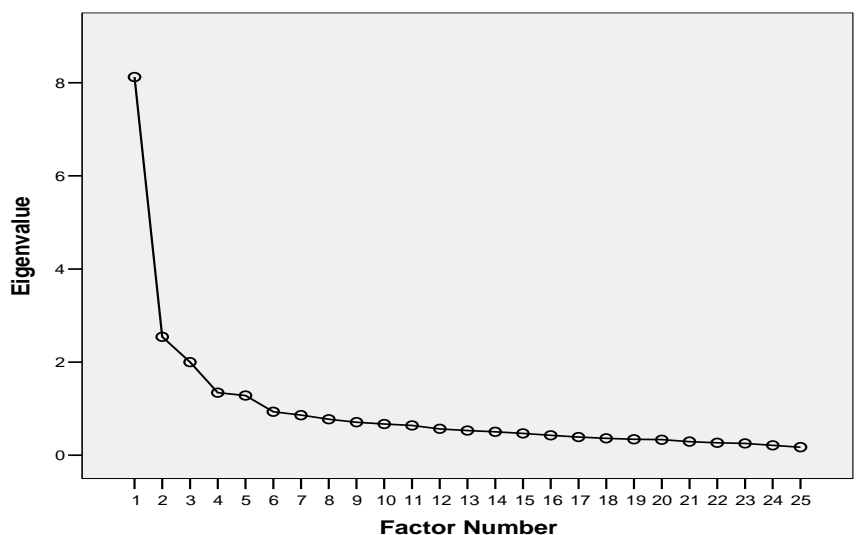

Figure 3: Scree Plot (Eigenvalues of Full Correlation Matrix)

This technique was complemented with the Cattell's scree test, which is generally considered to be more conservative than the Kaiser's criterion. This is a plot of the eigenvalues (and corresponding factors) of the full correlation matrix (Figure 3), and the maximum number of factors is indicated by the point before the plot levels off, that is, the point preceding the "elbow". Beyond this point, factors contribute negligible variance. The plot of the eigenvalues of the reduced correlation matrix was also examined, as the focus of this study in on common variance (Fabrigar et al., 1999). For these techniques, the first factor accounts for the largest amount of variance, and successive factors account for progressively smaller amounts of variance. Both the Kaiser's criterion and the scree test suggested a maximum of five factors. Therefore, one, two, three, four, and five-factor solutions were examined, and in the final analysis the five-factor solution was considered the most conceptually clear, theoretically sound, and interpretable.

**The Final 5-Factor Solution**

The final factor solution contained five (5) factors extracted by principal axis factor analysis (PAF), and was obliquely rotated to simple structure with the promax algorithm (kappa = 4). The percentage of variance explained by each rotated factor (Table 15) was calculated using Cattell's formula (Barrett & Kline, 1980; Cattell, 1978). That is, the variance explained by each rotated factor is equal to the sum of the product of the structure loadings and pattern coefficients for all variables. This 5-factor solution explained 51% of the total variance underlying the attitudinal items, and all (100%) of the common variance. A widely used and practical criterion for determining the acceptability of a factor solution, is that the solution should explain at least 50% of the total variance and close to 100% (usually not less than 80%) of the common variance (Floyd & Widaman, 1995; Streiner, 1994; Tinsley & Tinsley, 1987). For this study, the primary objective was to identify the latent factors underlying the common variance of the set of attitudinal items, and hence a more critical consideration is the proportion of common variance explained (Tinsley & Tinsley, 1987).

**Interpretation and Naming of the Factors (Subscales)**

Interpretation and naming of the factors (following oblique rotation) were based primarily on the factor pattern matrix coefficients (Table 15) and the marker items. The pattern matrix coefficients are equivalent to standardized regression coefficients (weights), and reflect the relative and independent contribution of each factor to the variance of the item on which it loads (Russell, 2002). Whereas the structure matrix loading is a measure of the association between each item and the factor on which it loads, when the factors are correlated (as in this study) there is overlap among the loadings, which makes the structure matrix loadings biased estimates of the independent relationship between the item and the factor (Russell, 2002). It is for this reason that interpretation of the factors was based on the pattern matrix coefficients.

**Table 15:  Factor Pattern Matrix (N = 221)**

| Attitudinal Items (see Table 13 for full text)<br>CBA = The Concept-Based Approach (Pedagogy) | * | \*Factor (overall alpha = .89) | | | | | |
|---|---|---|---|---|---|---|---|
| | | **1** | **2** | **3** | **4** | **5** | **H²** |
| 1. CBA makes students better prepared for further studies | U | **.79** | -.06 | .01 | -.02 | -.11 | .53 |
| 2. *CBA is a disservice to our students* | U | **.79** | .03 | .03 | -.04 | -.11 | .62 |
| 3. CBA enables students to understand research | U | **.79** | -.22 | .13 | -.08 | .09 | .51 |
| 4. CBA makes students better prepared for work | G | **.69** | .17 | -.05 | -.03 | .03 | .63 |
| 5. *CBA is for low achievers* | U | **.61** | -.03 | .31 | .07 | -.15 | .52 |
| 6. I am convinced that the CBA enhances learning | P | **.57** | **.39** | -.01 | -.14 | -.04 | .70 |
| 7. Using CBA is likely to be a positive experience for me | A | **.51** | .26 | -.10 | .02 | .06 | .51 |
| 8. I am interested in using CBA | A | **.42** | **.40** | -.03 | .10 | .04 | .63 |
| 9. I plan on teaching according to CBA | I | -.02 | **.86** | .07 | .04 | -.03 | .77 |
| 10. *CBA is not a priority for me* | I | -.01 | **.82** | .00 | -.05 | -.07 | .61 |
| 11. I am engaged in the CBA | P | .01 | **.76** | .20 | -.09 | -.16 | .63 |
| 12. I want to learn more about CBA | I | .13 | **.42** | -.28 | .26 | .17 | .46 |
| 13. *I do not understand how to organize my course to achieve literacy* | P | .00 | .20 | **.61** | .13 | -.08 | .49 |
| 14. Using CBA may be stressful for me | A | -.06 | .03 | **.54** | .16 | .10 | .36 |
| 15. *I am concerned that CBA may result in poor evaluation of me by students* | C | .29 | -.22 | **.49** | .12 | .07 | .34 |
| 16. *CBA may be problematic* | D | .08 | **.41** | **.44** | -.03 | .14 | .65 |
| 17. I will adjust easily to CBA | D | .05 | .23 | **.42** | -.09 | .27 | .50 |
| 18. *I will avoid using computers in my course* | I | -.16 | .08 | .08 | **.85** | -.13 | .70 |
| 19. *I am not comfortable using computers to teach statistics* | A | .00 | -.12 | .33 | **.56** | -.07 | .36 |
| 20. *I am hesitant to use computers without a teaching assistant* | C | -.07 | -.08 | .37 | **.45** | .00 | .29 |
| 21. Using CBA makes learning fun | A | .38 | -.07 | -.09 | **.43** | .14 | .40 |
| 22. I will incorporate active learning strategies into my course | I | .00 | .33 | .06 | **.37** | .09 | .38 |
| 23. *Using active learning strategies can make classroom management difficult* | D | .00 | -.07 | -.03 | .05 | **.69** | .45 |
| 24. *Emphasizing concepts and applications can be time consuming* | D | -.16 | -.09 | .17 | -.12 | **.61** | .44 |
| 25. *Preparation for CBA is burdensome* | D | -.03 | -.03 | .34 | -.09 | **.44** | .38 |
| **Eigenvalue** (reduced correlation matrix) following rotation | | 4.09 | 3.46 | 2.19 | 1.77 | 1.34 | 12.86 |
| **Percent of Total Variance** | | 16.36 | 13.84 | 8.76 | 7.08 | 5.36 | 51.4 |
| **Percent of Common Variance** | | 31.8 | 26.9 | 17.0 | 13.8 | 10.4 | 100 |
| **Cronbach's alpha** (internal consistency) | | .88 | .85 | .77 | .69 | .65 | |

*\*Initial classification of items:* **U** = Perceived Usefulness, **G** =General & Epistemological Belief, **A** =Affect, **I** = Intention, **D** = Perceived Difficulty, **C** = Perceived Behavioral Control, **P** = Beliefs about Preparation. **Grey shaded regions** indicate the pattern coefficients for the items forming each factor/subscale. **The horizontal line shading** indicates cross-loading.    **H²** = Final community estimate. **Italicized items were reverse-coded**. Extraction Method: **Principal Axis Factoring**. Rotation Method: **Promax with Kaiser Normalization**. The percent of variance explained by each rotated factor was calculated using Cattell's formula (Barrett & Kline, 1980; Cattell, 1978).

In general, Thurstone's criterion for simple structure (Thurnstone, 1956) was achieved, as 22 (96%) of the items loaded highly on only one factor (Tinsley & Tinsley, 1987). Pattern coefficients ("loadings") of .4 and higher (that is, a factor explaining at least 16% of an item's variance) were considered salient for a sample size of about 200, so as not to capitalize on chance (Field, 2000; Hair, et al., 1998; Stevens, 1992). However, both high and low pattern coefficients were noted as they tell us what the factor most likely is, and is not (Rummel, 1970). Items number 6, 8 and 16 (Table 15) cross-loaded, however, all instances of cross-loading provided valuable insight into the interpretation, and naming of the factors. Interpretation and naming of the subscales were also informed by Cronbach's alpha which is a measure of internal consistency (reliability).

While a Cronbach's alpha of .7 is the lower limit generally used to establish acceptable reliability, a recommended minimum standard for exploratory studies is .6 (Nunnally, 1974; Robinson, Shaver, & Wrightsman, 1991). Furthermore, according to Nunnally and Bernstein (1994), item-total correlations should be at least .3 in order for an item to be considered a meaningful contribution to the scale, and this standard was met for all the subscales. Deletion of any items did not appreciably improve the alpha, and although a marginal increase was noted for certain items (if deleted) each item tapped a specific and necessary element of each domain, and hence all were retained. The factor analysis solution (FA) was interpreted in conjunction with the MDS configurations (detailed later), and hierarchical cluster analysis (HC).

**Table 16: Descriptive Statistics of the Attitude Factors (Subscales)**

| Factors (Subscales) | Mean | Attitude Component | Std. Dev. | Alpha | N |
|---|---|---|---|---|---|
| 1. Perceived Usefulness | 4.15 | Cognition | .65 | .88 | 227 |
| 2. Intention | 3.95 | Intention | .77 | .85 | 225 |
| 3. Personal Teaching Efficacy | 4.04 | Cognition | .68 | .77 | 225 |
| 4. Avoidance-Approach | 4.12 | Affect | .64 | .69 | 225 |
| 5. Perceived Difficulty | 3.07 | Cognition | .87 | .65 | 226 |
| **Overall Attitude Scale** | **3.95** | **Tripartite** | **.50** | **.89** | **221** |

## FACTOR # 1 (PERECEIVED USEFULNESS)

This factor or subscale (1-7 in Table 15) explained the largest proportion of total (16.36 %) and common variance (31.8 %). This cluster of items depicts beliefs primarily about the value, usefulness or benefits of the concept-based approach to teaching introductory statistics with regard to students' preparation for further studies, work, and research, as well as general enhancement of teaching and learning. Seven items loaded on this factor, with pattern coefficients ("factor loadings") ranging between .51 and .79. Item number 8: *"I am interested in using the concept-based approach to teach introductory statistics"* cross-loaded on factor number 2, which reflects "intentionality", and therefore this item was included with that factor, with which it is more conceptually relevant and clear. Also, item number 6: *"I am convinced that the concept-based approach to teaching introductory statistics enhances learning"* tended to cross-load on factor number 2 with a pattern coefficient of .39, however, this item is clearly more substantively relevant to perceived usefulness, on which it loads higher.

This subscale demonstrated high internal consistency (reliability) with a Cronbach's alpha of .88, strongly suggesting that this set of items is tapping a common underlying concept. There was no meaningful change in Cronbach's alpha (internal consistency) if any of the items was deleted. The item-total correlations varied between .6 and .72, with most being close to or slightly above .7. Two marker items: *"The concept-based approach to teaching introductory statistics (rather than emphasizing calculations and formulas) makes students better prepared for further studies"*, and *"The concept-based approach to teaching introductory statistics enables students to understand research"* (1 and 3 in Table 15) loaded highest on this factor. This subscale was labeled **PERCEIVED USEFULNESS** (Table 16), which is theoretically and empirically established as a salient cognitive dimension of attitude, and a strong and significant predictor of intention and behavior (Ajzen & Fishbein, 1977; Davis, Bagozzi, & Warshaw, 1989; Taylor & Todd, 1995; Venkatesh & Davis, 2000). Davis (1989) defines perceived usefulness as **"**the degree to which a person believes that using a particular system would enhance his or her job performance".

### FACTOR # 2 (INTENTION)

This factor consists of five items (8-12 in Table 15), and explained 13.84 % of the total variance, and 26.9 % of the common variance. As mentioned above, item number 8: *"I am interested in using the concept-based approach to teach introductory statistics"* cross-loaded on factor number 1, but was retained for this factor, as it is clearly more pertinent to an expression of intention rather than usefulness. Also, item number 16: *"Concept-based teaching of introductory statistics may be problematic for me"* cross-loaded on this factor, but was not included, given its greater relevance to, and higher loading on the other factor. The retained items with stems such as: **"I Plan to"**, **"I am interested"**, **"I want to"**, and **"I am engaged"**, strongly suggest intentionality, and therefore this factor was labeled **INTENTION** (Table 16). Ajzen (1985,1991) defines intention as "an indication of a person's readiness to perform a given behavior", and he notes that "it is considered to be the immediate antecedent of behavior".

Intention is empirically and theoretically supported as a component of the tripartite attitude model (Breckler, 1984; Rosenberg & Hovland, 1960; Smith, 1947), and a strong, direct, and significant predictor of behaviors, especially those under volitional control (Ajzen, & Fishbein, 2005). Intention has also been established as a mediator of the attitude-behavior relationship in accordance with the theories of reasoned action and planned behavior (Ajzen & Fishbein, 2005; Kothandapani, 1971; MCGuire, 1969; Widaman, 1985). Of note is that the item: *"I plan on teaching introductory statistics according to the concept-based approach"* (number 6 in Table 15) has the highest loading on this factor, and was the marker item for intention. This subscale has a Cronbach's alpha of .85 (Table 16) indicating high internal consistency. There was no meaningful change in Cronbach's alpha if any of the items was removed. The item-total correlations ranged from .5 to .69, with four being between .65 and .78.

**FACTOR # 3 (PERSONAL TEACHING EFFICACY)**

Five items (13 -17 in Table 15) loaded on this factor, which explained 8.76 % of the total variance and 17 % of the common variance. As mentioned above, item number 16: *"Concept-based teaching of introductory statistics may be problematic for me",* cross-loaded on factor number 2 (intention), but was assigned to this factor because of greater conceptual clarity, and a higher loading (pattern coefficient). Notably, these items contain the referent **"I"** or **"me"** (Koul & Rubba, 1999), and reflect personal evaluation of, and concerns about one's ability to implement and effectively use concept-based pedagogy. This theme is analogous to the concept of self-efficacy (Bandura, 1977, 1989, 1998), defined as "belief in one's capabilities to organize and execute the courses of action required to produce given levels of attainments" (Bandura, 1998, p. 624).

More specifically, Bandura (1997) posits that self-efficacy is a context-specific and bi-dimensional construct, comprised of efficacy expectations (belief in personal capability to effect a behavior), and outcome expectations (belief that the behavior will effect a specific outcome). These dimensions have been translated, operationalized, and validated in the teaching context as "personal teaching efficacy" and "general teaching efficacy" respectively (Ashton, 1984; Enochs & Riggs, 1990; Gibson & Dembo, 1984). Given that the five items comprising this factor relate to internal or personal capacity to successfully use the concept-based approach to teach introductory statistics, this subscale was labeled **"PERSONAL TEACHING EFFICACY"** (Gibson & Dembo,1984; Nietfeld & Cao, 2003).

Self-efficacy was not an a priori dimension of attitude in this study, instead, a related construct, perceived behavioral control was used (Ajzen, 2002; Armitage & Conner, 1999), however, almost all of those items were dropped because of low communalities. It is well-established, that self-efficacy beliefs can serve as barriers to, and facilitators of action (Ajzen, 2002). Such beliefs can be linked to internal or personal factors (such as ability, skills, knowledge, competence, and sense of preparedness), as well as external factors (such as task demands, material resources, and the actions of others), a classification which is akin to the concept of locus of control (Ajzen, 2002; Rotter, 1966). Moreover, according to Leach, Hennessy and Fishbein (2001), perceived self-efficacy is a direct measure of attitude. The Cronbach's alpha for this subscale is .77 (Table 16) reflecting acceptable internal consistency. Additionally, the item-total correlations are between .41 and .64, with four being between .49 and .64. The deletion of any item did not appreciably alter the Cronbach's alpha.

**FACTOR # 4 (AVOIDANCE-APPROACH)**

This factor consists of 5 items (18-22 in Table 15), and explained 7.08% and 13.8% of the total and common variance respectively. Item number 22: *"I will incorporate active learning strategies (such as projects, hands-on data analysis, critiquing research articles, and report writing) into my introductory statistics course"* loaded at .37, slightly below the established minimum loading (pattern coefficient) of .4, however, given its closeness to .4, its core relevance to concept-based pedagogy, as well as the gain in interpretability, this item was retained. This set of seemingly disparate items reflects a single concept as explained below. In particular, item number 18 which has the highest loading on this factor, relates directly to **"avoidance"** (*I will avoid using computers in my introductory statistics course*), and at the other extreme, item number 22 (*I will incorporate active learning strategies, such as projects, hands-on data analysis, critiquing research articles, and report writing into my introductory statistics course*) suggests **"approach"**. The other items are predicated on affect, motivation, feelings or emotions (see "comfortable", "hesitant", and "fun").

Together, these items are consistent with the **"AVOIDANCE-APPROACH"** concept, and was so labeled. Townsend, Busemeyer, et al. (1989, p.116) describe this construct as, "the single goal approach-avoidance situation where a person is both attracted and repulsed by a single goal or choice object. Thus, her/his choice is basically whether the positive aspects of the goal outweigh the negative aspects". The authors further noted that the approach-avoidance theme reflects "a continuum of response possibilities." That is, varying degrees of psychological proximity to the behavior, which better captures reality and psychological predisposition, rather than "the more traditional all-or-none depiction".

The avoidance-approach construct is generally theorized as having an affective basis, and this is supported by researchers and theorist who have identified "positive emotionality/temperament" and "negative emotionality/temperament" as the underpinnings of avoidance and approach behavior respectively (Clark & Watson, 1999). Gray (1970) characterized the mechanism underlying this concept as "facilitative and inhibitory motivational systems", which he posited, produce positive and negative affect respectively. Consistent with this conceptualization, recent empirical evidence has shown that items measuring avoidance cluster with affect and motivation. Specifically, Yamashita (2004) in measuring affective reactions as a component of reading attitude among university students, reported the item: "*If it is not necessary, I prefer to* ***avoid*** *reading as much as possible*" as the strongest contributor (loading = .9) to the affective component of attitude. Also, Tapia and March (2004) in their study to design an instrument to measure attitudes toward mathematics, reported the item: "*I would like to* ***avoid*** *using mathematics in college*" as an item of the motivation subscale. Additionally, the items at the extremes of this subscale (items 18 and 22 in Table 15) are conceptually and structurally similar to the extremes of an avoidance-approach scale reported by Greeley et al. (1993a & b) in their work on craving and substance use. At one end of their scale was, "*definitely do not want a drink of alcohol*" (AVOIDANCE), and at the other end was, "*an extreme desire for a drink of alcohol*" (APPROACH).

The item- total correlations for this subscale, range from .37 to .58, and the Cronbach´s alpha is .69 (approximately .7) which exceeds the minimum acceptable level of .6 for exploratory studies (Nunnally, 1974; Robinson, Shaver, & Wrightsman, 1991). Deletion of any of the items would not have meaningfully improved the internal consistency of this subscale. Worthy of note is that both unidimensional and bi-dimensional scales have been implicated in measuring the concept of avoidance-approach (Breiner et al., 1999; Greeley et al., 1993). Also, avoidance-approach was not an a priori subscale in this study. Finally, given the theoretical and empirical rationalization and operationalization of the avoidance-approach concept, it is plausible to characterize this subscale as an affective component of attitude.

## FACTOR # 5 (PERCEIVED DIFFICULTY)

This factor consists of three items (22-23 in Table 15), and explained the least amount of total variance (5.36%), and common variance (10.4%), however, it is theoretically substantive and interpretable. All items reflect perceived difficulty (or ease) of effectively using the concept-based approach to teach introductory statistics, and were similarly classified a priori. Moreover, the marker item: *"Using active learning strategies can make class-room management difficult"* has the highest loading on this factor. Accordingly, this subscale was named **"PERCEIVED DIFFICULTY"**. Davis (1989) captures this concept in what he refers to as "**perceived ease-of-use",** defined as "the degree to which a person believes that using a particular system would be free from effort". The item-total correlations range from .44 to .48, and the Cronbach´s alpha is .65, above the acceptable minimum of .6 for exploratory studies (Nunnally, 1974; Robinson, Shaver, & Wrightsman, 1991). The deletion of any of the items would not have improved the internal consistency of the scale. Of the five factors or subscales reported in this study, this one has the lowest internal consistency, but also the fewest items, and this must be considered, given that Cronbach´s alpha is directly related to the number of items forming the scale (Field, 2005). Furthermore as Rummel (1970, p. 362) noted:

> Unfortunately it is the smaller factors that may tap conceptually new or unsuspected influences in a domain. The larger factors are usually already known by experienced observers aside from systematic research, while the smaller factors are masked by these larger interdependencies. Throwing away a strange factor, therefore, may toss out an important discovery.

**Inter-Factor (Subscale) Correlations**

This aspect of the study is important toward understanding the relationships among the factors. Such covariance provides insight into the plausibility of the factors, and appropriateness of the factor labels. In other words, the extent to which the factor structure is theoretically and empirically consistent can be determined (Cronbach & Meehl, 1955). Toward this end, the correlation matrix was examined for convergence of the factors (based on the strength and direction of the relationships), which is one dimension of construct validity (Bagozzi; 1979; Campbell & Fiske, 1959). Additionally, the extent and nature of the interrelationships are necessary for decisions regarding parsimony of the factor structure (Cronbach & Meehl, 1955), especially with regard to implications for higher-order or hierarchical factors. In order to explore these theoretical and conceptual relationships, Pearson's product moment correlation analysis of the factor scores was performed (Albarracin et al., 2001; Asher, 1997). These are bivariate correlations (which are symmetric), and hence, some of the relationships may not be readily clear because of the role of a third factor (mediator, moderator or confounding).

## Factor (Subscale) Scores

A composite factor score was calculated for each subject based on the score of items (with salient loadings) which constitute each factor (Cattell, 1957; Gorsuch, 1974; Kline, 2005). All items were equally weighted, and the mean of the item scores was obtained in order to retain the original scale properties (Hogue, Liddle, Singer, & Leckrone, 2005; Russel 2002). In particular, the factor scores were not automatically generated based on the factor loadings (weights), as this approach uses all the items that loaded (high and low) on a factor to generate the score. Factor loadings are sample-specific, and therefore, such scores may not be replicated in future studies (as items may not have the same loadings). The approach of summing the scores of items with salient loadings on each factor is generally recommended (Cattell, 1957; Gorsuch, 1974; Gorsuch, 1983; Kline, 2005) especially for exploratory factor analysis in scale development, as repeated studies for reliability and validation checks will be necessary.

**Table 17: Observed and Corrected Inter-factor Correlations along with Alpha Coefficients** [a,b]

|  | Factor # 1 | Factor # 2 | Factor # 3 | Factor # 4 | Factor # 5 |
|---|---|---|---|---|---|
| 1. Perceived Usefulness | **.88** | *.81* | *.59* | *.38* | *.10* |
| 2. Intention | .70[c] | **.85** | *.61* | *.52* | *.12* |
| 3. Personal Teaching Efficacy | .49[c] | .49[c] | **.77** | *.54* | *.58* |
| 4. Avoidance-Approach | .30[c] | .41[c] | .39 | **.69** | *.17* |
| 5. Perceived Difficulty | .07 | .09 | .41[c] | .11 | **.65** |

a. The alpha coefficients are shown on the diagonal, observed correlations below the diagonal, and correlations corrected for attenuation due to measurement error (italicized).

b.

This correction involves dividing the observed correlation between the two variables by the product of the square root of their reliability (Cronbach's alpha) (Lord & Novick, 1968).

c. **p < .01 (observed correlation coefficients)**

Table 17 shows the Pearson's correlation matrix of the five factors or subscales. Both the observed (below the diagonal) and the corrected[16] (above the diagonal) correlation coefficients are presented (Asher, 1997; Charles, 2005; Thurstone, 1947). The strength (effect size) of all the bivariate relationships was evaluated based on prevailing psychometric standards (Cohen & Cohen, 1975), and the interpretation that follows is based on the corrected correlation coefficients.

Perceived usefulness (anticipated benefits) is strongly and positively correlated with intention (r = .81), indicating that the more instructors perceived the concept-based pedagogy to be useful in enhancing teaching and learning, the stronger was their intention to adopt this approach. Among the factors, perceive usefulness has the strongest correlation with intention. Indeed, multiple major research studies (Ajzen & Fishbein, 1977; Davis, Bagozzi, & Warshaw, 1989; Taylor & Todd, 1995; Venkatesh & Davis, 2000) have concluded that perceived usefulness is the primary factor facilitating intention. While the size of the correlation coefficient suggests considerable overlapping of the factors, it does not rise to the generally acceptable level (of at least .9) for the factors to be considered redundant (Kline, 2005). Perceived usefulness and intention may therefore be considered related but separate constructs. Nonetheless, there exists the potential for a higher-order or hierarchical factor.

Additionally, perceived usefulness is moderately and positively correlated with personal teaching efficacy (r = .59) suggesting that as instructors perceived themselves to be more capable of successfully using the concept-based pedagogy, they were more likely to report favorable perceptions about the usefulness (beneficial outcomes) of this approach. This observation is supported by Bandura (1986, p.392) who noted that, *"the*

---

[16] Corrected for attenuation due to measurement error (Lord & Novick, 1968).

*types of outcomes people anticipate depend largely on their judgments of how well they will be able to perform in a given situation".* Specifically, Bandura (1986) stated that self-efficacy and expected outcomes (perceived usefulness) are different but related constructs. This relationship is also empirically supported in the context of technology acceptance (Compeau, Higgins, & Huff, 1999; Kulviwat et al., 2006). Albeit to a lesser degree, a meaningful positive relationship ($r = .38$) is noted between perceived usefulness and avoidance-approach, that is, higher levels of perceived usefulness of the concept-based approach is associated with a greater inclination, proclivity, or propensity toward, and desire to use this method.

Notably, perceived difficulty (ease of use) is significantly and moderately correlated with personal teaching efficacy only ($r = .58$). That is, higher levels of perceived capability to successfully use the concept-based approach were associated with the perception of greater ease of use (less difficulty). This finding is consistent with major attitude-behavior research. Specifically, Yzer, Hennessy, and Fishbein (2004) reported that perceived difficulty and self-efficacy are usually moderately to strongly correlated, and reflect distinct but related constructs. And Venkatesh (2000) surmised that personal self-efficacy is the strongest determinant of perceived ease of use (difficulty).

Furthermore, the construct of intention is moderately and positively correlated with avoidance-approach ($r = .52$) which could indicate that stronger inclination, propensity or proclivity toward, and desire to use the concept-based method is associated with higher levels of intention to act accordingly. A stronger correlation was observed between personal teaching efficacy and intention ($r = .61$) which is consistent with the expanded theory of reasoned action (Ajzen, 2002). In other words, the more instructors felt capable of using concept-based pedagogy successfully, the greater was their intention to use this method.

Finally, a moderate positive relationship ($r = .54$) was noted between personal teaching efficacy and avoidance-approach, implying that the more instructors felt capable of successfully using the concept-based approach, the greater was their inclination, propensity or proclivity toward, and desire to adopt this teaching approach. A similar relationship was reported by Johnson and Wardlow (2002) in the context of computer use and undergraduate students' perceived self-efficacy. Also, given the affective underpinning of the avoidance-approach construct, this finding parallels Compeau and Higgins (1995) observation in the context of computer use, that self-efficacy is directly related to affect (or feeling). In other words, if a person has a high perceived capability of successfully using a method, then that person is more likely to have a positive feeling toward using the method, and be less inclined to avoid (and more likely to approach) it.

In general, these bivariate relationships are all meaningful, plausible, logical, statistically significant, and consistent with the tripartite attitude theory. The general moderate and positive correlations strongly indicate that the factors are related but separate toward measuring different dimensions of a single global construct (attitude). Accordingly, convergent validity is established (and further addressed in the next section). The general moderate correlations among the factors could suggest the presence of higher order or hierarchical factors (Floyd & Widaman, 1995). Indeed, as Gorsuch (1983, pp. 254-255) noted:

> Implicit in all oblique rotations are higher-order factors….There is nothing sacred about either primary or higher-order factors. The importance of each lies in its relative merits for the theory under consideration.

Given the exploratory nature of this study, as well as the use of, and interest in the tripartite conceptualization of attitude, hierarchical factor analysis was not pursued.

**Validation Analysis**

The validity of an instrument is the extent to which it measures what it purports to measure (Nunnally & Berstein, 1994). The question addressed in this section is: Are these five factors (forming a scale) measuring attitude? Validity is a multifaceted concept, more appropriately referred to as "construct validity", which encompasses **content validity**, **criterion** (concurrent or predictive) **validity**, **convergent validity,** and **discriminant validity** (Cronbach and Meehl, 1955; Viswanathan, 1993). According to Anastasi (1950, p.67): "It is only as a measure of a specifically defined criterion that a test can be objectively validated at all….to claim that a test measures anything over and above its criterion is pure speculation."

Psychological constructs (for example attitude) are not directly observable, and in order to determine whether such hypothetical constructs are being measured, we must show that a measure of a given construct relates to a measure of another construct (the criterion) in a theoretically predictable way (Cronbach & Meehl, 1955). This approach is referred to as criterion validation or "known-groups validity" (Churchill, 1979; Cronbach & Meehl, 1955; Nunnally, 1978). In other words, does the measure differentiate between groups that are known to differ on the construct? Conventionally, if information on both constructs were ascertained in the same survey, then criterion validity is more appropriately termed concurrent validity, and if information on the criterion construct is collected later, it is referred to as predictive validity (Cronbach & Meehl, 1955).

**Criterion Validity**

For this study, the criterion variable (teaching practice score) was ascertained concurrently with the attitudinal items. A composite score was obtained for each respondent by summing their values on each of the ten practice items with a possible maximum score of 50 (higher scores reflect more favorable reform-based practice, consistent with constructivism). Respondents in the highest quartile (of practice score) were considered high-reform instructors (in tune with concept-based or constructivist teaching) whereas those in the lowest quartile (of practice score) were labeled low-reform instructors (more akin to the traditional, mathematical or behaviorist approach). A similar categorization (and pedagogical scale) was used by Riel and Becker (2000) in their study of teachers' beliefs, practices, and computer use. This dichotomous classification allowed for meaningful differentiation (for data analysis purposes) among teachers with respect to teaching practice.

One-way ANOVA, and the t-test, along with their non-parametric equivalent, the Kruskal-Wallis and Mann-Whitney tests respectively were the primary statistical analyses used to check for theoretically predictable differences between high and low-reform instructors with respect to total attitude and subscale scores. In other words, based on the theories of reasoned action and planned behavior, as well as general attitude-behavior relationship (Wallace et al., 2005), high-reform instructors (scored in highest quartile of practice score) should have more favorable attitude (higher attitude scores) toward concept-based teaching than low-reform instructors (scored in the lowest quartile of practice score).

**Table 18: Comparison of Overall Attitude and Subscale Scores by Teaching Practice Categorization**

| Attitude Subscales | Teaching Practice Score | N | Mean [c] | Std. Dev. | Level of Significance | | |
|---|---|---|---|---|---|---|---|
| | | | | | t | t-test | Mann-Whitney |
| Perceived Usefulness | Lowest Quartile [a] | 74 | 3.84 | .63 | 5.57 | .001 | .001 |
| | Highest Quartile [b] | 59 | 4.41 | .50 | | | |
| Intention | Lowest Quartile [a] | 74 | 3.57 | .77 | 6.90 | .001 | .001 |
| | Highest Quartile [b] | 59 | 4.36 | .55 | | | |
| Personal Teaching Efficacy | Lowest Quartile [a] | 74 | 3.71 | .66 | 7.47 | .001 | .001 |
| | Highest Quartile [b] | 59 | 4.43 | .45 | | | |
| Avoidance-Approach | Lowest Quartile [a] | 73 | 3.82 | .70 | 4.48 | .001 | .001 |
| | Highest Quartile [b] | 59 | 4.32 | .52 | | | |
| Perceived Difficulty | Lowest Quartile [a] | 73 | 2.93 | .76 | 1.16 | .248 | .211 |
| | Highest Quartile [b] | 59 | 3.10 | .92 | | | |
| **Overall Attitude (Total)** | Lowest Quartile [a] | 72 | 3.65 | .47 | 8.26 | .001 | .001 |
| | Highest Quartile [b] | 59 | 4.23 | .33 | | | |

a. Low-Reform Instructors

b. High-Reform Instructors

c. Higher values represent more favorable disposition toward reform-oriented instruction (concept-based pedagogy).

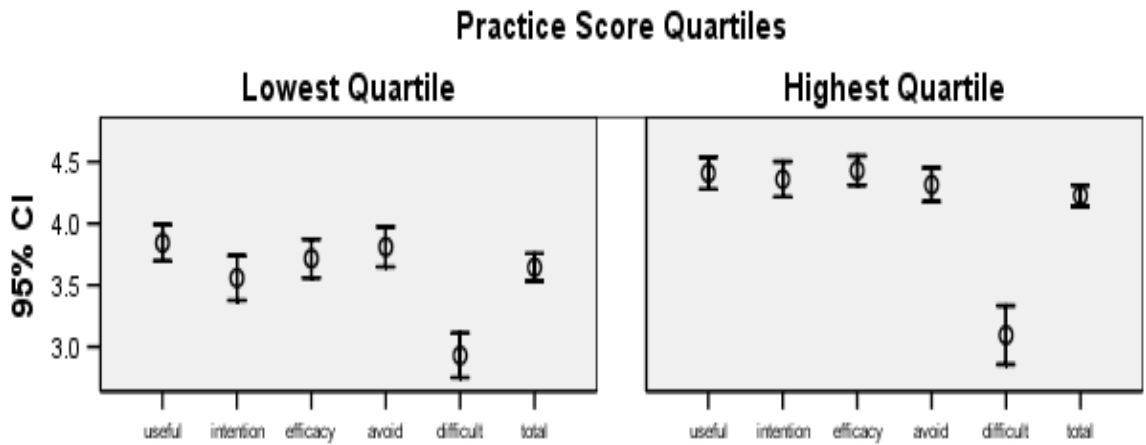

**Figure 4: 95% Confidence Interval of Mean Attitude (Total) and Subscale Scores by Practice**

As shown in Table 18, high-reform teachers, on average, did report higher scores (more favorable disposition toward reform-oriented or concept-based pedagogy) on the overall attitude scale, as well as each of the five subscales, and all but perceived difficulty (ease of use) were statistically significant. This information is illustrated in Figure 3. The lowest subscale score for both high and low-reform instructors was perceived difficulty (ease of use) with a value of about 3 (on a 5-point scale) for both groups, suggesting indecision or neutrality, in this regard. Correspondingly, the only known quantitative empirical study which focused on teachers' attitudes toward statistics (albeit pre-service primary school teachers) noted that, "future teachers' attitudes were little influenced by considering the topic either easy or difficult" (Estrada et al., 2005). Similar findings have also be been reported for undergraduate students' attitude toward research. According to Papanastasiou (2005), "there are different factors that can possibly influence the student's attitudes toward this subject, that have nothing to do with whether they consider a research methods course to be difficult or not".

Consistent with the theories of reasoned action and planned behavior (TRA and TPB), secondary analyses were conducted to determine and quantify the predictive value of the 5-factor attitude construct in relation to teaching practice, as well as the relative and unique contribution of each factor to explaining and predicting teaching practice. Also, given the separate role of intention in some conceptualizations of the attitude-behavior relationship, intention was analyzed as an outcome variable in relation to the other four factors. Toward this end, multivariate regression analysis was performed (Asher, 1997; Meehl, 1954). Although perceived difficulty was not statistically significant in the bivariate analyses (Tables 19 and 20) it was entered into the multiple regression models because of its noted conceptual and empirical relevance to both intention and behavior (Albarracin et al., 2000).

**Table 19: Bivariate Correlation between Attitude Subscales and Intention**

| Attitude Subscales | N | Intention (score) | |
|---|---|---|---|
| | | *Pearson's r* | *R-squared* |
| Perceived Usefulness | 225 | .704[a] | .496 |
| Personal Teaching efficacy | 223 | .495[a] | .245 |
| Avoidance-Approach | 224 | .405[a] | .164 |
| Perceived Difficulty | 224 | .090 | .008 |

a. $p < .01$

**Table 20: Bivariate Correlation between Attitude Subscales and Teaching Practice Score**

| Attitude Subscales | N | Teaching Practice Score | |
|---|---|---|---|
| | | *Pearson's r* | *R-squared* |
| Perceived Usefulness | 219 | .364[a] | .13 |
| Intention | 218 | .452[a] | .20 |
| Personal Teaching Efficacy | 217 | .421[a] | .18 |
| Avoidance-Approach | 218 | .387[a] | .15 |
| Perceived Difficulty | 218 | .073 | .01 |

a. $p < .01$

## Multiple Regression Analysis of Teaching Practice on Attitude Subscales

Teaching practice score was regressed (simultaneously) on the five subscale scores in a multiple linear regression model. The assumptions underlying this statistical technique were met (Cohen & Cohen, 1983; Pedhazur, 1997). As shown in table 21, intention, personal teaching efficacy, and avoidance-approach were significant independent predictors of teaching practice. Perceived usefulness, as well as perceived difficulty (ease of use) were not significant, however, the overall model was statistically significant ($F_{(5, 208)} = 17.3$, $p < .0001$) and explained 28% of the variance in teaching practice (adjusted $R^2 = .28$). Intention emerged as the strongest predictor ($\beta = .26$, $p < .005$) when the other variables (subscale scores) were held constant, an observation that is well-established both theoretically and empirically. Specifically, intention to perform a behavior is considered the strongest and most proximal antecedent of actual behavioral performance (Ajzen, 1991; Fishbein & Ajzen, 1975; Fisher & Fisher, 1992; Gollwitzer, 1993; Triandis, 1977).

**Table 21: Multiple Regression Analysis of Overall Teaching Practice Score on Attitude Subscale Scores[a]**

| Independent Variables (subscales)[b,c] | Unstandardized Coefficients | | Standardized Coefficients | | |
|---|---|---|---|---|---|
| | B | Std. Error | Beta | t | Sig. |
| (Constant) | 17.05 | 2.30 | | 7.42 | .001 |
| Perceived Usefulness | .04 | .61 | .01 | .07 | .946 |
| Intention | 1.58 | .52 | .26 | 3.01 | .003 |
| Personal Teaching Efficacy | 1.62 | .54 | .24 | 3.01 | .003 |
| Avoidance-Approach | 1.46 | .47 | .20 | 3.07 | .002 |
| Perceived Difficulty | -.42 | .35 | -.08 | -1.22 | .225 |

a. Dependent Variable: Overall Teaching Practice Score

b. Model Significance: F (5, 208) = 17.3, p<.001

c. Adjusted R-squared = .28 (28%)

Moreover, as reported by Armitage and Conner (2001; Cheung & Chan, 2000), and suggested by this study, personal teaching efficacy improved the prediction of teaching practice over intention (Tables 20 and 21). And according to Madden, Ellen and Ajzen (1992), this is associated with behaviors not under complete volitional control. Furthermore, the non-significance of perceive difficulty in this model is supported by a recent major meta-analysis which concluded that "one intriguing and unexpected finding" was that perceived difficulty did not moderate the attitude-behavior relationship in regard to both self-reported and directly observed behaviors (Wallace et al., 2005).

In evaluating this multiple regression model, the substantial correlation between perceived usefulness and intention, as well as the correlation of each with the teaching practice scale much be considered. Specifically, the fact that both are highly correlated, and intention has a higher correlation with practice than perceived usefulness, quite likely explains why perceived usefulness is not statistically significant. Multicollinearity was ruled out. Consequently, a distinction must be made between predicting and explaining behavior (Asher, 1997; Cook & Campbell, 1979). The results of this multivariate regression analysis should not be interpreted as indicating that perceived usefulness is "unimportant" in understanding the theory underlying these variables. What it indicates is simply that it is not very useful in predicting behavior in this context (Ajzen, 2002; Asher, 1997). One logical explanation is that the construct of intention may be hierarchical, with perceived usefulness as a necessary component, in this context. In other words, perceived usefulness may be nested within intention.

**Multiple Regression Analysis of Intention on Attitude Subscales**

When intention was regressed on the other four factors or subscale scores (perceived usefulness, personal teaching efficacy, perceived difficulty, and avoidance-approach), the resulting adjusted $R^2$ was .55 ($F_{(4, 216)} = 67.7$, $p < .0001$), indicating that 55% of the variance in intention was explained by (the other) attitudinal factors (Table 22). In particular, perceived usefulness was the strongest significant independent predictor of intention ($\beta = .58$, $p < .0001$), followed by personal teaching efficacy ($\beta = .16$), and avoidance-approach ($\beta = .17$). Perceived difficulty was not statistically significant in this model. The finding of perceived usefulness being the strongest predictor of intention is consistent with a large body of research on the technology acceptance model (TAM) guided by the TRA, and the TPB. Specifically, Ajzen and Fishbein (1977; Davis et al., 1989; Taylor & Todd, 1995; Venkatesh & Davis, 2000) have established that perceived usefulness is the primary belief underlying intention.

**Table 22: Multiple Regression Analysis of Intention Score on the Other Attitude Subscale Scores[a]**

| Independent Variables (Subscales)[b,c] | Unstandardized Coefficients | | Standardized Coefficients | | |
|---|---|---|---|---|---|
| | B | Std. Error | Beta | t | Sig. |
| (Constant) | -.41 | .30 | | -1.39 | .167 |
| Perceived Usefulness | .69 | .06 | .58 | 10.94 | .001 |
| Personal Teaching Efficacy | .19 | .07 | .16 | 2.73 | .007 |
| Avoidance-Approach | .21 | .06 | .17 | 3.42 | .001 |
| Perceived Difficulty | -.03 | .04 | -.03 | -.66 | .508 |

a. Dependent Variable: Intention Score

b. Model Significance: $F_{(4, 216)} = 67.7$, $p < .001$

c. Adjusted R-squared = .55 (55%)

In general, these findings are consistent with prevailing theories, and major empirical studies (meta-analyses) on the attitude-intention-behavior model, especially with reference to the TRA and TPB. For example, Armitage and Conner (2001) reported that the TPB explained, on average, 27% of the variance in behavior which is almost identical to the percentage (28%) obtained in this study for the attitude-practice relationship (Table 21). And Godin and Kok (1996) reported an average of 34% for the variance explained in health-related behavior. These two studies also documented the average variance explained in intention as 41% and 39%, respectively, compared to the finding of 55% in this study (Table 22). Note however, that the values from these meta-analytic studies are averages, and indeed some individual study values were much higher (Ajzen & Fishbein, 2004). Another important theoretical relationship is the intention-behavior correlation, which emerged as .45 in this study (Table 20), the exact value reported by Randall and Wolff (1994) in their meta-analysis of studies addressing a broad spectrum of behavioral domains. Other meta-analyses have reported intention-behavior correlation coefficients of .47 (Armitage & Conner, 2001; Notani, 1998), .53 (Shepherd, Hartwick, & Warshaw, 1988), and .62 (van den Putte, 1993).

When evaluating these results, it must be considered that this particular research design (cross-sectional) where attitudinal items were measured concurrently with teaching practice, can result in inflated correlations due to shared error variance and "hypothesis guessing" (Albarracin et al., 2001; Wallace et al., 2005). In other words, respondents could have attempted to give attitude-practice responses in order to be consistent. Finally, although the factors used in this study are not exactly analogous to the TRA and TPB, the model concept is similar. In this regard, it is well-established that the underlying attitude-intention-behavior theory is considered confirmed if just one of the predictors of intention or behavior is significant and meaningful (Ajzen & Fishbein, 2004). The rationale for this is that the relevance and importance of each antecedent factor is dependent on the research context, including the behavior and population being studied (Ajzen, 2002; Ajzen & Fishbein, 2004; Triandis, et al., 2001).

**Content Validity**

In addition to the criterion validity of the 5-factor attitude structure, other dimensions of validity were established. Particularly, content validity (Nunnally, 1978) was achieved by beginning with a representative sample of attitude items, that could adequately and plausibly reflect the theoretical and empirical dimensions of attitude in relation to reform-oriented teaching. This was facilitated by a thorough review of the related literature, consultation with experts, and a multidisciplinary team approach to item generation and item analysis. In keeping with a best practice in this regard (Nunnally & Bernstein, 1994; Haynes, Richard, & Kubany, 1995), the initial list of facets of the attitude construct was presented to multiple experts (teaching, content, and measurement), for their evaluation of the relevance, and salience of each to the study context. There was also an open-ended section. Moreover, each of the final factors (ranging between 3 and 7 items) reflects adequate coverage of the specific attitude domain, as is theoretically and empirically established.

**Convergent Validity**

Convergent and discriminant validity analyses were performed (Campbell & Fiske, 1959). For convergent validity, a modified version of the multi-trait multi-method (MTMM) approach (Campbell & Fiske, 1959) used by Bagozzi et al. (1979) to examine the construct validity of the tripartite classification of attitude was applied. According to the principle of convergent validity, measures of theoretically similar constructs should be substantially intercorrelated. Each of the five subscales was considered a "method" for measuring the "trait", attitude, given the conceptual relatedness of the factors. In general the correlations are moderate and statistically significant (Table 17), which is evidence of convergence. As Campbell and Fiske (1959, p. 82) notes, in order to establish convergent validity, the relevant correlations "should be significantly different from zero and sufficiently large".

Also, in general, items reflecting cognition (perceived usefulness, personal teaching efficacy, and perceived difficulty), affect (avoidance-approach) and intention clustered together with statistically significant pattern coefficients, which according to Anderson and Gerbing (1988) establishes convergent validity. A minimum pattern coefficient of .4 was used (for a sample of about 200), to guard against retaining items and factors by chance (Field, 2000; Hair, et al., 1998; Stevens, 1992). Convergent validity can be assessed from yet another perspective. For a measure to be valid it must first be reliable. In this regard, and according to Spearman's correction of correlation (for attenuation due to measurement error) (Lord & Novick, 1968), as the reliability (Cronbach's alpha) of either measure decreases, so does the observed convergent validity evidence (Fishbein & Ajzen, 1975). Given that the alpha levels of the overall scale and subscales range from .65 to .89, this suggests an acceptable to high level of covariance among the items of each measure, and hence acceptable convergent validity.

**Discriminant Validity**

The general moderate correlations, and simple structure (general absence of overlapping of items) support the distinctiveness and discriminant validity of the factors or subscales (Bagozzi et al., 1979; Simms et al., 2005). According to the principle of discriminant validity, measures of theoretically different but related constructs should not correlate highly with each other. However, a more rigorous test of discriminant validity based on the average variance extracted (AVE) for each construct, was applied. Fornell and Larcker (1981) recommended that in order to demonstrate discriminant validity, the AVE for each construct (within construct variance) should be greater than the squared correlation (variance) between that construct and another. The results are mixed.

**Table 23: Discriminant Validity Analysis [a]**

| Factors | Factor # 1 | Factor # 2 | Factor # 3 | Factor # 4 | Factor # 5 |
|---|---|---|---|---|---|
| 1. Perceived Usefulness | **.53** | *.66* | *.35* | *.14* | *.01* |
| 2. Intention | .49 | **.55** | *.37* | *.27* | *.01* |
| 3. Personal Teaching Efficacy | .24 | .24 | **.36** | *.29* | *.34* |
| 4. Avoidance-Approach | .09 | .17 | .15 | **.32** | *.03* |
| 5. Perceived Difficulty | .01 | .01 | .17 | .01 | **.36** |

a. The bold diagonal values are the average variance extracted (AVE) for each factor/subscale. The lower off-diagonal values are the observed (uncorrected) squared correlations between factors (shared variance). The upper off-diagonal values are the squared correlations (corrected for attenuation due to measurement error) between factors. For acceptable discriminant validity, the diagonal elements should be larger than the corresponding off-diagonal values.

As detailed in table 23, all the AVE values exceeded the observed squared correlations (between construct variance) thereby demonstrating acceptable discriminant validity. However, as shown in the very table, for the corrected squared correlations, there was some discrepancy.  Perceived usefulness and intention did not emerge statistically as distinct constructs albeit they are conceptually so, but may be operating differently in this specific context.  Otherwise discriminant validity was noted, albeit marginally in a few instances, and slightly partial for intention and personal teaching efficacy. It must be noted that the avoidance-approach subscale has the potential to be bi-dimensional, and therefore the corrected correlation coefficient associated with this factor may be overcorrected, resulting in a higher between construct variance. According to Schmitt (1996, p.353):

> In correcting for attenuation due to unreliability, use of alpha as an estimate of reliability is based on the notion that the measures involved are unidimensional. When this is not the case, the corrected coefficients will be overcorrected.

Both the corrected and uncorrected squared correlations are presented. However, as this is an initial exploratory study, the uncorrected values may be more relevant at this stage, especially given the unresolved debate about the value of correction for attenuation due to measurement error (Charles, 2005; Schmitt, 1996). From these analyses and evaluations, it can be concluded that this five-factor attitude structure demonstrated acceptable construct validity. Indeed, greater evidence of discriminant validity was seen with the observed (rather than corrected) correlations, which is the more widely used approach in scale construction. Furthermore, criterion-related validity which is clearly established in this study, is generally considered adequate and the most concrete and important dimension to establishing validity in theory testing and scale development (Anastasi, 1950; Bolton, 2001; Flannelly, Ellison, & Strock, 2004; Lewis 1999).

**Selected Personal and Sociodemographic Factors as Determinants of Attitude**

One-way ANOVA (analysis of variance) along with the independent samples t-test, and their non-parametric equivalents; the Kruskal-Wallis and Mann-Whitney tests respectively were used to examine overall attitude and subscale scores for possible subgroup differences. Where appropriate, post-hoc analyses; Fisher's LSD (least significant difference) and Bonferroni (Bonferroni, 1936; Shaffer, 1995) were conducted. As this is supposedly an initial, and exploratory study, Bonferroni corrected results are presented but not emphasized (Bender & Lange, 1999; Sequeira, Hollins, & Howlin, 2003), as this adjustment tends to be overly conservative, and may conceal important differences (Perneger, 1998; Rothman, 1990) which can constitute plausible hypotheses for confirmatory studies. Moreover, these subgroup analyses do not constitute "data dredging" or a "fishing expedition" for significant findings, but are planned or pre-specified comparisons, and hence were stated and conceptualized a priori (Cook & Farewell, 1996; Perneger, 1998).

No statistically significant difference was observed in overall attitude and subscale scores with respect to gender, employment status, membership status in professional organizations, geographic location, ethnicity, highest academic qualification, and degree concentration. Nonetheless, two of these variables approached significance ($p = .07$). That is, a marginal difference in intention was noted, being higher for instructors who reported membership in professional organizations, as well as non-USA instructors.

**Table 24: Comparison of Overall Attitude and Subscale Scores by Teaching Area**

| Attitude Subscales and Overall Score | Teaching Area | N | Mean | Std. Dev. | F | Sig |
|---|---|---|---|---|---|---|
| Perceived Usefulness | Health Sciences | 94 | 4.05 | .65 | 1.79 | .169 |
| | Psychology/Behavioral Sciences | 102 | 4.23 | .67 | | |
| | Health and Behavioral Sciences | 31 | 4.18 | .56 | | |
| Intention | Health Sciences | 93 | 3.89 | .85 | .58 | .561[a] |
| | Psychology/Behavioral Sciences | 101 | 4.00 | .69 | | |
| | Health and Behavioral Sciences | 31 | 3.97 | .80 | | |
| Personal Teaching Efficacy | Health Sciences | 93 | 3.91 | .72 | 4.25 | .015 |
| | Psychology/Behavioral Sciences | 101 | 4.18 | .62 | | |
| | Health and Behavioral Sciences | 31 | 3.93 | .69 | | |
| Avoidance-Approach | Health Sciences | 92 | 4.06 | .52 | 2.76 | .066 |
| | Psychology/Behavioral Sciences | 102 | 4.23 | .67 | | |
| | Health and Behavioral Sciences | 31 | 3.97 | .80 | | |
| Perceived Difficulty | Health Sciences | 93 | 3.02 | .97 | .59 | .555[a] |
| | Psychology/Behavioral Sciences | 102 | 3.13 | .78 | | |
| | Health and Behavioral Sciences | 31 | 2.98 | .83 | | |
| Overall Attitude | Health Sciences | 90 | 3.87 | .52 | 2.76 | .066 |
| | Psychology/Behavioral Sciences | 100 | 4.04 | .49 | | |
| | Health and Behavioral Sciences | 31 | 3.90 | .46 | | |

a. Kruskal Wallis test (p > .05) was performed because of unequal variances.

**Table 25: Post-Hoc Analysis of Personal Teaching Efficacy Score by Teaching Area**

| Dependent Variable | (I) Teaching area | (J) Teaching area | Mean Difference (I-J) | Sig | |
|---|---|---|---|---|---|
| | | | | LSD | Bonferroni |
| Personal Teaching Efficacy | Health Sciences | Psychology/ Behavioral Sciences | -.27 | .01 | .02 |
| | | Health and Behavioral Sciences | -.02 | .91 | 1.00 |
| | Psychology/ Behavioral Sciences | Health and Behavioral Sciences | .25 | .07 | .21 |

Fisher's LSD (Least Significant Difference) uses the t-test to perform all pairwise comparisons. No adjustment is made to the error rate for multiple comparisons.
For Bonferroni correction, the observed significance is adjusted for the fact that multiple comparisons are being made.

With regard to teaching area (Table 24), instructors in psychology and the behavioral sciences, on average, reported a significantly more favorable level of personal teaching efficacy (in relation to reform-oriented or concept-based teaching) compared to instructors in the health sciences. This difference persisted following Bonferroni correction (Table 25). Albeit not statistically significant, instructors in psychology and the behavioral sciences also had the highest scores (more reform-oriented) on the remaining measures (overall attitude, perceived usefulness, intention, avoidance-approach, and perceived difficulty).

Statistically significant differences in some measures were observed with respect to age group (Table 26). Specifically, younger instructors (40 years and less) reported lower and less favorable levels of overall attitude, perceived usefulness, personal teaching efficacy, and intention than both groups of older instructors (41 –50, and 51 + years). Some of these subgroup differences remained significant or approached significance following Bonferroni correction (Table 27).

Table 26: Comparison of Overall Attitude and Subscale Scores by Age Group

| Attitude Subscales and Overall Score [a] | Age Groups | N | Mean | Std. Deviation | F | Sig |
|---|---|---|---|---|---|---|
| Perceived Usefulness | 40 years and less | 84 | 3.99 | .65 | 3.63 | .03 |
| | 41 -50 years | 62 | 4.24 | .63 | | |
| | 51+ years | 73 | 4.23 | .65 | | |
| Intention | 40 years and less | 83 | 3.77 | .80 | 3.57 | .03 |
| | 41 -50 years | 62 | 4.10 | .76 | | |
| | 51+ years | 72 | 4.00 | .72 | | |
| Personal Teaching Efficacy | 40 years and less | 83 | 3.87 | .66 | 4.14 | .02 |
| | 41 -50 years | 61 | 4.10 | .71 | | |
| | 51+ years | 73 | 4.17 | .67 | | |
| Avoidance/Approach | 40 years and less | 83 | 4.07 | .53 | .96 | .39 |
| | 41 -50 years | 61 | 4.10 | .72 | | |
| | 51+ years | 73 | 4.21 | .67 | | |
| Perceived Difficulty | 40 years and less | 84 | 3.06 | .92 | .04 | .96 |
| | 41 -50 years | 62 | 3.09 | .88 | | |
| | 51+ years | 72 | 3.05 | .82 | | |
| Overall Attitude | 40 years and less | 82 | 3.83 | .50 | 3.61 | .03 |
| | 41 -50 years | 60 | 4.03 | .56 | | |
| | 51+ years | 71 | 4.02 | .45 | | |

a. The assumption of homogeneity of variances was met for all measures.

Table 27: Post-Hoc Analysis of Overall Attitude and Subscale Scores by Age Group

| Attitude Subscales and Overall Score | (I) Age group of respondents | (J) Age group of respondents | Mean Difference (I-J) | Sig | |
|---|---|---|---|---|---|
| | | | | LSD [b] | Bonferroni [a] |
| Value | 40 years and less | 41 -50 years | -.25 | .02 | .06 |
| | | 51+ years | -.23 | .02 | .07 |
| | 41 -50 years | 51+ years | .02 | .88 | 1.00 |
| Intention | 40 years and less | 41 -50 years | -.33 | .01 | .03 |
| | | 51+ years | -.23 | .06 | .19 |
| | 41 -50 years | 51+ years | .10 | .46 | 1.00 |
| Personal Teaching Efficacy | 40 years and less | 41 -50 years | -.23 | .05 | .14 |
| | | 51+ years | -.30 | .01 | .02 |
| | 41 -50 years | 51+ years | -.07 | .56 | 1.00 |
| Overall Attitude | 40 years and less | 41 -50 years | -.19 | .02 | .07 |
| | | 51+ years | -.19 | .02 | .07 |
| | 41 -50 years | 51+ years | .01 | .91 | 1.00 |

a. For Bonferroni correction, the observed significance is adjusted for the fact that multiple comparisons are being made.

b. Fisher's LSD (Least Significant Difference) uses the t-test to perform all pairwise comparisons. No adjustment is made to the error rate for multiple comparisons.

Finally, according to Pearson's correlation (and the non-parametric equivalent, Spearman's rank correlation, reported herein) significant positive but weak relationships were noted between the duration of teaching (years) and personal teaching efficacy (Spearman's rho = .21, df = 219, p < .01), as well as avoidance-approach (Spearman's rho = .17, df = 219, p < .05). These relationships suggest that there is a tendency for more experienced instructors to perceived greater capability for reform-oriented (or concept-based) teaching, and greater approach (less avoidance) in this regard. It must be considered that, in general, the significant differences and relationships observed are weak, but indeed plausible (including being in the expected direction).

## Multidimensional Scaling (MDS) of the Attitudinal Items

Thus far, the exploration of the structure underlying the attitudinal items[17] was based on exploratory factor analysis[18] which requires the assumption of linear relationship, and the use of interval or ratio-scaled data (Floyd & Widaman, 1995). For other factor analytic approaches (such as maximum likelihood estimation), the additional requirement of multivariate normality must also be met (Floyd & Widaman, 1995). In this study, Likert-type scales were used to measure attitude, and such data are generally considered quasi-interval and subjected to factor analysis (Floyd & Widaman, 1995; McKinley et al., 1997). However, strictly speaking, Likert-type data are ordinal, and assuming the properties of an interval scale may mask critical information about the relationships among the variables, and hence misrepresent the construct being measured.

---

[17] Items are also referred to as variables, objects or stimuli.
[18] Principal Axis Factor Analysis

Furthermore, given the complex nature of psychological phenomena, including attitude, it seems reasonable to assume that non-linear relationships between attitudinal items may exist, and that the rank order of the measures may yield important information (Donald & Cooper, 2001). Moreover, given that this is an initial exploratory study, a more comprehensive and formative assessment of the data could be achieved by using multiple statistical methods, under different assumptions about the data. Toward this end, the factor analytic model was augmented with multidimensional scaling (MDS) techniques. Compared to factor analysis, there is a clear underutilization of MDS for scale development and data exploration in psychology and the health sciences, albeit its unique benefits and advantages have been long recognized (Fitzgerald & Hubert, 1987). This could be attributed, in large part, to the apparent complexity of MDS procedures, the limited range of models available in common statistical packages, and the lack of consensus among educators that MDS should be a core topic of the typical graduate level multivariate statistics course in the behavioral and health sciences.

**The Logic of Multidimensional Scaling (MDS)**

MDS is a set of multivariate analytical techniques aimed at reducing and organizing data so as to elucidate how and why items are related. MDS seeks to achieve a spatial representation (geometric map or configuration usually in 2-dimensions) of the latent or hidden structure that underlies and explains the relationships among the items which constitute the map (Coxon, 1982; Fitzgerald & Hubert, 1987; Kruskal & Wish, 1978)**.** By using MDS, we can therefore discern the dimensions of the perceptual space of subjects used to evaluate the set of items (Coxon, 1982; Fitzgerald & Hubert, 1987; Kruskal & Wish, 1978; Pinkley, Gelfand, & Duan, 2005).Unlike factor analytic techniques which require the assumption of interval or ratio-scaled (metric) data, and linear relationship among the items, all MDS models do not impose such restrictions on the data, as there are both metric (linear transformation) and non-metric (ordinal transformation) variants of MDS (Coxon, 1982; Kruskal & Wish, 1978; Schiffman, Reynolds, & Young, 1981;Young & Lewyckyj, 1979). Moreover, compared to factor analysis, MDS can result in more parsimonious and interpretable solutions (Fitzgerald & Hubert, 1987). Indeed, as these attitudinal data are truly non-metric (measured using Likert-type scales), MDS is quite suitable for this study (Pinkley, Gelfand, & Duan, 2005).

The input information for MDS is a numerical measure of distance indicating how similar (or dissimilar) each item is to every other item. Integral to the use of MDS is the assumption that the items in the dataset are related through some underlying psychological concept (Fitzgerald & Hubert, 1987). The data subjected to MDS are the Likert-type responses to the attitudinal items which were also explored with factor analysis. Both metric and non-metric MDS were carried out, and for each approach both Pearson's correlation coefficient (based on the interval properties of the data) and Kendall's tau (based on the rank order of the data) were used as measures of similarity of the items, as depicted in the table 28.

For both metric and non-metric MDS, two-dimensional solutions were obtained. In metric MDS, the algorithm used (MRSCAL) sought coordinates in 2-dimensional space so that the Euclidean distance (simple straight-line distance) between any pair of items is linearly related to the corresponding input proximity value (correlation coefficient). This involves a linear transformation of the data, and the relationship is negative, as the correlations represent similarities. That is, higher correlation coefficients (Pearson's and Kendall's) would indicate greater similarity between the items, and hence they will be closer in the spatial map (that is a shorter inter-item distance). For non-metric MDS (using the MINISSA algorithm) the Euclidean distance between any pair of items in the spatial map matches the rank-order of the corresponding input proximity, and hence a monotonic, ordinal or non-metric transformation of the data is carried out.

**Table 28: MDS Approaches and Measures of Similarity**

| MEASURES OF SIMILARITY | MDS APPROACHES | |
|---|---|---|
| | METRIC | NON-METRIC |
| *Pearson's Correlation Coefficient* (assumes interval level data) Parametric | *MRSCAL[19]* | *MINISSA[20]* |
| *Kendall's tau-b* (assumes ordinal level or rank-ordered data) Non- Parametric | *MRSCAL* | *MINISSA* |

---

[19] MRSCAL refers to **M**et**R**ic **SCAL**ing (see Roskam, 1972).
[20] MINISSA refers to **S**mallest **S**pace **A**nalysis (see Roskam and Lingoes, 1970).

INDSCAL[21] (INDividual SCALing) was not considered necessary for this initial exploratory study as there were no significant or meaningful subgroup differences[22] in attitude with respect age, gender, ethnicity, geographic location, teaching experience, highest academic degree, degree concentration, teaching area, membership status in professional organizations, and employment status (Coxon, 1982). In other words, there was no compelling evidence to suggest that the different subgroups of instructors (based on the abovementioned variables) would attach different weights or levels of salience to the overall underlying dimensions of the attitude construct.

**Criteria for Interpretation of the MDS Maps**

There are two basic approaches to interpreting MDS maps or configurations, and while these complement each other (Pinkley, Gelfand, & Duan, 2005), each is generally used separately. Dimensional interpretation (assigning meaning to the dimensions) is the more common approach which is akin to identifying factors in factor analysis (Coxon 1982, Kruskal & Wish, 1978). This generally involves multiple regression analysis, seeking statistical support for patterns within the maps. However, these dimensions are based solely on mathematical properties of the data, and therefore may not be plausible or theoretically relevant. Nonetheless, this technique may afford greater objectivity in interpretation (Pinkley, Gelfand, & Duan, 2005). For this approach, the respondents are generally required to rate each item or stimulus on a number of attributes (considered external), and then the attributes are regressed on the coordinate values of each stimulus in the map (Coxon, 1982). A more subjective approach to dimensional interpretation is to attempt to give meaning (qualitatively) to the dimensions based on the clusters at the ends of the dimension or axis in the map or configuration (Kruskal & Wish, 1978; Pinkley, Gelfand, & Duan, 2005; Schiffman, Reynolds, & Young, 1981).

---

[21] INDSCAL (**IND**ividual **SCAL**ing analysis), developed by Carroll and Chang (1970).
[22] Based on bivariate statistical analyses (t-test and ANOVA).

The other approach to interpretation of the MDS maps or configuration involves identifying and assigning meanings to patterns, neighborhoods or regions with related characteristics, and which plausibly differ from other regions, and for this, "a two-dimensional configuration is far more useful than one involving three or more dimensions" (Kruskal & Wish, 1978, p.58). This is called an internal analysis, as only the original data (measures of similarity, in this study) are used in the interpretation (Coxon, 1982). Notably, it has been established that the neighborhood or regional interpretation approach can give rise to different and more informative maps or solutions, as its focus is primarily on the small distances (large similarities), whereas, the dimensional approach attends mostly to the large distances (Kruskal & Wish, 1978; Guttman, 1965). The neighborhood or regional approach is commonly used as the primary technique for interpreting MDS maps (Bilsky & Jehn, 2002; Cooper & Donald, 2003; Fitzgerald & Lawrence, 1987; Guttman, 1965; Papanastasiou, 2005; Pfleiderer, 2003), and was used in this study. According to Guttman (1965) this approach is more advantageous, and should be preferred over dimensional interpretation.

In order "to guard against the human tendency to find patterns whether or not they exist" (Kruskal & Wish, 1978, p.36), hierarchical cluster analysis was used to guide the identification of structures within the spatial configurations or maps (Coxon, 1982). Also, to achieve interpretability, all the configurations were rotated to "simple structure". Indeed, all ordinary MDS solutions (as opposed to individual differences scaling - INDSCAL) are subject to rotation (Kruskal & Wish, 1978). Following rotation, the projection of points on the axes change, however, the distance between points or items (which the configuration is based on) does not change (Kruskal & Wish, 1978). Rotation allows for identifying meaningful clusters if these in fact exist. Regions or clusters with conceptually and theoretically related items are delineated in the maps to show potentially separable structures underlying the attitude construct. For both the metric and non-metric MDS approaches, measures of goodness of fit (between the Euclidean distances and the input proximities) were generated and used in conjunction with evidence of interpretability, parsimony, stability, plausibility, and construct validity to evaluate the adequacy of the MDS solutions.

**The MDS Results**

Table 29 shows the twenty-five attitudinal items which were subjected to MDS (and also used for factor analysis). The metric (Figures 5 and 6) and non-metric (Figures 7 and 8) maps are largely similar, capturing the dimensions or factors obtained with factor analysis (Table 15). The loops delineate items in relatively close proximity which are conceptually and theoretically related. The identification of these clusters was guided by the dendrogram obtained from hierarchical cluster analysis of these data (Coxon, 1982). As mentioned above, the MDS solution parallels the factor analysis solution, identifying five underlying dimensions of attitude which can be classified and reduced to three attitude components: **cognitive** (perceived usefulness, perceived difficulty, and perceived teaching efficacy), **affective** (avoidance-approach[23]), and behavioral **intention**. These solutions are consistent with the tripartite conceptualization of attitude (Breckler, 1984; Rosenberg & Hovland, 1960; Smith, 1947).

---

[23] Gray (1970) characterized the mechanism underlying this concept as "facilitative and inhibitory motivational systems", which he posited, produce positive and negative affect respectively. See also, Clark & Watson (1999).

With MDS, valuable information can be obtained by evaluating each cluster with regard to regions of high and low density points (reflecting low and high similarity of the stimuli within each cluster), as well as inter-cluster separation (Borg & Shye, 1995; Coxon, 1982; Lingoes, 1977). Small dense regions may indicate strong relationship, that is, high scores on all items, indicating that they are closely related, and also consistent in rank order. Larger, diffuse regions suggest weaker relationships among the items. The degree of inter-cluster separation reflects the magnitude of the relationship between the clusters, with smaller distances reflecting stronger relationship.

**Table 29: The Twenty-Five Attitudinal Items Used for Multidimensional Scaling**

1. The concept-based approach to teaching introductory statistics (rather than emphasizing calculations and formulas) makes students better prepared for work.
2. Teaching introductory statistics with emphasis on concepts and applications rather than calculations and formulas, can be time consuming.
3. I will adjust easily to teaching introductory statistics using the concept-based approach.
4. The preparation required to teach introductory statistics using the concept-based approach is burdensome.
5. Using active learning strategies (such as projects, group discussions, oral and written presentations) in the introductory statistics course can make classroom management difficult.
6. The concept-based approach to teaching introductory statistics (rather than emphasizing calculations and formulas) makes students better prepared for further studies.
7. Emphasizing concepts and applications in the introductory statistics course (rather than calculations and formulas) is a disservice to our students.
8. The concept-based approach to teaching introductory statistics is for low achievers only.
9. The concept-based approach to teaching introductory statistics enables students to understand research.
10. Concept-based teaching of introductory statistics may be problematic for me.
11. I do not understand how to organize my introductory statistics course to achieve statistical literacy.
12. I am engaged in the teaching of introductory statistics using the concept-based approach.
13. I am convinced that the concept-based approach to teaching introductory statistics enhances learning.
14. Teaching introductory statistics using the concept-based approach is likely to be a positive experience for me.
15. I am not comfortable using computer applications to teach introductory statistics.
16. Teaching introductory statistics with emphasis on concepts and their applications (rather than calculations and formulas) may be stressful for me.
17. Using computers to teach introductory statistics makes learning fun.
18. I am interested in using the concept-based approach to teach introductory statistics.
19. I want to learn more about the concept-based approach to teaching introductory statistics.
20. Using the concept-based approach to teach introductory statistics is not a priority for me.
21. I plan on teaching introductory statistics according to the concept-based approach.
22. I will avoid using computers in my introductory statistics course.
23. I will incorporate active learning strategies (such as projects, hands-on data analysis, critiquing research articles, and report writing) into my introductory statistics course.
24. I am hesitant to use computers in my introductory statistics class without the help of a teaching assistant.
25. I am concerned that using the concept-based approach to teach introductory statistics may result in me being poorly evaluated by my students.

**Characteristics of the MDS Clusters**

Of the seven items in the **PERCEIVED USEFULNESS** cluster (Figures 5 to 8), six are relatively close, forming a region of high density (items 1, 6, 7, 8, 13 and 14), and these (Tables 15 and 16) reflect perceived benefits of concept-based teaching of introductory statistics; relating to work, further studies, and general enhancement of teaching and learning. However, item 9, which is part of this cluster is remote from the other items (on all the MDS maps), and reflects perceived benefit with regard to facilitating students' understanding of research. This pronounced and consistent intra-cluster pattern may be indicating that instructors view the benefits of reform-oriented or concept-based teaching along two dimensions, (1) general enhancement of student life, teaching and learning, and (2) improvement in research understanding. Furthermore, instructors may not necessarily perceive these two benefits simultaneously, as the specific perceived benefit(s) could be related to the teaching context (including the characteristics of the student population).

A similar pattern is seen with the **INTENTION** cluster, with item 19 (*I want to learn more about the concept-based approach to teaching introductory statistics*) being distant from the other items. This can plausibly suggest that instructors' increased engagement in (item 12), interest in using (item 18), giving priority to (item 20), and planning to use (item 21) reform-oriented or concept-based teaching strategies, were not directly related to a desire to learn more about this teaching approach (item 19). This observation seems to indicate that two groups of instructors exists, one that is engaged in, and ready to adopt reform-oriented or concept-based instructional methods versus those for whom this teaching approach is new, and they simply just want to learn more about it, at this stage.

Furthermore, in all the MDS maps, the **PERCEIVED USEFULNESS** and **INTENTION** clusters are the closest (marginally separated) indicating that they are highly related. Indeed, these two factors had the highest inter-factor correlation (Table 17), which is theoretically and empirically supported. In other words, perceived usefulness is generally considered to be the primary factor facilitating intention (Ajzen & Fishbein, 1977; Davis, Bagozzi, & Warshaw, 1989; Taylor & Todd, 1995; Venkatesh & Davis, 2000). Viewed another way, the higher the level of perceived usefulness about reform-oriented or concept-based pedagogy, the greater was the instructors' intention to use it.

Another seemingly complex cluster is **AVOIDANCE-APPROACH** which depicts two smaller clusters at the extremes. Items 17 (*Using computers makes learning fun*), and 23 (*I will incorporate active learning strategies into my statistics course*) are consistent with APPROACH, that is, inclination or proclivity toward adopting reform-oriented or concept-based instructional strategies. Whereas, items 15 (I am not comfortable…..), 22 (I will avoid…..), and 24 (I am hesitant…..) connote AVOIDANCE or being less inclined to use this teaching approach. This apparent bi-dimensional cluster is theoretically and empirically supported (Breiner et al., 1999; Clark & Watson, 1999; Greeley et al., 1993a & b; Gray, 1970).

Noteworthy also, is that in general, the **PERCEIVED DIFFICULTY** cluster is most proximal to **PERCEIVED TEACHING EFFICACY** than the other clusters, which is consistent with the inter-factor correlations, (Table 17) and is theoretically and empirically plausible. Yzer, Hennessy, and Fishbein  (2004) have noted that perceived difficulty and self-efficacy are usually moderately to strongly correlated, and reflect distinct but related constructs, whereas Venkatesh (2000) surmised that personal self-efficacy is the strongest determinant of perceived ease of use (difficulty). Furthermore, from the relative positions of the clusters in all maps, is can be deduced that perceived difficulty about successfully using reform-oriented or concept-based instructional strategies, is seemingly unrelated to perceived usefulness, avoidance-approach (feelings or affect) and intention to use this approach. Indeed, these relationships were all non-significant with Pearson's correlation (Table 17) and very weak to almost non-existent. In seems reasonable to assume that perceived difficulty is not a salient determinant of overall attitude. A similar finding was noted by Estrada et al. (2005), among pre-service teachers and their attitudes toward statistics, as well as Papanastasiou (2005), in a study of undergraduate students' attitude toward research.

**Adequacy of the MDS Solutions**

The adequacy of the MDS solutions is evidenced by the stress[24] and R-squared values[25]. Both stress and R-squared (RSQ) are measures of fit, indicating how well the solution or model (MDS map or configuration) fits the data, in terms of the percentage of variance in the proximity data that is explained or remains unaccounted for by the MDS model (Coxon, 1982; Kruskal & Wish, 1978). Smaller stress values (approximate amount of variance not accounted for by the MDS model) indicate better fit, whereas larger RSQ values (the amount of variance explained) reflect better fit. For both of the non-metric solutions[26] (Figures 7 and 8), the stress value is approximately .15, which is smaller than stress based on simulation approximation to random data (Spence, 1979). This observation suggests an acceptable and stable solution. Moreover, for both the metric and non-metric solutions (Figures 5 to 8), the RSQ values vary between .82 and .83. That is, approximately 82% to 83% of the variance in the data is explained by the MDS solutions or models, reflecting a very good fit.

---

[24] Stress 1 = Residual sum of squares from monotonic regression.
[25] Stress and R-squared values range from 0 to 1.

Additionally, the stability of the solutions is in keeping with the empirical guideline of at least $4k + 1$ objects (items) for a $k$-dimensional solution (Kruskal & Wish, 1978, p. 34), in this case, two dimensions ($k =2$), and 25 objects (or items). Stability or reliability of the 5-cluster solution is further supported by the consistency of this structure across all the MDS maps or configurations (Kruskal & Wish, 1978). Furthermore, these MDS-maps are highly interpretable, and plausible with regard to empirical and theoretical knowledge about the attitude construct, in particular, the tripartite conceptualization of attitude (Breckler, 1984; Rosenberg & Hovland, 1960; Smith, 1947). This observation also supports the validity of the solutions. In other words, the identified 5-cluster structure is measuring what was intended to be measured, that is, attitude. Validity is enhanced by the convergence or triangulation of the solutions from metric and non-metric multidimensional scaling, exploratory factor analysis, and hierarchical cluster analysis. These analytical techniques represent different methodologies, data perspectives, and assumptions (Hill et al., 1997; Mitchell, 1986; Stiles, 1993), and include both quantitative and qualitative approaches to interpretation of the results. This five-factor structure identified in this study can therefore be considered a reliable and valid measure of instructors' attitude toward the teaching of introductory statistics in the health and behavioral sciences. These five dimensions influence instructors' attitudes.

---

[26] Not consistent with a degenerate, trivial or artifactual solution.

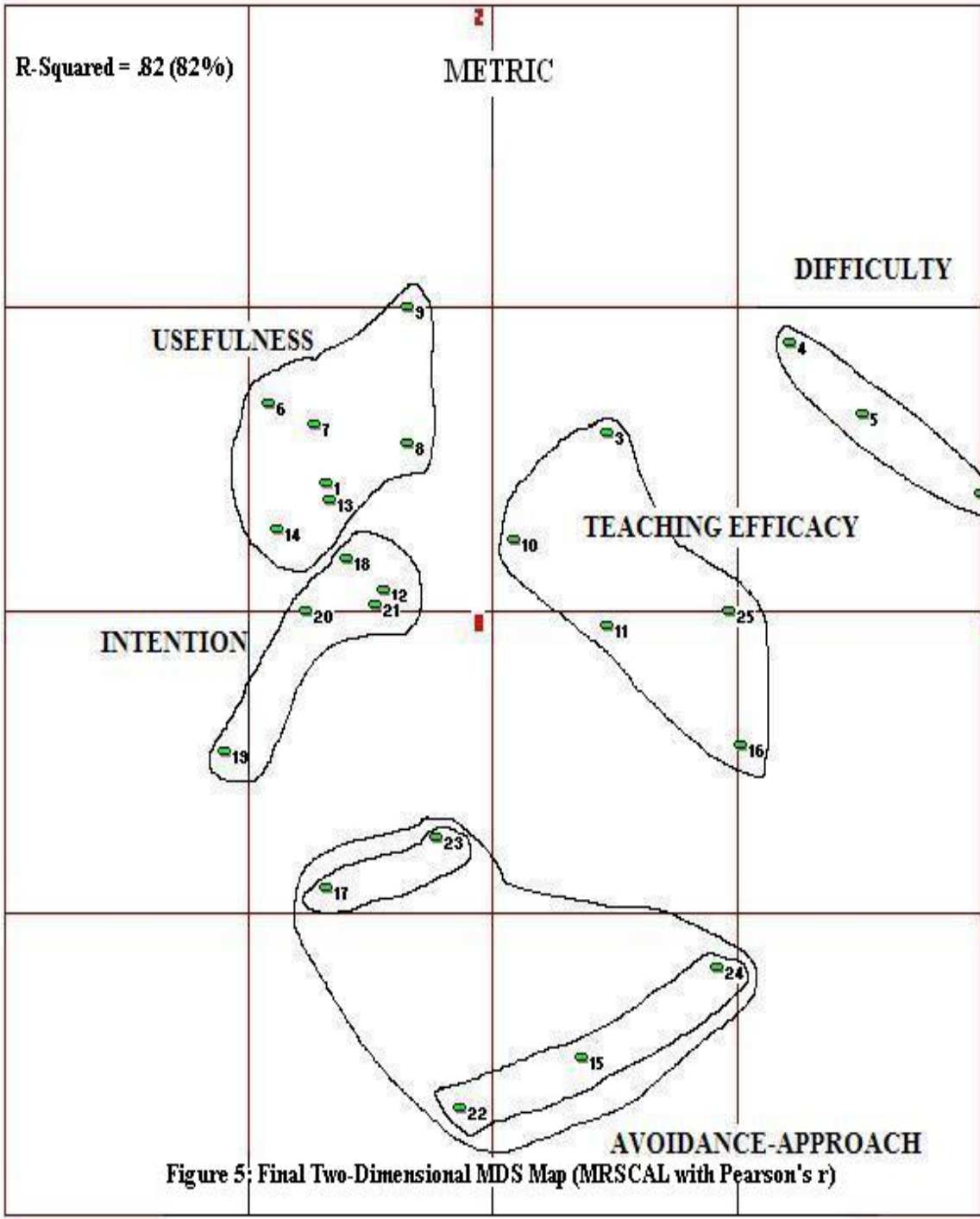

Figure 5: Final Two-Dimensional MDS Map (MRSCAL with Pearson's r)

Please refer to Table 29 for number labels.

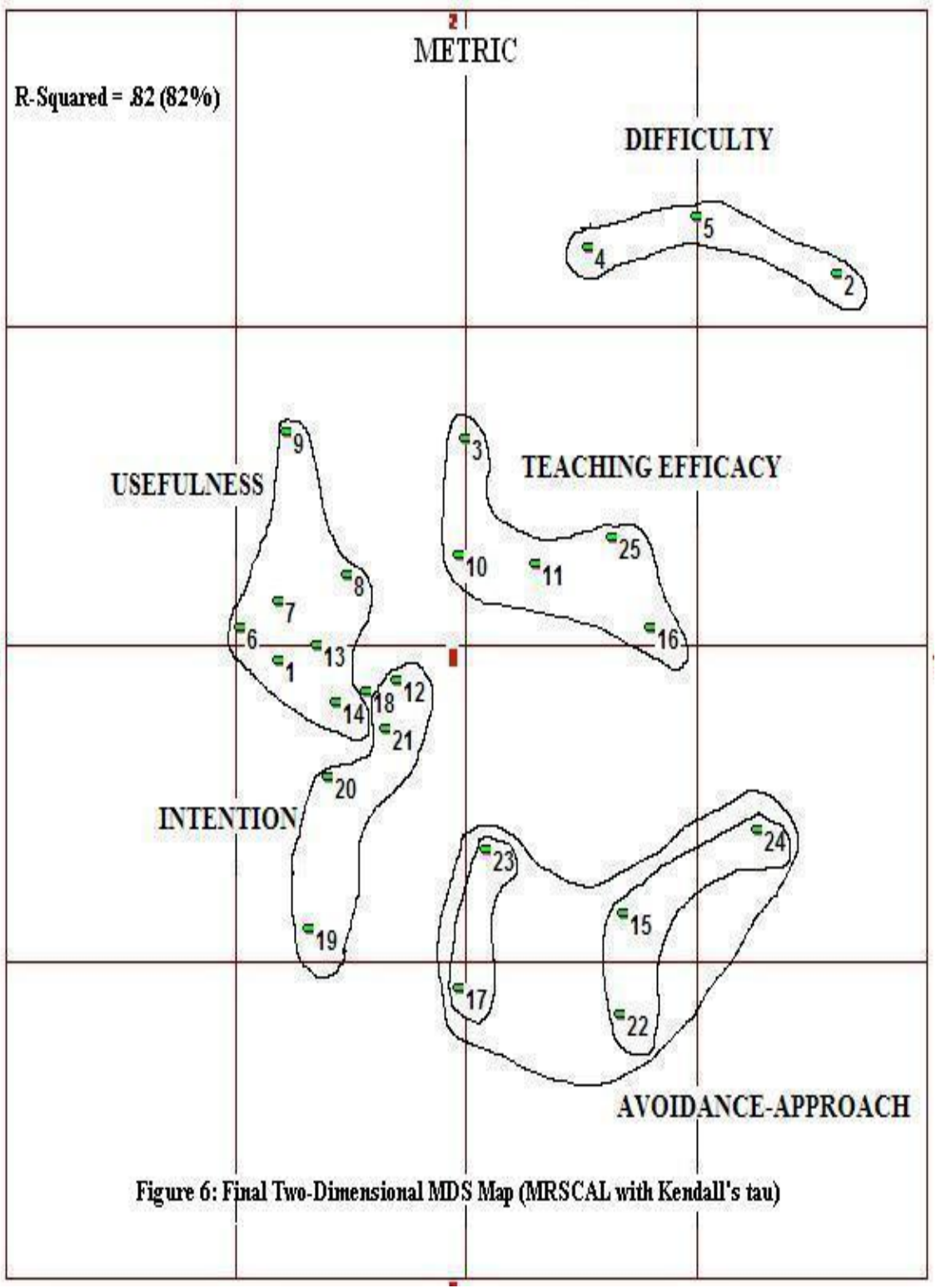

Figure 6: Final Two-Dimensional MDS Map (MRSCAL with Kendall's tau)

Please refer to Table 29 for number labels.

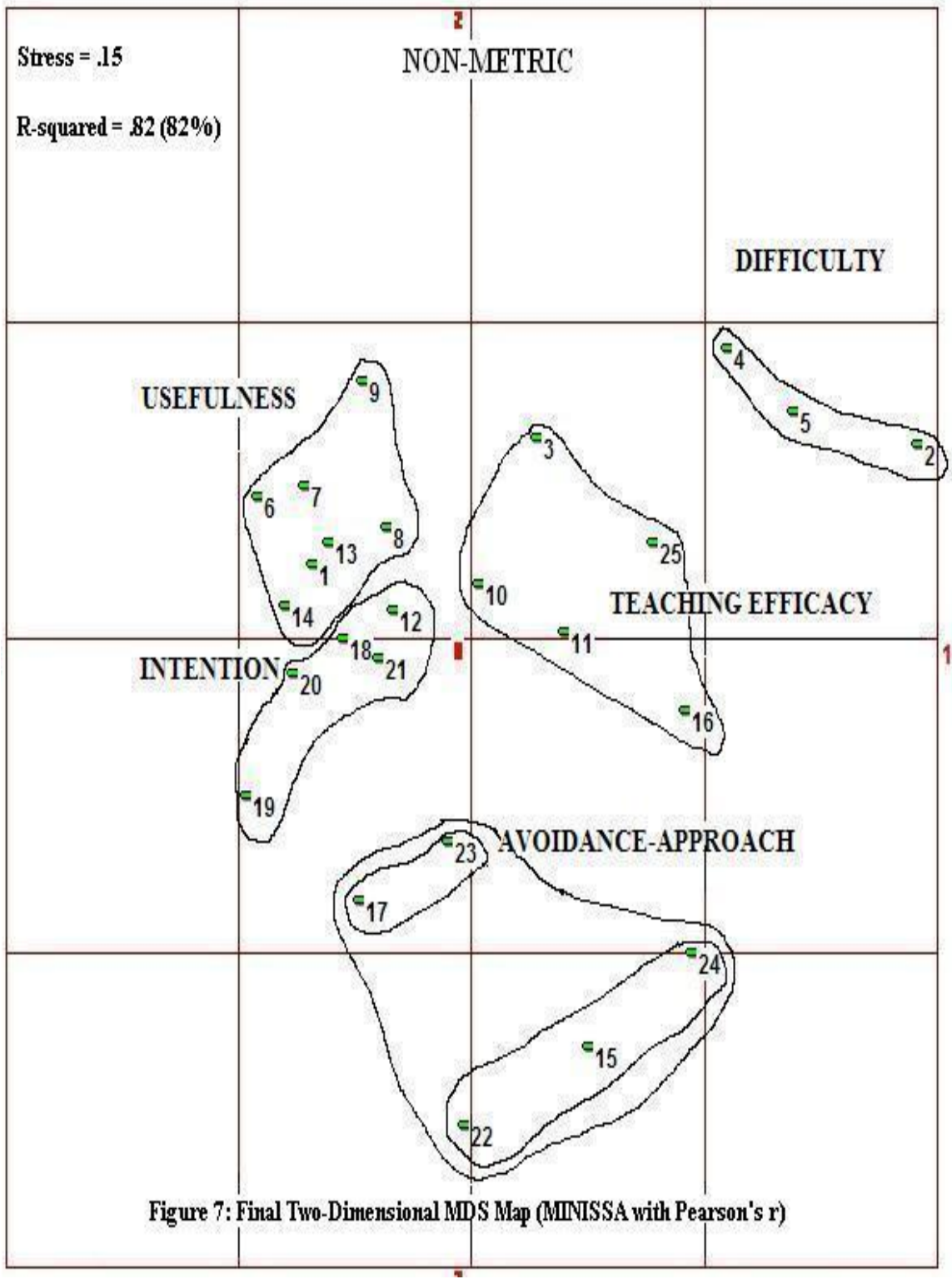

Figure 7: Final Two-Dimensional MDS Map (MINISSA with Pearson's r)

Please refer to Table 29 for number labels.

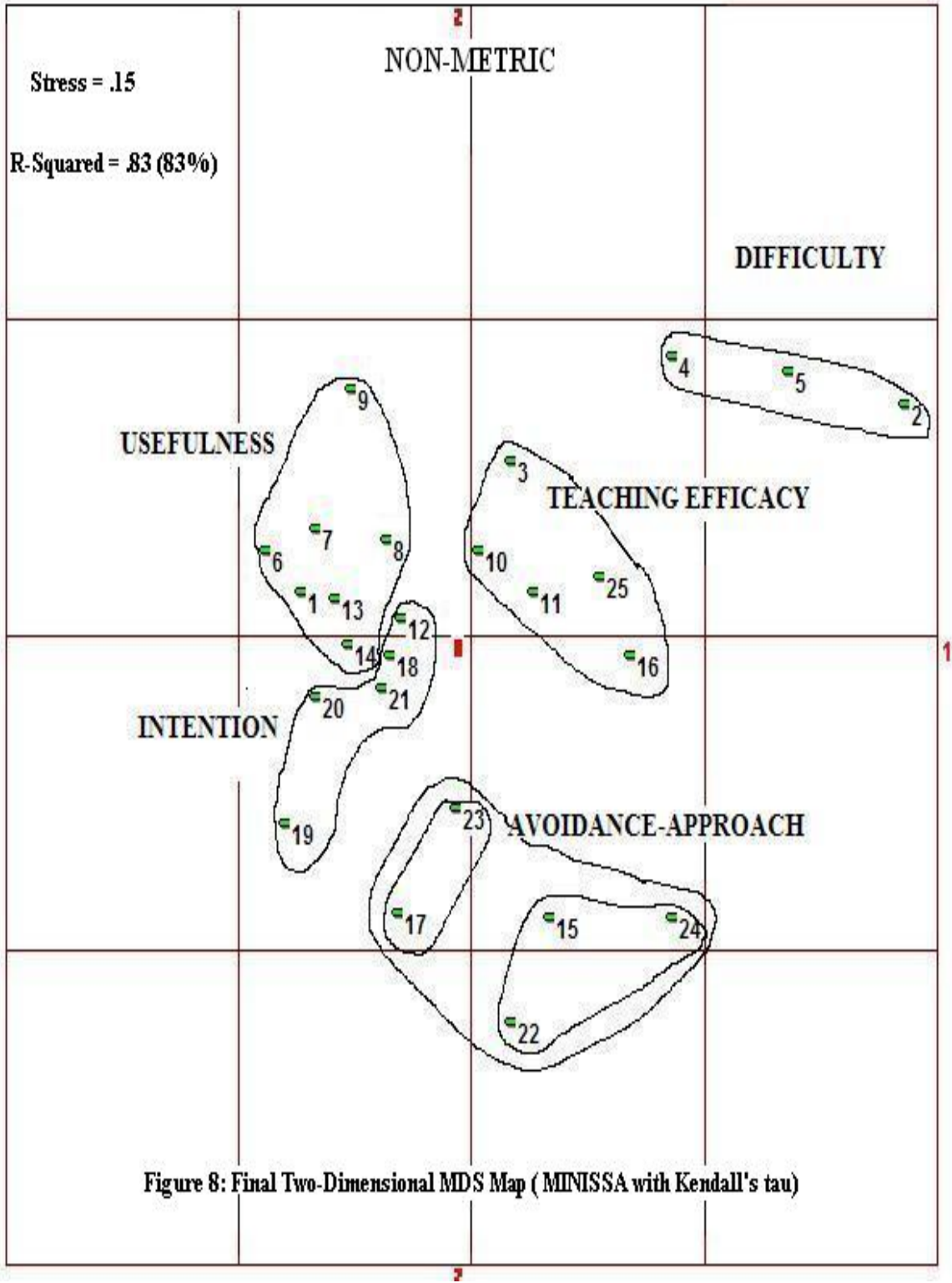

Figure 8: Final Two-Dimensional MDS Map ( MINISSA with Kendall's tau)

Please refer to Table 29 for number labels.

# Scoring and Use of the Attitude Scale

The components or subscales are detailed below with the number of items (refer to Table 26), and the level of internal consistency or reliability (alpha) obtained in this initial exploratory study. Subscale scores are calculated based on a five-point Likert-type scale: 1= strongly disagree, 2= disagree, 3=undecided, 4=agree, 5=strongly agree. All negatively worded items (indicated with an asterisk) must be reverse-coded so that for all items, higher values indicate more positive attitude. Summing the values within each subscale will provide that component score. Either reform-oriented or constructivist can be substituted for concept-based.

**PERCEIVED USEFULNESS: <u>Beliefs</u>** about the value, benefits and worth of the concept-based approach to teaching introductory statistics (7 items, alpha = .85).

1. The concept-based approach to teaching introductory statistics (rather than emphasizing calculations and formulas) makes students better prepared for work.

6. The concept-based approach to teaching introductory statistics (rather than emphasizing calculations and formulas) makes students better prepared for further studies.

7. *Emphasizing concepts and applications in the introductory statistics course (rather than calculations and formulas) is a disservice to our students.

8. *The concept-based approach to teaching introductory statistics is for low achievers only.

9. The concept-based approach to teaching introductory statistics enables students to understand research.

13. I am convinced that the concept-based approach to teaching introductory statistics enhances learning.

14. Teaching introductory statistics using the concept-based approach is likely to be a positive experience for me.

**PERSONAL TEACHING EFFICACY: <u>Beliefs</u>** about one's capability to successfully used the concept-based approach to teach introductory statistics (5 items, alpha = .77).

3. I will adjust easily to teaching introductory statistics using the concept-based approach.

10. *Concept-based teaching of introductory statistics may be problematic for me.

11. *I do not understand how to organize my introductory statistics course to achieve statistical literacy.

16. *Teaching introductory statistics with emphasis on concepts and their applications (rather than calculations and formulas) may be stressful for me.

25. *I am concerned that using the concept-based approach to teach introductory statistics may result in me being poorly evaluated by my students.

**PERCEIVED DIFFICULTY (EASE OF USE):** <u>**Beliefs**</u> about the ease of use, or effort involved in using the concept-based approach to teach introductory statistics (3 items, alpha = .65).

2.  *Teaching introductory statistics with emphasis on concepts and applications rather than calculations and formulas, can be time consuming.
4.  *The preparation required to teach introductory statistics using the concept-based approach is burdensome.
5.  *Using active learning strategies (such as projects, group discussions, oral and written presentations) in the introductory statistics course can make classroom management difficult.

**AVOIDANCE-APPROACH[27]** (<u>**Affect**</u>): Positive and negative <u>**feelings**</u>, inclination, proclivity or propensity toward using the concept-based pedagogy to teach introductory statistics (5 items, alpha = .69).

15. *I am not comfortable using computer applications to teach introductory statistics.

17. Using computers to teach introductory statistics makes learning fun.

22. *I will avoid using computers in my introductory statistics course.

23. I will incorporate active learning strategies (such as projects, hands-on data analysis, critiquing research articles, and report writing) into my introductory statistics course.

24. *I am hesitant to use computers in my introductory statistics class without the help of a teaching assistant.

**BEHAVIORAL INTENTION:** Likelihood of using the concept-based pedagogy to teach introductory statistics (5 items, alpha = .85).

12. I am engaged in the teaching of introductory statistics using the concept-based approach.

18. I am interested in using the concept-based approach to teach introductory statistics.

19. I want to learn more about the concept-based approach to teaching introductory statistics.

20. *Using the concept-based approach to teach introductory statistics is not a priority for me.

21. I plan on teaching introductory statistics according to the concept-based approach.

---

[27] Gray (1970) characterized the mechanism underlying this concept as "facilitative and inhibitory motivational systems", which he posited, produce positive and negative affect respectively. See also, Clark & Watson (1999).

The attitude scale should be presented as follows. Either reform-oriented or constructivist can be substituted for concept-based.

**DIRECTIONS:** The items below are intended to measure your attitude toward the teaching of introductory statistics using the concept-based (constructivist or reform-oriented) approach **(refer to definition above)**. For each item there are 5 possible responses, 1 (strongly disagree = SD), 2 (disagree = D), 3 (undecided = U), 4 (agree = A) and 5 (strongly agree = SA). For each item, circle one number that best represents your position.  Quickly read each item and respond. TRY not to think too deeply.

## Table 30: The Final Attitude Scale

| Attitudinal Items | SD | D | U | A | SA |
|---|---|---|---|---|---|
| 1. The concept-based approach to teaching introductory statistics (rather than emphasizing calculations and formulas) makes students better prepared for work. | 1 | 2 | 3 | 4 | 5 |
| 2. Teaching introductory statistics with emphasis on concepts and applications rather than calculations and formulas, can be time consuming. | 1 | 2 | 3 | 4 | 5 |
| 3. I will adjust easily to teaching introductory statistics using the concept-based approach. | 1 | 2 | 3 | 4 | 5 |
| 4. The preparation required to teach introductory statistics using the concept-based approach is burdensome. | 1 | 2 | 3 | 4 | 5 |
| 5. Using active learning strategies (such as projects, group discussions, oral and written presentations) in the introductory statistics course can make classroom management difficult. | 1 | 2 | 3 | 4 | 5 |
| 6. The concept-based approach to teaching introductory statistics (rather than emphasizing calculations and formulas) makes students better prepared for further studies. | 1 | 2 | 3 | 4 | 5 |
| 7. Emphasizing concepts and applications in the introductory statistics course (rather than calculations and formulas) is a disservice to our students. | 1 | 2 | 3 | 4 | 5 |
| 8. The concept-based approach to teaching introductory statistics is for low achievers only. | 1 | 2 | 3 | 4 | 5 |
| 9. The concept-based approach to teaching introductory statistics enables students to understand research. | 1 | 2 | 3 | 4 | 5 |
| 10. Concept-based teaching of introductory statistics may be problematic for me. | 1 | 2 | 3 | 4 | 5 |
| 11. I do not understand how to organize my introductory statistics course to achieve statistical literacy. | 1 | 2 | 3 | 4 | 5 |
| 12. I am engaged in the teaching of introductory statistics using the concept-based approach. | 1 | 2 | 3 | 4 | 5 |
| 13. I am convinced that the concept-based approach to teaching introductory statistics enhances learning. | 1 | 2 | 3 | 4 | 5 |
| 14. Teaching introductory statistics using the concept-based approach is likely to be a positive experience for me. | 1 | 2 | 3 | 4 | 5 |
| 15. I am not comfortable using computer applications to teach introductory statistics. | 1 | 2 | 3 | 4 | 5 |
| 16. Teaching introductory statistics with emphasis on concepts and their applications (rather than calculations and formulas) may be stressful for me. | 1 | 2 | 3 | 4 | 5 |
| 17. Using computers to teach introductory statistics makes learning fun. | 1 | 2 | 3 | 4 | 5 |
| 18. I am interested in using the concept-based approach to teach introductory statistics. | 1 | 2 | 3 | 4 | 5 |
| 19. I want to learn more about the concept-based approach to teaching introductory statistics. | 1 | 2 | 3 | 4 | 5 |
| 20. Using the concept-based approach to teach introductory statistics is not a priority for me. | 1 | 2 | 3 | 4 | 5 |
| 21. I plan on teaching introductory statistics according to the concept-based approach. | 1 | 2 | 3 | 4 | 5 |
| 22. I will avoid using computers in my introductory statistics course. | 1 | 2 | 3 | 4 | 5 |
| 23. I will incorporate active learning strategies (such as projects, hands-on data analysis, critiquing research articles, and report writing) into my introductory statistics course. | 1 | 2 | 3 | 4 | 5 |
| 24. I am hesitant to use computers in my introductory statistics class without the help of a teaching assistant. | 1 | 2 | 3 | 4 | 5 |
| 25. I am concerned that using the concept-based approach to teach introductory statistics may result in me being poorly evaluated by my students. | 1 | 2 | 3 | 4 | 5 |

# CHAPTER 5

## CONCLUSION AND RECOMMENDATIONS

> Too often, however, the emphasis within teacher education programs is placed on the infusion of content knowledge, pedagogy, and pedagogical content knowledge, with only a cursory treatment of the beliefs that, for better or for worse, will govern the eventual application of what has been acquired within these programs. (Liljedahl, 2005, p.1)

**Overview of the Results**

The objective of this study was to develop and initially validate a scale for measuring instructors' attitudes toward reform-oriented[28] teaching of introductory statistics in the health and behavioral sciences at 4-year degree-granting institutions in the USA (and the equivalent internationally). This is an exploratory study, and based on the accessible literature, and consultation with pioneer educators and researchers in this field, it is the first of its kind. The acronym FATS (**F**aculty **A**ttitudes **T**oward **S**tatistics) will be used to refer to this preliminary scale. All data were obtained from a maximum variation sample of 227 (two hundred and twenty seven) tertiary level instructors of introductory statistics in the health and behavioral sciences, using an internet-based survey[29]. In general, missing data (item non-response) was minimal (less than 4% for almost all analyses), and no imputation was performed[30]. Data were analyzed using primarily factor analysis (FA), multidimensional scaling (MDS) and hierarchical cluster analysis (HC).

Throughout this study, the **concept-based** (reform-oriented or constructivist) approach (Erickson, 2001; Rumsey, 2002) to teaching introductory statistics is operationally defined as a set of related strategies intended to promote statistical thinking and literacy by emphasizing concepts and their applications rather than calculations, procedures and formulas. It involves active learning strategies such as projects, group discussions, data collection, hands-on computer data analysis, critiquing of research articles, report writing, oral presentations, and the use of real-life data. **Statistical literacy** is the ability to understand, critically evaluate, and use statistical information and data-based arguments (Gal, 2000; Garfield, 1999). **Attitude** was conceptualized as an evaluative disposition toward some object, based upon cognitions, affective reactions, and behavioral intentions. In other words, attitude is an informed predisposition to respond, and is comprised of three components, namely beliefs, feelings, and a readiness or intent for action (Jaccard & Blanton, 2004; Zimbardo & Leippe, 1991).

---

[28] Also referred to as the constructivist or concept-based approach.
[29] Internet connectivity (and usage) is now a basic expectation and requirement of faculty at regionally accredited institutions in the USA (and the equivalent internationally).
[30] See Tables 3-8 and 15.

The overall scale (alpha of .89) consists of 25 items and five subscales (**PERCEIVED USEFULNESS, PERSONAL TEACHING EFFICACY, PERCEIVED DIFFICULTY, AVOIDANCE-APPROACH**[31] and **INTENTION**). The structure of this scale is consistent with the tripartite attitude theory[32] (Breckler, 1984; Rosenberg & Hovland, 1960; Smith, 1947). These five subscales (alphas between .65 and .85) represent cognition, affect, and intention[33], and based on the factor analysis solution[34], explained 51% of the total variance, and all of the common variance underlying the twenty-five (25) attitudinal items. Construct validity (comprising of face, content, criterion, convergent and discriminant dimensions) was established. In particular, for criterion validity, the overall scale, and all subscales (except perceived difficulty) plausibly and significantly differentiated between low-reform and high-reform practice instructors (Table 18). In other words, high-reform practice instructors reported more positive attitudes compared to low-reform practice instructors.

---

[31] Gray (1970) characterized the mechanism underlying this concept as "facilitative and inhibitory motivational systems", which he posited, produce positive and negative affect respectively. See also, Clark & Watson (1999).
[32] As well as the model concept of the Theories of Planned Behavior and Reasoned Action
[33] INTENTION has been characterized as a cognitive antecedent of behavior. David Velleman (2002) has argued that intentions reflect beliefs about what one plans to do. This would imply that intentions have a cognitive underpinning, and raises the question: Are cognition and intention separable constructs?
[34] Promax (oblique) rotation was conducted.

The teaching practice scale was developed for this study, and it consists of two dimensions (behaviorist and constructivist) with acceptable internal consistency (reliability), as well as construct validity (Tables 9 and 12). In total, the psychometric evidence establishes that these two teaching practice subscales are independent measures of different dimensions of teaching practice. Furthermore, all instructors reported some degree of each practice (behaviorist and constructivist), and hence can be considered eclectic in their teaching (Lenski, Wham, & Griffey, 1998; Ravitz, Becker, & Wong, 2000; Woolley, Benjamin, & Woolley, 2004). That is, these teaching practices (behaviorist and constructivist) coexist. Together the five attitude subscales explained 28% of the variance in teaching practice, with perceived difficulty being the least salient and predictive in this context.

No statistically significant difference was observed in overall attitude and subscale scores with respect to gender, employment status, membership status in professional organizations, geographic location, ethnicity, highest academic qualification, and degree concentration. With regard to teaching area (Table 24), instructors in psychology and the behavioral sciences, on average, reported a significantly more favorable level of personal teaching efficacy (in relation to reform-oriented or concept-based teaching) compared to instructors in the health sciences. This difference persisted following Bonferroni correction (Table 25). Statistically significant differences in some measures[35] were observed with respect to age group (Table 26). Specifically, younger instructors (40 years and less) reported lower and less favorable levels of overall attitude, perceived usefulness, personal teaching efficacy, and intention, than both groups of older instructors (41 –50, and 51 + years).

Also, statistically significant positive, but weak relationships were noted between the duration of teaching (years) and personal teaching efficacy (Spearman's rho = .21, df = 219, p < .01), as well as avoidance-approach (Spearman's rho = .17, df = 219, p < .05). These relationships suggest that there is a tendency for more experienced instructors to perceive greater capability to use the concept-based pedagogy, and exhibit greater approach (less avoidance) in this regard. In general, the significant differences and relationships observed, are of low magnitude, but plausible. This scale can be considered a reliable and valid measure of instructors' attitudes toward reform-oriented (or concept-based) teaching of introductory statistics in the health and behavioral sciences, at the tertiary level.

Figure 9 summarizes the relationships among the five attitude subscales, and their relative contribution to predicting (and explaining) teaching practice. While these relationships are based on the observed correlation coefficients, corrected correlation coefficients are presented and discussed elsewhere in this dissertation.

---

[35] Intention and Personal Teaching Efficacy remained significant following Bonferroni correction.

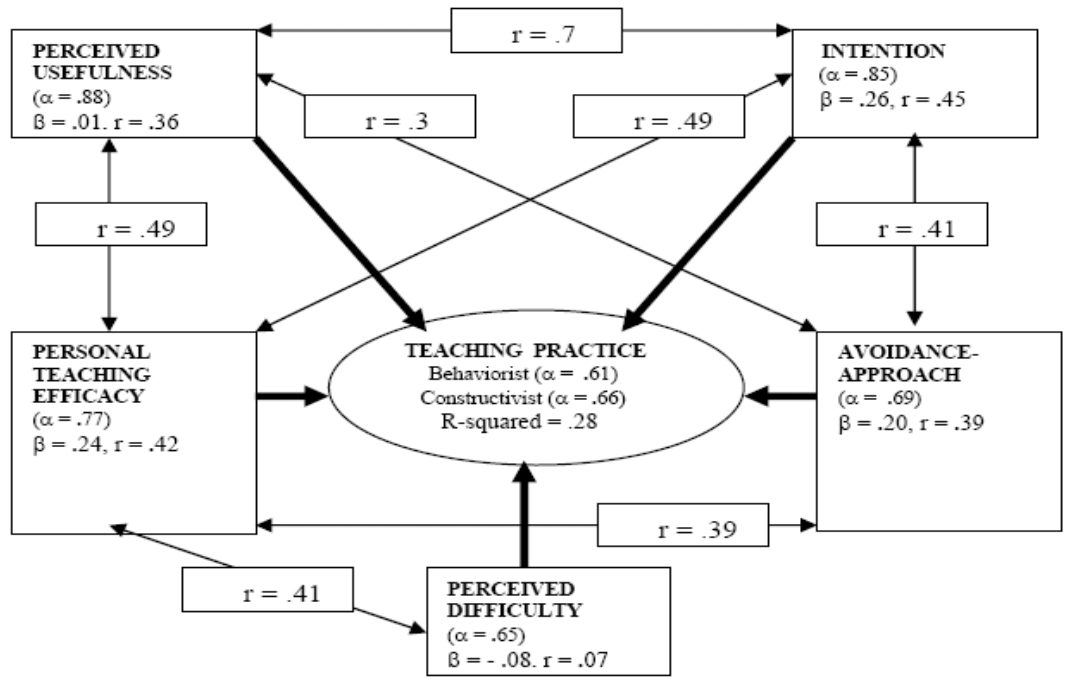

**Figure 9:** Schematic of the salient observed relationships (r) among the attitude subscales, and their relative contribution (β) to teaching practice

## Methodological Considerations (Strengths & Limitations)

The interpretation of these data, and use of this preliminary attitude scale must be guided and tempered by the following considerations, each of which reflects either a strength or limitation of this study.

(1) The conceptualization of attitude and practice is epistemologically restrictive to this study. That is, it applies only to the teaching of introductory statistics in the health and behavioral sciences at the tertiary level, using the concept-based or reform-oriented pedagogy (considered an innovative teaching approach).

(2) The very nature of exploratory factor analysis (EFA) and multidimensional scaling (MDS) in formulating and validating a scale, is at best, a method that allows for a preliminary or tentative outcome. Tucker and MacCallum (1997) note that researchers should anticipate issues such as items which do not define factors as intended, the finding of unanticipated factors, and the absence of anticipated factors. They further state that:

> The achievement of the objective of factor analytic research requires a series of studies, proceeding from initial studies where hypotheses are only loosely formed and analyses are exploratory, to final studies where confirmatory analyses are conducted to test well-developed hypotheses. (Tucker & MacCallum, 1997, p.132)

While the 5-factor attitude scale that emerged from this study is theoretically and empirically plausible, additional studies are required to further explore, and possibly confirm this structure and its psychometric properties.

(3) It is recognized that statistical techniques such as EFA and MDS will always extract factors and generate patterns, which may or may not be plausible or salient with regard to the study context. In other words, the "garbage in, garbage out" phenomenon can apply. To guard against this, a structured and meticulous process was followed for item generation and analysis. This included, a comprehensive review of the literature (including related instruments), consultation with teaching, research and content experts (including psychometricians), structured item analysis sessions, and pre-testing, with attention to attitude theory[36].

(4) Data were analyzed[37] using EFA and MDS. Unlike EFA techniques which require the assumption of interval- or ratio-scaled (metric) data, and linear relationship among the items, all MDS models do not impose such restrictions on the data, as there are both metric (linear transformation) and non-metric (ordinal transformation) variants of MDS (Coxon, 1982; Kruskal & Wish, 1978;

---

[36] Theories of Reasoned Action and Planned Behavior, as well as the Concerns Based Adoption Model
[37] Hierarchical cluster analysis was also used to guide the identification of clusters and patterns within the data.

Schiffman, Reynolds, & Young, 1981;Young & Lewyckyj, 1979). This mixed data analytic approach allowed for examining the data using different models (under different assumptions), and this was considered appropriate given that this is an initial exploratory study. Also, given the complex nature of psychological phenomena, including attitude, it was plausibly assumed that non-linear relationships between attitudinal items (measured using Likert-type scales) may exist, and that the rank order of the measures may yield important information. Hence, using MDS (specifically the non-metric model) was appropriate and helpful.

(5)  All data were obtained from a maximum variation sample of 227[38] (two hundred and twenty seven) tertiary level instructors of introductory statistics in the health and behavioral sciences. This type of "convenience sampling is suited for these studies rather than probabilistic sampling because the aim is not to establish population estimates, but rather to use correlational analysis to examine relationships between items and measures" (Viswanathan, 2005, p.70). Seeking maximum variation in the measures of interest (in this case, attitude and practice) through targeted[39] and purposeful sampling, was intended to counter the potential effect of attenuation of correlation coefficients[40] (and factor loadings) associated with a restricted range in measurement (Comrey & Lee, 1992; Fabrigar et al., 1999; Gorsuch, 1983; Tucker & MacCallum, 1997).

---

[38] In the absence of empirical data for comparison, personal communication with pioneer educators and researchers has established that these data are plausible and quite likely representative of the population distribution of statistics instructors with respect to age, gender, ethnicity, and highest academic degree.
[39] Include institutions and individual faculty engaged in reform-oriented and traditional teaching, as well as listservs with similar members. Characterization was based on knowledge of their engagement in the reform movement, nature of publications, research interests, course outlines, and other materials.
[40] Inter-factor correlation coefficients were corrected for attenuation due to measurement error.

(6) Also, with regard to exploratory factor analysis, the generally recommended conservative sample size of at least 200, and a participant to variable ratio of 5:1 for meaningful results, were met (Floyd & Widaman, 1995; Gorsuch, 1983). There were a few institutions with multiple respondents, and this could have resulted in some correlated data, and consequently a reduction in the effective sample size. Theoretically, this could have decreased the power of the study, especially to detect true differences in attitude and practice with respect to personal and sociodemographic variables. Hence the external validity of such observed correlates and predictors of attitude and practice may be limited.

(7) There were respondents (albeit a few) from international locations where English is not the primary language, however, in using the data, it was assumed that they understood the items as intended by the researcher.

(8) All attitudinal and teaching practice data were self-reported, and the exact validity of these is not known. Furthermore, these are cross-sectional data, and concerns may arise about the temporality of observed relationships between attitudes and practices (which could be symmetric). That is, did the attitude precede the practice or vice versa (considering that both were reported at the same point in time)? Therefore, no claims of causality are made herein.

(9) The concurrent measurement of attitudinal items with teaching practice could have resulted in inflated correlations due to shared error variance, and respondents' "hypothesis guessing" (Albarracin et al., 2001; Wallace et al., 2005). In other words, respondents could have attempted to give attitude-practice responses in order to be consistent.

(10)    It must be noted that the operationalization of concept-based (or reform-oriented) pedagogy as detailed in this study, reflects one variant of this teaching approach, and that other combinations of strategies and curricular materials used by instructors, may also result in effective teaching and learning. Additionally, the definition of concept-based pedagogy as used herein, could have inaccurately implied a dichotomy in teaching approach, that is, concept-based versus mathematical. Furthermore, it cannot be completely discounted that some participants may have responded to the attitude items without fully understanding or referring to the definition (of concept-based pedagogy) detailed in the research instrument. Also, the phraseology of some of the attitudinal items could have been biased toward potential future users of the concept-based (or reform-oriented) pedagogy.

(11)    The size and heterogeneity of the class taught by instructors were not ascertained in this study. Specifically, a large class, and a mix of abilities and academic majors (particularly statistics and non-statistics) could serve as barriers to reform-oriented or concept-based teaching, even though instructors may find it favorable. It is likely therefore, that given this situation, some instructors may have reported negative intentions about using concept-based (or reform-oriented) strategies although they possess an overall positive or favorable attitude toward this teaching approach.

(12)     The offer of a financial incentive could have lead to participation by instructors who were motivated to receive this award, and hence, this could have created a possible selection bias. However, as the incentive was not automatically given to each respondent (but a chance in a lottery to win one of three cash awards), the issue of selection bias, in this regard, is less of a concern.

(13)     Notably, the dimensions of attitude employed by respondents and the importance (salience) they attach to these may vary over time. This five-factor attitude scale is therefore, context, as well as time dependent.

## Translating the Results into Action

The current links between technology in education and constructivist learning environments will succeed more favourably if teachers' beliefs are considered and confronted. (Handal, 2004, p.6).

Preliminarily, these results indicate that instructors' attitude toward the teaching of introductory statistics (according to the reform-oriented, constructivist or concept -based pedagogy) in the health and behavioral sciences at four-year degree-granting institutions, is multidimensional. Specifically, this attitude is a composite of selected beliefs (cognition), feelings (affect), and intentions. The primary evidence of the validity of this attitude scale is its ability to discriminate between low-reform and high-reform practice instructors. Additionally this attitude scale explains 28% of the variability in teaching practice, which is consistent with major attitude-behavior meta-analytic research (Armitage & Conner, 2001). In particular, intention has the greatest explanatory (and predictive) value, making it a prime focus for practice change interventions.  However, in order to influence intention, it is necessary to target the underlying beliefs, as well as feelings (generated by beliefs) from which it is formed.

The statistical analyses also establish that perceived usefulness (beliefs about the benefits and value of reform-oriented pedagogy) is the strongest determinant of intention, followed by personal teaching efficacy (beliefs about one's capability to effectively use the reform-oriented pedagogy), and avoidance-approach (which has an affective underpinning). Accordingly, perceived usefulness is the single most important predictor of intention to act, which in turn, is the primary determinant of teaching practice. The role of perceived difficulty in attitude formation, and predicting teaching practice in this context, is not clear, albeit its moderate and plausible relationship with personal teaching efficacy suggests a possible higher-order or hierarchical factor. The primary target for intervention toward reinforcing and facilitating positive attitudes is therefore, cognition or beliefs, which can be robust, and difficult to change. Interventions will be also be required to achieve attitude-behavior consistency (Fazio, 1990; Snyder, 1982), that is, to facilitate positive attitudes to translate into the adoption and use of the reform-oriented or concept-based pedagogy.

## Implications of the Results and Specific Recommendations

I feel strongly, for example, that statistics is not a subfield of mathematics, and that in consequence, beginning instruction that is primarily mathematical, or even structured according to an underlying mathematical theory, is misguided. (Moore, 1997, p.136)

1. This is an initial exploratory cross-sectional study, and further research of this kind will be required in order to be conclusive about the structural and psychometric properties of this attitude scale. Furthermore, this study examined internal consistency, and not test-retest reliability, which should also be assessed in order to determine the stability of the scale over time. It is strongly recommended that future studies also use factor analysis rather than principal components analysis, as the aim is to identify latent constructs underlying a set of items, and not merely to achieve data reduction. Additionally, non-linear modeling of attitudinal items should be conducted with methods such as multidimensional scaling, as used in this study. Also, noting that the attitude scale in this study explained about 28% of the variance in teaching practice (leaving a large proportion of the variance unexplained), other attitudinal dimensions and items should be explored, as well as other (non-attitudinal) factors, toward a model with greater explanatory (and predictive) value, with attention to parsimony and theoretical plausibility. Two-way and higher-order interactions between and among the personal and sociodemographic factors in relation to attitude should also be examined, as should possible confounding. Consideration must also be given to possible higher-order or hierarchical factors with regard to factor analysis. For example, perceived personal teaching efficacy and difficulty may have the potential to form a single higher-order factor. The teaching practice scale should be similarly explored.

2. Reform-oriented pedagogy is relatively innovative, and the underlying principles may therefore be in conflict with instructors' assumptions and beliefs about teaching and learning. This is particularly so, given that a large proportion of the instructors would have come from the traditionally taught system. Hence, resistance to this approach is to be expected and carefully and tactfully addressed, in order to facilitate and maintain positive change. It is evident from this research that instructors' beliefs about the benefits of reform-oriented teaching, as well as self-efficacy are salient in their decision-making process regarding the use of this teaching approach. And theoretically, such beliefs can result in a disposition of avoidance or approach in relation to reform-oriented pedagogy.

In order to facilitate changing beliefs that are counter to reform-oriented teaching, instructors must be facilitated to reflect on, and analyze their own beliefs, feelings and practices. This can take the form of faculty development seminars and workshops where experienced and successful reform-oriented instructors share their teaching strategies, curricular materials and learning outcomes. The rationale for reform-oriented pedagogy should be made clear, emphasizing the benefits to the instructor, the students, and the institution. In general, best practices from empirical research should be detailed and discussed. Presenters must also discuss the challenges and barriers they faced, and explain how these were overcome or are being addressed.

3. These seminars or workshops should be conducted in a friendly, informal and non-threatening manner, and interaction encouraged. Peer facilitators must also address possible doubts that instructors may entertain about their own ability to use reform-oriented pedagogy successfully. It must be made clear that there are various combinations of active learning strategies that may give rise to the same positive learning outcomes. Furthermore, consistent with the guided mastery approach (Bandura, 1994), an incremental and manageable implementation of reform-oriented teaching strategies should be advocated, that is, gradual change, rather than attempting a complete overhaul of a course, all at once. Discipline-specific facilitators may bring more authority to this process, and prove more persuasive, as the instructor-participants are more likely to identify with them as true peers. Specialized and extended sessions may be required to address the use and integration of technology in the teaching of statistics. In particular, some instructors may have never used a computer for statistics, or may not be familiar with a particular software being recommended. At every stage, timely debriefing sessions should be conducted with constructive feedback.

4. Once instructors are aware of the details of reform-oriented teaching and learning (particularly the potential benefits), and have developed an interest in this approach, self-efficacy comes into question (**Can I do it?).** At this stage, modeling of best practices by faculty peers who have been successful with this approach could help to foster beliefs of capability to adopt and effectively use reform-oriented strategies. Beyond this, attitude-behavior consistency becomes the primary focus (translating attitudes to action). Instructors should be provided with formal peer mentorship. Collaboration in course development, as well as team-teaching with experienced and successful reform-oriented instructors should also be facilitated.

5. Essential to initiating and maintaining the desired teaching approach is appropriate and adequate ongoing support, such as the availability of, and accessibility to appropriate resources, in the form of mentors, books and other materials, as well as time off for preparation and attending professional conferences, adequate financial support, and recognition by the administration. Additionally, there needs to be a formal dialogue between instructors of introductory statistics courses and those teaching advanced courses, so that the respective course objectives and outcomes can be understood and appreciated, and greater harmony achieved.

6. The American Statistical Association, and other bodies concerned with the teaching and learning of statistics should advocate that graduate statistics programs provide students with at least a formal comprehensive introduction to reform-oriented or concept-based pedagogy. This should also become a core focus of professional development and other continuing education programs for instructors. Greater awareness and importance of this pedagogy may be achieved by offering a US-based graduate certificate program in the teaching of statistics, similar to the Royal Statistical Society Certificate in Teaching Statistics in Higher Education, administered by Nottingham Trent University (UK). This is especially so, given the recognized nuances and challenges of statistical reasoning. In particular, some statistical methods and procedures have underlying concepts and assumptions that may be viewed as difficult and counterintuitive.

7. Introductory statistics courses with central and unifying themes such as "variability" (Moore, 1997), "prediction" (McLean, 2000), "data management," and "decision-making" (Hassad, 2002) have been suggested as effective for facilitating statistical thinking and literacy (Wild & Pfannkuch, 1999; Chance, 2002). Research-oriented statistics courses have also been shown to be advantageous in this regard. Toward this end, the epidemiological model has been posited as a practical framework for designing introductory statistics courses to achieve quantitative reasoning (Stroup et al., 2004). This is especially applicable to the evidence-based disciplines (such as health and behavioral sciences).

Instructors should adopt the epidemiological model to guide course development, as this encompasses the themes of variability, prediction, data management, research, and decision-making, with reference to real-world data, as well as salient and universal issues (health). This can motivate students to explore data, discover meaning, and achieve deep and conceptual learning.

8.  Other strategies for enhancing awareness and the importance of reform-oriented pedagogy is to rename the introductory statistics course so that it reflects in a direct, friendly and less threatening way, precisely what we want to achieve. That is, statistical thinking and literacy. Two possible names are "reasoning with data", and "concepts and methods of data analysis".

9.  The importance of reform-oriented pedagogy can also be promoted by making available to instructors, by way of a dedicated online resource, the works and endorsements of the following national and international bodies, projects, and conferences which are dedicated solely or primarily to reform-oriented teaching and learning: The **ASA** (American Statistical Association, Sections on Statistics Education, and teaching of Statistics in the Health Sciences), the **NSF** (National Science Foundation), **CAUSE** (Consortium for the Advancement of Undergraduate Statistics Education), **ARTIST** (Assessment Resource Tools for Improving Statistical Thinking), **USCOTS** (United States Conference On Teaching Statistics), the **GAISE** Report (Guidelines for Assessment and Instruction in Statistics Education), **StatLit** (W.M. Keck Statistical Literacy Project),  the **RSS** (The Royal Statistical Society, Center for Statistics Education), **ICOTS** (International Conference On the Teaching of Statistics), **IASE** (International Association for Statistical Education), and **ISI** (International Statistical Institute).

## Conclusion

Our teaching must therefore avoid the "professional's fallacy" of imagining that our first courses are a step in the training of statisticians. We should ask whether traditional introductions to statistics for general students are too narrow. (Moore, 1997, p.124)

Statistics is increasingly becoming a core requirement for most college majors, and especially in the evidence-based disciplines such as psychology, it is regarded by some academics as "the single most important course in terms of admittance into graduate schools" (Alder & Vollick, 2000). Beyond the academic domain, statistical literacy is necessary for effective decision-making in many aspects of everyday life. Yet, empirical reports abound that students are experiencing difficulties with introductory statistics, and emerge with a lack of understanding of critical concepts, attributed primarily to passive teaching-learning strategies. Moreover, the introductory statistics course may be the only formal exposure that most students will ever have to statistics, and they may go on to become professionals and leaders in the health and behavioral sciences, which now demand evidence-based practice.

Consequently, there has been a long-held consensus among educators that improvement of the introductory statistics course is necessary so that students can emerge with useful and transferable knowledge and skills. While this has been the core focus of the national reform movement for the teaching and learning of introductory statistics, the emphasis has remained largely on subject content, pedagogy, assessment, and integration of technology with little or no attention to instructor characteristics, in particular, their beliefs and attitudes about teaching and learning, which can significantly influence what and how they teach, and how students learn. This study has established that instructors' attitudes are important in the teaching and learning process, and must be given key emphasis if reform efforts are to be successful.

The aforementioned recommendations address further attitude-practice research, instructor education, skills building, teaching practice, appropriate ongoing faculty support and development, as well as timely evaluation and feedback. A comprehensive and multi-level intervention approach to reinforcing and facilitating favorable attitudes and reform-oriented teaching practices is advocated. In this regard, attention must be given to both initiating, and maintaining the desired practices, with focus on the individual (instructor), the academic department, the wider institution, and discipline-specific organizations. Moreover, intervention programs must be cognizant of the institutional culture, in order to identify contextual factors and dynamics that may serve as barriers to, and facilitators of reform.

Critical to the success of such interventions is that instructors must robustly perceive reform-oriented pedagogy as important and beneficial to them, their students, and the institution. Formal commitment to reform-oriented pedagogy by the departmental and institutional administration should prove helpful in this regard, especially amidst the almost universal trend of statistical literacy being a required competency for graduation in most academic majors, in particular, the health and behavioral sciences. The administration must also be committed to promoting and supporting a departmental and wider institutional culture of cooperation and collaboration rather than competition. This is particularly necessary given the many personal concerns and doubts that are known to be associated with the adoption and implementation of innovative practices, such as reform-oriented pedagogy.

Finally, unless and until substantial weighting is given to curricular development and pedagogy, as is given to peer-reviewed publications and grant awards by tenure and promotion committees, national teaching reform programs may prove to be unsuccessful. The pervasive lack of recognition of curricular development and pedagogy as scholarly and scientific endeavors, in this regard, borders on being an act of duplicity that is plaguing higher education, and undermining its core mission of effective teaching and learning.

# LIST OF REFERENCES

Abelson, R. P., & Prentice, D. A. (1989). Beliefs as possessions: A functional perspective. In A. R. Pratkanis, S. J. Breckler, & Ag. G. Greenwald (Eds.), Attitude Structure and Function (pp. 361-381). Hillsdale, NJ: Erlbaum.

Ajzen, I.  (2002). Perceived behavioral control, self-efficacy, locus of control, and the theory of planned behavior. Journal of Applied Social Psychology, 32, 665-683.

Ajzen, I. (1985). From intentions to actions: A theory of planned behavior. In J. Kuhl & J. Beckman (Eds.), Action-control: From cognition to behavior (pp. 11-39). Heidelberg: Springer.

Ajzen, I. (1988). Attitudes, personality, and behavior. Chicago: Dorsey.

Ajzen, I. (1991). The theory of planned behavior. Organizational Behavior and Human Decision Processes, 50(2), 179-211.

Ajzen, I. (2001). Nature and operation of attitudes. Annual Review of Psychology, 52, 27-58.

Ajzen, I., & Fishbein, M. (1977). Attitude-behavior relations: A theoretical analysis and review of empirical research. Psychological Bulletin, 84, 888-918.

Ajzen, I., & Fishbein, M. (1980). Understanding attitudes and predicting social behavior. Englewood Cliffs, NJ: Prentice Hall.

Ajzen, I., & Fishbein, M. (2004). Questions raised by a reasoned action approach: Comment on Ogden (2003). Health Psychology, 23, 431-434.

Ajzen, I., & Fishbein, M. (2005). The influence of attitudes on behavior. In D. Albarracín, B. T. Johnson, & M. P. Zanna (Eds.), The handbook of attitudes (pp. 173-221). Mahwah, NJ: Erlbaum.

Ajzen, I., & Madden, T. J. (1986). Prediction of goal-directed behavior: Attitudes, Intentions, and perceived behavioral control. Journal of Experimental and Social Psychology, 22, 453-474.

Albarracin, D., Johnson, B.T., Fishbein, M., and Muellerleile, P.A. (2001). Theories of Reasoned Action and Planned Behavior as Models of Condom Use: A Meta-Analysis. Psychology Bulletin, 127, 142-161.

Albert, J. (200). Using a Sample Survey Project to Assess the Teaching of Statistical Inference. Journal of Statistics Education v.8, n.1.

Alder, A.G., & Vollick, D. (2000). Undergraduate statistics in psychology: A survey of Canadian institutions. Canadian Psychology, 41, 149-151.

Allport, G. W. (1935). Attitudes. In C. Murchison (Ed.), Handbook of social psychology. Clark University Press, Worcester, Mass, 796-834.

Anastasi, A. (1986). Evolving concepts of test validation. Annual review of psychology, 37, 1-15.

Anastasi, A. (1950). The concept of validity in the interpretation of test scores. Educ. psychol. Measmt, 10, 67-78.

Anastasiado, S. (2002). An instrument for identification of secondary mathematics teachers' attitudes towards statistics. Proceedings of the fifth Cyprus Conference on Mathematics Education.

Anderson, J. C., & Gerbing, D. W. (1988). Structural equation modeling in practice: A review and recommended two-step approach. Psychological Bulletin, 103 (3), 411-423.

Anderson, S. E. (1997). Understanding teacher change: Revisiting the concerns based adoption model [Electronic version]. Curriculum Inquiry, 27(3), 331-367.

Anderson, T. W. & Rubin, H. (1956). Statistical inference in factor analysis. Proceedings of the Third Berkeley Symposium on Mathematical Statistics and Probability, 5, 111-150.

Armitage, C. J., & Conner, M. (1999). The theory of planned behaviour: Assessment of predictive validity and "perceived control." British Journal of Social Psychology, 38, 35-54.

Asher, J.W. (1997). The role of measurement, some statistics, and some factor analysis in family psychology research. Journal of Family Psychology, 11(3) 351-360.

Ashton, P. (1984). Teacher efficacy: A motivational paradigm for effective teacher education. Journal of Teacher Education, 35 (5), 28-32.

Askew, M., Rhodes, V., Brown, M., William, D. & Johnson, D. (1997) Effective Teachers of Numeracy. Report of a Study Carried Out for the Teacher Training Agency. London: King's College London, School of Education.

Aukstakalnis, S., & Mott, W. (1996). Transforming Teaching and Learning through Visualization. Syllabus, 9, no. 6, 14-16.

Bagozzi, R.P., A.M. Tybout, C.S. Craig, and B. Sternthal (1979). The Construct Validity of the Tripartite Classification of Attitudes. Journal of Marketing Research, 17 88-95.

Bailey, D., and Palsha, S. (1992). Qualities of the Stages of Concern Questionnaire and implications for educational innovations. Journal of Educational Research, 85(4): 226-232.

Bakker, A., & Gravemeijer, K. (2004). Learning to reason about distribution. In D. Ben-Zvi and J. Garfield (Eds.), The Challenge of Developing Statistical Literacy, Reasoning, and Thinking (pp. 147-168). Dordrecht, The Netherlands: Kluwer.

Bandura, A. (1986). Social foundations of thought and action. Englewood Cliffs, NJ: Prentice Hall.

Bandura, A. (1989). Social cognitive theory. In E. Barnouw (Ed.), International encyclopedia of communications (Vol. 4, pp. 92-96). New York: Oxford University Press.

Bandura, A. (1994). Self-efficacy. In V. S. Ramachaudran (Ed.), Encyclopedia of human behavior (Vol. 4, pp. 71-81). New York: Academic Press. (Reprinted in H. Friedman [Ed.], Encyclopedia of mental health. San Diego: Academic Press, 1998).

Bandura, A. (1997). Self-efficacy: The exercise of control. New York: Freeman.

Bandura, A. (1998). Health promotion from the perspective of social cognitive theory. Psychology and Health, 13, 623-649.

Bandura, A., & Adams, N. E. (1977). Analysis of self-efficacy theory of behavioral change. Cognitive Therapy and Research, 1, 287-308.

Bandura, A., Adams, N. E., & Beyer, J. (1977). Cognitive processes mediating behavioral change. Journal of Personality and Social Psychology, 35, 125-139.

Baron, A. & Byrne D. (1984). Social psychology: understanding human interaction. (4th edition) Boston: Allyn and Bacon.

Barrett, P. and Kline, P. (1980). Personality factors in the Eysenck Personality Questionnaire. Personality and Individual Differences, 1, 317-333.

Barrett, P., and Kline, P. (1980) The location of superfactors P, E and N within an unexplored personality factor space. Personal and Individual Differences, 1, 239-247.

Barrows, H. S., and Tamblyn, R. M. (1980). Problem-Based Learning: An Approach to Medical Education, New York: Springer.

Bartholomew, D. J. (1995). What is Statistics? J. Roy. Stat. Soc. A, 158(3), 1-16.

Basom, M., R. Timothy Rush, and James Machell (1994). Pre-Service Identification of Talented Teachers Through Non-Traditional Measures: A Study of the Role of Affective Variables as Predictors of Success in Student Teaching. Teacher Education Quarterly, Volume 21 Number 2.

Batanero, C., Godino, J. D., Vallecillos, A., Green, D. R., and Holmes, P. (1994). Errors and difficulties in understanding elementary statistical concepts. International Journal of Mathematical Education in Science and Technology, 25(4), 527-547.

Batanero, C., Godino J. D., Steiner, H. G., & Wenzelburger, E. (1994). The training of researchers in mathematics education: Results from an international study. Educational Studies in Mathematics, 26, 95-102.

Belar, C. (2003). Training for evidence-based practice. APA Monitor on Psychology, ` Volume 34, No. 5., p56.

Bender, D., and Bruno, L.(Ed.) (1991). Cognitive Dissonance: Opposing viewpoints. San Diego: Greenhaven Press, Inc.

Bender, R., Lange, S. (1999). Multiple test procedures other than Bonferroni's deserve wider use. BMJ 318:600–601.

Biggs, J. (1989). Approaches to the enhancement of teaching. Higher Education Research and Development, 8,7-25.

Bilsky, W., & Jehn, K. (2002). Organizational Culture and Individual Values: Evidence for a Common Structure. In M. Myrtek (Ed.), The Person in Biological and Social Context, pp. 211-228. Goettingen, Germany: Hogrefe Press.

Bolton, P. (2001). Cross-cultural validity and reliability testing of a standard psychiatric assessment instrument without a gold standard.  J Nerv Ment Dis., 189(4):238-42.

Bonferroni, C. E. (1936). Teoria statistica delle classi e calcolo delle probabilità. Pubblicazioni del R Istituto Superiore di Scienze Economiche e Commerciali di Firenze, 8, 3-62.

Borg, I., & Shye, S. (1995). Facet theory: Form and content. Newbury Park: Sage (Advanced Quantitative Methods Series).

Borich, G. (1996). Effective Teaching Methods (Third Edition). New York: Macmillan

Boyer Commission on Educating Undergraduates in the Research University (1998). Reinventing Undergraduate Education: A Blueprint for America's Research Universities.

Boyle, C.R. (1999). A problem-based learning approach to teaching biostatistics. Journal of Statistics Education, v.7, n.1.

Breckler, S. J. (1984). Empirical validation of affect, behavior and cognition as distinct components of attitude. Journal of Personality and Social Psychology, 47, 1191-1205.

Breiner, M. J., Stritzke, W. G. K., & Lang, A. R. (1999). Approaching avoidance– A step essential to the understanding of craving. Alcohol Research and Health, 23, 197-206.

Broers, N. J. (2001). Analyzing Propositions Underlying the Theory of Statistics. Journal of Statistics Education, Volume 9, Number 3.

Brooks, J. G., & Brooks, M. G. (1993). In search of understanding: The case for constructivist classrooms. Alexandria, VA: Association for Supervision and Curriculum Development.

Brown, C.A. & Cooney. T.J. (1982). Research on teacher education: A philosophical orientation. Journal of Research and Development in Education, 15 (4), pp. 13-18.

Bruner, J. (1973). Going Beyond the Information Given. New York: Norton.

Bruner, J. (1983). Child's Talk: Learning to Use Language. New York: Norton.

Bruner, J. (1990). Acts of Meaning. Cambridge, MA: Harvard University Press.

Buche, D. D., and Glover, J. A. (1988), Teaching Students to Review Research as an Aid for Problem-Solving, in Handbook For Teaching Statistics and Research Methods, Hillsdale, NJ: Lawrence Erlbaum Associates, 126-129.

Burstein, L., McDonnell, L., Van Winkle, J., Ormseth, T., Mirocha, J., and Guitton, G. (1995). Validating National Curriculum Indicators. The RAND Corporation, Santa Monica, CA.

Butler, R. S. (1998). On the failure of the widespread use of statistics. Amstat News, No. 251, 84.

Campbell, D. T., & Fiske, D. W. (1959). Convergent and discriminant validation by The multitrait-multimethod matrix. Psychological Bulletin, 56, 81-105.

Caprio, M. (1994). Easing into constructivism. Journal of College Science Teaching, 23(6), pp. 210-212.

Carroll, J. D., & Wish, M. (1976). Multidimensional scaling: models, methods and relevance to DELPHI. In H. A. Linstone & M. Turoff (Eds.), The Delphi Method: Techniques and Applications (pp. 402–431). Reading, MA: Addison-Wesley.

Carroll, J. D., & Chang, J. J. (1970a). Analysis of individual differences in multidimensional scaling via an N-way generalization of "Eckart-Young" decomposition. Psychometrika, 35, 283–319. [Reprinted (1984) in P. Davies & A. P. M. Coxon (Eds.), Key Texts in Multidimensional Scaling (pp. 229–252).

Cattell, R. B. (1957). Personality and motivation structure and measurement. New York: World Book.

Cattell, R. B. (1978). The scientific use of factor analysis in behavioral and life sciences. New York: Plenum.

Cattell, R. B., & Jaspers, J.A. (1967). A general plasmode (No. 30-10-5-2) for factor analytic exercises and research. Multivariate Behavioral Research Monographs, 67, 211.

Cattell, R. B. (1966). The Scree test for the number of factors. Multivariate Behavioral Research, 1, 245.

CEBMH (2002). Workshop Manual on Evidence Based Mental Health. Centre for Evidence-Based Mental Health: Oxford University (U.K.).

Chance, B. L. (2002). Components of Statistical Thinking and Implications for Instruction and Assessment. Journal of Statistics Education [Online], 10(3).

Chance, B., delMas, R., & Garfield, J. (2004). Reasoning about sampling distributions. In D.Ben-Zvi and J. Garfield (Eds.), The Challenge of Developing Statistical Literacy, Reasoning, and Thinking (pp. 295-323). Dordrecht, The Netherlands: Kluwer.

Charles, E.P. (2005). The correction for attenuation due to measurement error: Clarifying concepts and creating confidence sets. Psychological Methods, 10, 206-226.

Cheung, S.F., & Chan, D.K.S. (2000). The role of perceived behavioral control in predicting human behavior: A meta-analytic review of studies on the theory of planned behavior. Unpublished manuscript. Chinese University of Hong Kong.

Cho, H., Kim, J., & Choi, D (2003). Early Childhood Teachers' Attitudes Toward Science Teaching: A Scale Validation Study. Education research Quarterly, 27, 2.

Christensen, R., and Knezek, G. (1998). Parallel Forms for Measuring Teachers' Attitudes Toward Computers. Paper Presented at The Society of Information Technology & Teacher Education (SITE) 9th International Conference, Washington, DC, March 13, 1998.

Christensen, R., & Knezek, G. (1999). Stages of adoption for technology in education. Computers in New Zealand Schools, 11(3), 25-29.

Christensen, R. (2002). Effects of Technology Integration Education on the Attitudes of Teachers and Students. Journal of Research on Technology and Education, Volume 34, Number 4.

Churchill, G. A. (1979).A Paradigm for Developing Better Measures of Marketing Constructs. Journal of Marketing Research, Vol. 16, No. 1:64-73.

Clark, L.A., & Watson, D. (1999). Temperament: A new paradigm for trait psychology. In L.A. Pervin & O.P. John (Eds.), Handbook of personality: Theory and research (2nd ed., pp. 399-423). New York: Guilford Press.

Clark, M. S., & Fiske, S. T. (1982). Affect and cognition. Hillsdale, NJ: Erlbaum.

Cobb, G. W. (1992). Teaching statistics. In L. Steen (Ed.), Heeding the call for change: Suggestions for curricular action, MAA Notes, No. 22, 3-43.

Cobb, G.W. (1993). Reconsidering Statistics Education: A National Science Foundation Conference. Journal of Statistics Education, v.1, No. 1.

Cobb, G. W., & Moore, D. S. (1997). Mathematics, Statistics and Teaching. The American Mathematical Monthly, Volume 104, Number 9.

Cobb, P. (1994). Where is the Mind? Constructivist and Sociocultural Perspectives on Mathematical Development. Educational Researcher, 23, 13-20.

Cohen, J., & Cohen, P. (1983). Applied multiple regression/correlation analysis for the behavioral sciences. Hillsdale, NJ: Lawrence Erlbaum Associates, Inc.

Cohen, J., & Cohen, P. (1975). Applied multiple regression/correlation analysis for the behavioral sciences. Hillsdale, NJ: Erlbaum.

Compeau, D., Higgins, C. A., & Huff, S. (1999). Social cognitive theory and individual reactions to computing technology: A longitudinal study. MIS Quarterly, 23 (2), 145-158.

Comrey, A. L. (1988). Factor-analytic methods of scale development in personality and clinical psychology. Journal of Consulting and Clinical Psychology, 56, 754-761.

Comrey, A. L., & Lee, H. B. (1992). A first course in factor analysis (2nd ed.). Hillsdale, NJ: Erlbaum.

Contreras, L.C. (1998). Resolución de problemas: Un análisis exploratorio de las concepciones de los profesores acerca de su papel en el aula. Tese doctoral, Universidad de Huelva.

Cook, R.J., & Farewell, V.T. (1996). Multiplicity Considerations in the Design and Analysis of Clinical Trials. Journal of the Royal Statistical Society, Series A, 159, 93-110.

Cook, T.D., & Campbell, D.T. (1979). Quasi-experimentation: Design and analysis issues for field settings. Chicago, IL: Rand McNally.

Coombs, C. H. (1964). The Theory of Data. New York: Wiley.

Coombs, C. H., Dawes, R. M., & Tversky, A. (1970). Mathematical psychology: An elementary introduction. Englewood Cliffs, NJ: Prentice Hall.

Cooper, S. & Donald, I. (2003). What can multidimensional scaling and facet theory offer health psychology? Health Psychology Update (The British Psychological Society), Vol 12, Issue 1.

Cordery, J. L., & Sevastos, P. P. (1993). Responses to the original and revised job diagnostic survey: Is education a factor in responses to negatively worded items? Journal of Applied Psychology, 78(1), 141-143.

Costello, A. B., & Osborne, J.(2005). Best practices in exploratory factor analysis: four recommendations for getting the most from your analysis. Practical Assessment Research & Evaluation, 10(7).

Cox, D. (2004). ISI honorary member interview. ISI Newsletter, 28(1), 9-10.

Cox, D.R. (1997). The current position of statistics: a personal view. International Statistical Review, 65, 3.

Coxon, A.P.M. (1982). The user's guide to multidimensional scaling. London: Heinemann Educational Books.

Crano, W. D., & Meese, L.A. (1982). Social Psychology: Principles and themes of interpersonal behavior. Dorsey Press, Homewood, Ill.

Crites, S. L., Jr., Fabrigar, L. R., & Petty, R. E. (1994). Measuring the affective and cognitive properties of attitudes: Conceptual and methodological issues. Personality and Social Psychology Bulletin, 20, 619-634.

Cronbach, L. J. (1951). Coefficient alpha and the internal structure of tests. Psychometrika, 16(3), 297-334.

Cronbach, L.J. & Meehl, P.C. (1955). Construct validity in psychological tests. Psychological Bulletin, 52: 281-302.

Cudeck, R., & O'Dell, L.L. (1994). Applications of standard error estimates in unrestricted factor analysis: Significance tests for factor loadings and correlations. Psychological Bulletin, 115, 475-487.

Dallal, G.E. (1990). Statistical Computing Packages: Dare We Abandon Their Teaching to Others? (Editor's Invited Column). The American Statistician, 44, 265-266.

Daugherty, T., Lee, W.N., Gangadharbatla, H., Kim, K., & Outhavong, S. (2005). Organizational virtual communities: Exploring motivations behind online panel participation. Journal of Computer-Mediated Communication, 10(4), article 9.

Davis, F., Bagozzi, R., & Warshaw, R. (1989). User Acceptance of Computer Technology: A Comparison of Two Theoretical Models. Management Science, Volume 35, 1989, pp. 982-1003.

Davis, F.D. (1989). Perceived Usefulness, Perceived Ease of Use and User Acceptance of Information Technology. MIS Quarterly, Vol. 13, No. 3 (September), pp. 319-340.

Dawes, R. M., & Smith, T. L. (1985). Attitude and opinion measurement. In G. Lindzey & E. Aronson (Eds.), Handbook of social psychology (3rd Ed., Vol. 1, pp. 509-566). New York: Random House.

De Flora, S., Quaglia, A., Bennicelli, C., & Vercelli, M.(2005). The epidemiological revolution of the 20th century. FASEB J., 19: 892-897.

delmas, R., Garfield, J., Ooms, A., & Chance, B. (2006). Assessing Students' Conceptual Understanding After a First Course in Statistics. Paper presented at the Annual Meeting of The American Educational Research Association (AERA) San Francisco, CA.

delMas, R. C., Garfiled, J. B., & Chance, B. L., (1998). Exploring the role of computer simulations in developing understanding of sampling distributions. Paper presented at the AERA Annual Meeting.

delMas, R., & Liu, Y. (2005). Exploring students' conceptions of the standard deviation. Statistics Education Research Journal, 4(1), 55-82.

Donald, I., & Cooper, S. R. (2001). A Facet Approach to Extending the Normative Component of the Theory of Reasoned Action. British Journal of Social Psychology, 40, 599-621.

Eagley, A.H. (1992). Uneven Progress: Social Psychology and the Study of Attitudes. Journal of Personality and Social Psychology, 6(5): 693-710.

Edwards, K., & von Hippel, W. (1995). Hearts and minds: The priority of affective versus cognitive factors in person perception. Personality and Social Psychology Bulletin, 21, 996-1011.

Engel, G.L. (1977). The need for a new medical model: a challenge for biomedicine. Science, 196:129–136.

Enochs, L.G. & Riggs, I.M. (1990). Further development of an elementary science teaching efficacy belief instrument. School Science and Mathematics, 93(5), 264-270.

Erickson, H. L. (2001). Stirring the head, heart and soul: Redefining curriculum and instruction (2nd ed.). Thousand Oaks, CA: Corwin Press, Inc.

Erlandson, D.A., Harris, E.L., Skipper, B., & Allen, S.D. (1993). Doing Naturalistic Inquiry: A Guide to Methods. Newbury Park, CA: Sage Publications.

Estrada, A., Batanero, C., Fortuny, J. M., & Díaz, C. (2005). A structural study of future teachers' attitudes towards statistics. Paper presented at the Fourth European Conference on Mathematics Education, Spain.

Fabrigar, L.R., Wegener, D.T., MacCallum, R.C., & Strahan, E.J. (1999). Evaluating the use of exploratory factor analysis in psychological research. Psychological Methods, 4, 272-299.

Fazio, R. H. (1990). Multiple processes by which attitudes guide behavior: The MODE model as an integrative framework. In M. P. Zanna (Ed.), Advances in experimental social psychology (Vol. 23, pp. 75–109). New York: Academic Press.

Festinger, L. (1957). A Theory of Cognitive Dissonance. Evanston, IL: Row, Perterson & Company.

Field, A. P. (2000). Discovering Statistics using SPSS for Windows. London: Sage publications.

Field, A. P. (2005). Discovering Statistics using SPSS (2nd edition). London: Sage publications.

Fishbein, M., & Ajzen, I. (1975). Belief, attitude, intention, and behavior: an introduction to theory and research. Reading, MA: Addison-Wesley Pub.

Fisher, N.I., & Lee, J.J. (2005): The Vital Role of Statistical Science in Assuring National Prosperity. Internat. Statist. Rev., Volume 73, Number 2, p.153.

Fisher, J. D., & Fisher, W. A. (1992). Changing AIDS-risk behavior. Psychological Bulletin, 111, 455–474.

Fitzgerald, L. F., & Hubert, L. J. (1987). Multidimensional scaling: Some possibilities for counseling psychology. Journal of Counseling Psychology, 34, 469-480.

Flannelly, K. J., Ellison, C. G., & Strock, A. L. (2004). Methodologic Issues in Research on Religion and Health. Southern Medical Journal. 97(12): 1231-1241.

Floyd, F.J., & Widaman, K.F. (1995). Factor analysis in the development and refinement of clinical assessment instruments. Psychological Assessment, 7(3), 286-299.

Fornell, C., & Larcker, D.F. (1981). Evaluating structural equation models with unobservable variables and measurement error. Journal of Marketing Research, 18, 39-50.

Franklin, C., & Garfield, J. (2006). Guidelines for Statistics Education Endorsed by ASA Board of Directors. Amstat News (Education), Issue No. 348.

Freeman, C., & Tyrer, P. (1995). Research Methods in Psychiatry. , Gaskell, London. Fried, Y., & Ferris, G. R. (1986). The dimensionality of job characteristics: Some neglected issues. Journal of Applied Psychology, 71, 419-426.

Froman, R.D. (2001). Elements to consider in planning the use of factor analysis: Southern Online Journal of Nursing Research, Issue 5, Vol. 2.

Fullagar, C. (1986). A factor analytic study of the validity of a union commitment scale. Journal of Applied Psychology, 71, 129-136.

Fuller, F.F. (1969). Concerns of teachers: A developmental conceptualization. American Educational Research Journal, 6 (2), 207-226.

Gal, I. (2000). Statistical literacy: Conceptual and instructional issues. In D. Coben, J.

O'Donoghue & G. E. Fitzsimons (Eds.), Perspectives on adults learning mathematics (pp. 135-150). Boston: Kluwer Academic Publishers.

Garfield, J., & Ahlgren, A.(1988). Difficulties in Learning Basic Concepts in Probability and Statistics: Implications for Research. Journal for Research in Mathematics Education, 1988, Vol 19, No 1, 44-63.

Garfield, J. (2002).  The challenge of developing statistical reasoning. Journal of
Statistics Education, Volume 10, Number 3.

Garfield, J. (1993). Teaching Statistics Using Small-Group Cooperative Learning.
Journal of  Statistics Education [Online], 1(1).

Garfield, J. (1995). How Students Learn Statistics. International Statistical Review, 63,
25–34.

Garfield, J. (1999). Thinking about statistical reasoning, thinking and literacy. Paper
presented at First Annual Roundtable on Statistical Thinking, Reasoning and
Literacy (STRL-1).

Garfield, J. (2000). An Evaluation of the Impact of Statistics Reform: Final Report for
National Science Foundation project REC – 9732404.

Garfield, J. (2003). Assessing statistical reasoning. Statistics Education Research Journal,
2(1), 22-38.

Garfield, J., Hogg, B., Schau, C., & Whittinghill, D. (2002). First Courses in Statistical
Science: The Status of Educational Reform Efforts. Journal of Statistics
Education [Online], 10(2).

Garfiled J., delMas R. C., Chance, B. L. (1999). The Role of Assessment in Research on
Teaching and Learning Statistics. Paper presented at the AERA Annual Meeting.

Ghiselli, E.E., Campbell, J.P., & Zedeck, S. (1981). Measurement theory for the
behavioral sciences. San Francisco: W.H. Freeman and Co.

Gibson, S., & Dembo, M.H. (1984). Teacher Efficacy: a construct validation. Journal of
Educational Psychology, 76(4), 569-582.

Giesbrecht, N. (1996). Strategies for Developing and Delivering  Effective Introductory
Level Statistics and Methodology Courses. ERIC Document Reproduction
Service, No. 393-668, Alberta. BC.

Gilmore, E. (1998) Impact of training on the information technology attitudes of
university  faculty. Doctoral dissertation, University of North Texas, Denton.

Glencross, M. J., & Kamanzi W. B. (1996). The Role of Technology in Statistics
Education: A View From a Developing Region. The International Association for
Statistical Education.

Goertz, M. E., Floden, R. E., O'Day, J. (1995). Studies of Education Reform: Systemic Reform, Volume I: Findings and Conclusions (No. RR-035A) U.S. Department of Education.

Gollwitzer, P. M. (1993). Goal achievement: The role of intentions. In W. Stroebe & M. Hewstone (Eds.), European review of social psychology, Vol. 4 (pp. 141–185). Chichester, UK: Wiley.

Gorsuch, R. L. (1983). Factor analysis. Hillsdale, NJ: Lawrence Erlbaum Associates.

Gorsuch, R.L. (1974), Factor Analysis, Philadelphia: W.B. Saunders Co.

Gray, J. A. (1970). The Psychophysiological basis of introversion-extroversion. Behavior Research and Therapy, 8, 249-266.

Greeley, J.D., Swift, W., & Heather, N (1993a). To drink or not to drink? Assessing conflicting desires in dependent drinkers in treatment. Drug and Alcohol Dependence, 32:169-179.

Greeley, J.D., Swift, W., Prescott, J., & Heather, N (1993b). Reactivity to alcohol-related cues m heavy and light drinkers. Journal of Studies on Alcohol, 54:359-368.

Guttman, L. (1965). Introduction to facet design and analysis. Proceedings of the 15th International Congress of Psychology. Amsterdam, Holland.

Guttman, L. (1971). Measurement as a structural theory. Psychometrika, 36:329-347.

Haas, P. F. & Keeley, S. M. (1998). Coping with faculty resistance to teaching critical thinking. College Teaching, 46(2), 63-68.

Hair, J.F.J., Anderson, R.E., Tatham, R.L. & Black, W.C. (1998). Multivariate data analysis, 5th edn. New Jersey: Prentice Hall.

Hall, M.A., Camacho F., Dugan E., & Balkrishnan, R. (2002). Trust in the Medical Profession: Conceptual and Measurement Issues. Health Serv Res., 37(5): 1419-39.

Hall, B., Wallace, R.C., & Dosset, W. A. (1973). A development conceptualization of the adoption process with educational institutions. Austin, TX: Research and Development Center for Teacher Education, The University of Texas.

Hall, G. (1976). The study of individual teacher and professor concerns about innovations. Journal of Teacher Education, 27(1), 22-23.

Hall, G. E., Georg, A.A., & Rutherford, W.L. (1977). Measuring stages of concern about the innovation: A manual for use of the SoC Questionnaire. Austin, TX: Southwest Educational Development Laboratory.

Hall, G., & Hord, S. (1987). Change in schools: Facilitating the process. Albany, NY: State University of New York Press.

Hall, G., George, A.A., & Rutherford, W.L. (1978). Stages of concern about the innovation: The concept, verification, and implications. Austin, TX: Southwest Educational Development Laboratory.

Hall, G.E., George, A.A., & Rutherford, W.A. (1986). Measuring stages of concern about the innovation: A manual for use of the SoC questionnaire. Austin: Southwest Educational Development Laboratory (SEDL).

Hammersley, P. (2004). Using Statistics. Nurse Researcher, Volume 11, Number 3.

Handal, B. (2004). Teachers' instructional beliefs about integrating educational technology. e-Journal of Instructional Science and Technology, 17(1).

Handal, B. (2003). Profiling Teachers: Constructivist- and Behaviorist-Oriented Mathematics. The International Online Journal on Science and Mathematics Education, Volume 3.

Hannah, A., & English, S. (1999). Why innovate: some preliminary findings from a research project on innovations in teaching and learning. Studies in Higher Education, 24(3), 279-289.

Hassad, R. A. (2002). Link and Think - A Model for Enhancing the Teaching and Learning of Statistics in the Behavioral Sciences. Proceedings of the Joint Statistical Meetings (American Statistical Association).

Haynes, S. N., Richard, D. R., & Kubany, E. S. (1995). Content validity in psychological assessment: A functional approach to concepts and methods. Psychological Assessment, 7, 238-247.

Haller, H., & Krauss, S. (2002). Misinterpretations of Significance: A Problem Students Share with Their Teachers? Methods of Psychological Research Online, Vol.7, No.1.

Henderson, C. R. (2002). Faculty conceptions about the teaching and learning of problem solving in introductory calculus-based physics. Ph.D. Dissertation, University of Minnesota.

Hodgson, T. R. (1996). The Effects of Hands-On Activities on Students' Understanding of Selected Statistical Concepts. In E. Jakubowski, D. Watkins, & H. Biske (Eds.), Proceedings of the Eighteenth Annual Meeting of the North American Chapter of the International Group for the Psychology of Mathematics Education (pp. 241–246). Columbus, OH.

Hofer, B. (2002). Personal epistemology as a psychological and educational construct: An introduction. In B. Hofer & P. Pintrich (Eds.), Personal epistemology: The psychology of beliefs about knowledge and knowing (pp. 3-15). Mahwah, NJ: Erlbaum.

Hogg, R. (1992). Report of workshop on statistics education. In L. Steen (Ed.), Heeding the call for change: Suggestions for Curricular Action (MAA Notes No. 22, pp. 34-43). Washington: Mathematical Association of America.

Hogg, R. (1991). Statistical Education: Improvements Are Badly Needed. The American Statistician, 45, 4, 342-343.

Hogue, A., Liddle, H. A., Singer, A., & Leckrone, J. (2005). Intervention fidelity in family-based prevention counseling for adolescent problem behaviors. Journal of Community Psychology, 33(2), 191-211.

Iversen, G.R. (1992). Mathematics and statistics: An uneasy marriage. In F. Gordon and S. Gordon (Eds.), Statistics for the Twenty-First Century. Washington DC: The Mathematical Association of America. MAA Notes No. 26, pp. 37-44.

Jaccard, J., & Blanton, H. (2004). The Origins and Structure of Behavior: Conceptualizing Behavior in Attitude Research: In D. Albarracin, B.T. Johnson, & M.P, Handbook of Attitudes and Attitude Change. Mahwah, NJ: Erlbaum.

Janiak, R. (1997). Blueprint 2000 Student Performance Standards: What Variables Correlate with Teacher Perceptions of Goal 3? Florida Journal of Educational Research, Vol. 37 (1).

Johnson, D.M., & Wardlow, G.W. (2004). Computer experiences, self-efficacy, and knowledge of students entering a land-grant college of agriculture by year and gender. Journal of Agricultural Education, 45 (3), 53 - 64.

Julka, D. L., & Marsh, K. L. (2000). Matching persuasive messages to experimentally induced needs. Current Research in Social Psychology, 5(21), 299-319.

Katz, D. (1960). The functional approach to the study of attitudes. Public Opinion Quarterly, 24 (2), 163-204.

Kember, D. (1997). A reconceptualisation of the research into university academics' conceptions of teaching. Learning and Instruction, 7(3), 255-275.

Keyser, B. B., Broadbear, J. T. (1999). The Paradigm Shift Toward Teaching for Thinking: Perspectives, Barriers, Solutions and Accountability. IEJHE (The International Electronic Journal of Health Education), Vol. 2(3), 111-117.

Kim, J.O., & Mueller, C.W. (1978). Introduction to factor analysis. Beverly Hills: Sage Publications.

King, C. V. (POPULUS). Factor Analysis and Negatively Worded Items. Retrieved from: http://www.populus.com/techpapers/download/fa&_neg_worded.pdf

Kippax, S., Crawford, J. (1993). Flaws in the theory of reasoned action. In D.J. Terry, C. Gallois, and M. McCamish (Eds.), The theory of reasoned action: Its application to AIDS-preventive behavior (pp. 253-269). New York: Pergamon Press.

Kline, T. (2005). Psychological Testing: A Practical Approach to Design and Evaluation. Sage: Thousand Oaks, CA.

Koballa, J.R., & Crawley, F.E. (1985). The influence of attitude on science teaching and learning. School Science and Teaching, 20(4), 222-232.

Koszalka, T. (2001). Effect of Computer-Mediated Communications on Teachers' Attitudes toward using Web Resources in the Classroom. Journal of Instructional Psychology, (28)2, 95-103.

Koszalka, T. (2000) The Validation of a Measurement Instrument: Teachers' Attitudes Toward the use of Web Resources in the Classroom. Quarterly Review of Distance Education. 1(2), 139-144.

Kothandapani, V. (1971). Validation of feeling, belief, and intention to act as three components of attitude and their contribution to prediction of contraceptive behavior. Journal of Personality and Social Psychology, 19, 321–333.

Koul, R., & Rubba, P. A. (1999). An Analysis of the Reliability and Validity of Personal Internet Teaching Efficacy Beliefs Scale. Electronic Journal of Science Education, 4(1)

Krathwohl, D.R., Benjamin S.B., & Bertram, B. M. (1964). Taxonomy of Educational Objectives: The Affective Domain. New York: McKay.

Kruskal, J. B., & Wish, M. (1978). Multidimensional Scaling. Newbury Park, CA: Sage.

Kulviwat, S., Bruner, G., & Neelankavil, J. (2005). The Role of Self-efficacy in Predicting Technology Acceptance. Proceedings of the Academy of Marketing Science Annual Conference.

Le, V., Lockwood, J.R., Stecher, B., Hamilton, L., Williams, V., Robyn, A., Ryan, G., & Alonzo, A. (2004). Is Reform-Oriented Teaching Related to Mathematics and Science Achievement? RAND working paper series. Retrieved from http://www.rand.org/education/pubs/reports.html

Leach, M., Hennessy, M., & Fishbein, M. (2001). Perception of easy-difficult: Attitude or self-efficacy?. Journal of Applied Social Psychology 3, 1–20.

Lenski, S. D., Wham, M. A., & Griffey, D. C. (1998). Literacy orientation survey: A survey to clarify teachers' beliefs and practices. Reading Research and Instruction, 37(3), 217-236.

Leont'ev, A. N. (1972). The problem of activity in psychology. Voprosy filosofii, 9, 95-108. Reprinted in J. V. Wertsch (Ed. & Trans.), The concept of activity in soviet psychology. Armonk: M. E. Sharpe.

Lewin, K. (1951). Problems of research in social psychology. In D. Cartwright (Ed.), Field theory in social science: Selected theoretical papers (pp. 155-169). New York: Harper & Row.

Lewis R, J. (2004). Reliability and Validity: Meaning and Measurement. Annual Meeting of the Society for Academic Emergency Medicine.

Likert, R. (1932). A Technique for the Measurement of Attitudes. Archives of Psychology, 140, 55.

Liljedahl, P. (2005). Re-educating Preservice Teachers of Mathematics: Attention to the affective domain. Proceedings of the 27th International Conference for Psychology of Mathematics Education - North American Chapter. Roanoke, Virginia.

Lingoes, J. C. (1977). Geometric representations of relational data. Ann Arbor, MI: Mathesis.

Lord, F. M., & Novick, M. R. (1968). Statistical theories of mental test scores. Reading, MA: Addison-Wesley.

Lumpe, A. T., Haney, J., & Czerniak, C. (2000). Assessing Teachers' Beliefs about Their Science Teaching Context. J Res Sci Teach, 37: 275-292.

MacCallum, R. C., Widaman, K. F., Zhang, S., & Hong, S. (1999). Sample size in factor analysis. Psychological Methods, 4, 84-99.

Maclnnes, D. (2003). Evaluating an assessment scale of irrational beliefs for people with mental health problems. Nurse Researcher, Volume 10, Number 4.

Macnaughton, D. (1996).  The Introductory Statistics Course: A New Approach. MathStat  Research Consulting, Inc.  Retrieved from: http://www.matstat.com/teach/p0000.htm.

Madden, T. J., Ellen, P.S., & Ajzen, I. (1992). A Comparison of the Theory of Planned Behavior and the Theory of Reasoned Action. Personality and Social Psychology Bulletin, 18, 3-9.

Maney, D.W., Monthley, H. L., & Carner, J (2000). Preservice teachers' attitudes toward teaching health education. American Journal of Health Studies, 16 ( 4), 185-192.

Marsh, H. W. (1996). Positive and negative global self-esteem: A substantively meaningful distinction or artifacts? Journal of Personality and Social Psychology, 70(4), 810-819.

Masse, M.H., Popovich, M.N. (1998). Assessing faculty attitudes towards the teaching of writing. Journalism & Mass Communication Educator, 53 (3), 50-64.

Matthew, S. (2001). An analysis of selected variables to determine faculty attitudes toward adoption of computer-based instruction at Oklahoma State University (Doctoral dissertation, Oklahoma State University). Dissertation Abstracts International, 63/01A, 0664.

Mayer, D. (1999). Measuring Instructional Practice: Can Policy Makers Trust Survey Data? Educational Evaluation and Policy Analysis, 21(1): 29-45.

McGuire,W. J. (1969). The nature of attitudes and attitude change. In G. Lindzey & E. Aronson (Eds.), The handbook of social psychology (Vol. 3, pp. 136–314). Reading, MA: Addison-Wesley.

McKinley, R. K., Manku-Scott, T., Hastings, A., French, D., & Baker, R. (1997). Reliability and validity of a new measure of patient satisfaction with out of hours primary medical care in the United Kingdom: development of a patient questionnaire. British Medical Journal, 314:193.

McLean, A. L. (2000). The predictive approach to teaching statistics. Journal of Statistics Education, 8(3).

Mcleod, S. (1991). The Affective Domain and the Writing Process: Working Definitions. JAC 11 (1).

Mcnamara, O., Jaworski, B., Rowland, T., Hodgen, J., & Prestage, S. (2002). Developing Mathematics Teaching and Teachers. (A Research Monograph). Retrieved from: http://www.maths-ed.org.uk/mathsteachdev/pdf/mdevc2.pdf

MCPHS (2004). Massachusetts College of Pharmacy and Health Sciences, Online Catalog. Retrieved from: http://www.mcp.edu

McQueen, D.V. (2007). Continuing efforts in global chronic disease prevention. Preventing Chronic Disease, Volume 4, Number 2. Retrieved from: http://www.cdc.gov/pcd/issues/2007/apr/07_0024.htm.

Medawar, P.B. (1979). Advice to a Young Scientist. Pan Books, London and Sydney.

Meehl, P. E. (1954). Clinical versus statistical prediction: A theoretical analysis and a review of the evidence. Minneapolis: University of Minnesota Press.

Messick, S. (1995). Validity of Psychological Assessment: Validation of Inferences from Persons' Responses and Performances as Scientific Inquiry into Score Meaning. American Psychologist, 50, 741-749.

Miller, G.A. (1956). The Magical Number Seven, Plus or Minus Two: Some Limits on Our Capacity for Processing Information. The Psychological Review, 1956, vol. 63, pp. 81-97.

Mills, J.D. (2002), Using Computer Simulation Methods to Teach Statistics: A Review of the Literature. Journal of Statistics Education [Online], 10(1).

Moore, D. (1997). New pedagogy and new content: the case of statistics. International Statistical Review, 65(2), 123-165

Moore, D. S. (1990). Uncertainty. In L. A. Steen (Ed.), On the shoulders of giants: New Approaches to numeracy (pp. 95-137). Washington, DC: National Academy Press.

Moore, D.S. (1988). Should Mathematicians Teach Statistics? (with discussion). The College Mathematics Journal, 19, 3-35.

Moore, D. S. (1988). Statistics Among the Liberal Arts. Journal of the American Statistical Association, Vol. 93, No. 444, pp. 1253-1259.

Moore, D. S. (1993). A Generation of Statistics Education: An Interview with Frederick Mosteller. Journal of Statistics Education, v.1, n.1.

Moore. D. S. (2005). Quality and Relevance in the First Statistics Course: International Statistical Review, Vol. 73, Issue 2, p. 205.

Mosteller, F. (1989). Foreword. In J. Tanur, F. Mosteller, W. Kruskal, E. Lehmann, R. Link, R. Pieters, & G. Rising (Eds.), Statistics: a Guide to the Unknown, (Third ed., pp. ix-x). Pacific Grove, C. A.: Wadsworth & Brooks/Cole Advanced Books & Software.

Muijs, D., & Reynolds, D. (2001). Effective teaching: Evidence and practice. London: Paul Chapman Publishing.

Muldoon, M. F., Barger, S., Flory, J., & Manuck, S. (1998). What are quality of life measurements measuring? British Medical Journal, Vol. 316, No. 7130, 542 –545.

Nietfeld, J. L., & Cao, L. (2003). Examining instructional strategies that promote pre-service teachers' personal teaching efficacy. Current Issues in Education [Online], 6(11).

NIH (2003). Report of the National Institutes of Health - Program in Biomedical and Behavioral Nutrition Research and Training.

NIH (1991). National Institutes of Health - Toxicology Outreach Panel Report.

Notani, A. S. (1998). Moderators of perceived behavioral control's predictiveness in the theory of planned behavior: A meta-analysis. Journal of Consumer Psychology, 7, 247–271.

Nunnally, J. C., & Bernstein, I. H. (1994). Psychometric theory (3rd Ed.). New York: McGraw-Hill.

Nunnally, J.C. (1967). Psychometric Theory. New York: McGraw Hill.

Nunnally, J.C. (1978). Psychometric Theory (2nd ed.). New York: McGraw Hill.

O'Connell, A. (2002). Student Perceptions of Assessment Strategies in a Multivariate Statistics Course. Journal of Statistics Education [Online], 10(1).

Ogden, J. (2003). Some problems with social cognition models: a pragmatic and conceptual analysis. Health Psychology, 22, 424–428.

O'Keefe, D. J. (2002). Persuasion: Theory & Research, (2nd Edition). Thousand Oaks, CA: Sage Publications, Inc.

Onwuegbuzie, A.J., & Leech, N.L. (2003). Assessment in statistics courses: More than a tool for evaluation. Assessment & Evaluation in Higher Education, 28(2), 115-127.

Open Learning Technology Corporation (1996). Learning with Software - Pedagogies and Practice - Learning Concepts. Retrieved from: http://www.oltc.edu.au/CP/05.htm

Osgood, C.E., Suci, G., & Tannenbaum, P. (1957). The measurement of meaning. Urbana, IL: University of Illinois Press.

Ottaviani, M-G. (1998). Developments and Perspectives in Statistical Education. Proceedings of the Joint IASS/IAOS Conference: Statistics for Economic and Social Development, Mexico.

Ottaviani, M-G.(1999). A Note on Developments and Perspectives in Statistics Education. Invited paper at CLATSE4 (IV Congreso Latinoamericano De Sociedades de Estadistica), Argentina.

Pakenham-Walsh, N. (2002). Local capacities to create and adapt information for healthcare workers in developing countries: An information explosion - with little impact? Bulletin von Medicus Mundi Schweiz, Nr. 85.

Papanastasiou, E. C. (2005). Factor Structure of the Attitudes toward Research Scale. Statistics Education Research Journal, 4(1), 16-26.

Patton, M. Q. (1990). Qualitative evaluation and research methods (2nd ed.). Newbury Park, CA: Sage Publications.

Pedhazur, E. J. (1997). Multiple regression in behavioral research. Harcourt Brace: Orlando, FL.

Perez, M., Luquis, R., & Allison, L. (2004). Instrument development for measuring teachers'attitude and comfort in teaching human sexuality. American Journal of Health Education, 35(1), 24-29.

Perneger, T. V. (1998). What's wrong with Bonferroni adjustments. British Medical Journal (Education and Debate), 316:1236-1238.

Perney, J., & Ravid, R. (1991). The Relationship Between Attitudes Towards Statistics, Math Self-Concept, Test Anxiety and Graduate Students' Achievement in an Introductory Statistics Course. Unpublished manuscript, National College of Education, Evanston, IL.

Pfleiderer, E.M. (2003). Development of an empirically-based index of aircraft mix. Office of Aviation Medicine, Report No. DOT/FAA/AM-03/8. Washington, DC.

Piaget, J. (1921). Une forme verbale de la comparaison chez l'enfant. Archives de Psychologie, 141-172.

Piaget, J. (1929). The Child's Conception of the World. NY: Harcourt, Brace Jovanovich.

Piaget, J. (1932). The Moral Judgement of the Child. NY: Harcourt, Brace Jovanovich.

Pinkley, R. S., Gelfand, M. J., & Duan, L. (2005). When, where and how: The use of multidimensional scaling methods in the study of negotiation and social conflict. International Negotiation, 10, 79-96.

Polit, D.F., Beck, C.T., & Hungler, B.P. (2001). Essentials of Nursing Research: Methods, appraisal and Utilization (5th Ed). Philadelphia: Lippincott Williams & Wilkins.

Prosser, M., & Trigwell, K. (1999). Understanding learning and teaching: The experience in higher education. Buckingham: SRHE/Open University press.

Rakes, G. C., & Casey, H.B. (2002). An analysis of teacher concerns toward instructional technology. International Journal of Educational Technology v3, n1.

Ramsden, P. (1992). Learning to Teach in Higher Education. London: Routledge.

Ravitz, J.L., Becker, H.J., & Wong, Y. (2000). Constructivist compatible beliefs and practices among U.S. teachers. Teaching, Learning & Computing Report 4. Irvine, CA: Center for Research on Information Technology and Organizations, University of California. Retrieved from: http://www.crito.uci.edu/TLC/findings/report4/

Reading, C., & Shaughnessy, J. M. (2004). Reasoning about variation. In D. Ben-Zvi and J. Garfield (Eds.), The Challenge of Developing Statistical Literacy, Reasoning, and Thinking (pp. 201-226). Dordrecht, The Netherlands: Kluwer.

Riel, M., & Becker, H. (2000). The Beliefs, Practices, and Computer Use of Teacher Leaders. Paper presented at the American Educational Research Association Conference (AERA), New Orleans.

Roberts, H. V. (1992). Student-Conducted Projects in Introductory Statistics Courses. In F. and S. Gordon (Eds.), Statistics for the Twenty-First Century (MAA Notes No. 26), pp. 109-121. Washington, DC: Mathematical Association of America.

Robinson, J.P., Shaver, P.R., & Wrightsman, L.S. (1991). Measures of personality and social psychological attitudes. San Diego: Academic Press, pp. 527-534.

Rogers, E. M. (1983). Diffusion of innovations (3rd ed.). New York: The Free Press.

Rogers, E.M. (1995). Diffusion of innovations (4th ed.). New York: The Free Press.

Rosenberg, M., & Hovland, C. (1960). Cognitive, affective and behavioral components of attitudes. In Carl I. Hovland & Milton J. Rosenberg (Eds.), Attitude organization and change (pp. 1–14). New Haven, CT: Yale University Press.

Rosenthal, R., & Rosnow, R. (1991). Essentials of Behavioral Research: Methods and Data Analysis. McGraw-Hill.

Roskam, E. E., & Lingoes, J.C. (1970) Minissa-I:  A FORTRAN IV (G) program for the smallest space analysis of square symmetric matrices. Behavioral Science, 15, 204-205.

Roskam, E.E. (1972). M.D.S. by metric transformation of the data. Psychologie, 27.

Rothman, K. (1990). No Adjustments Are Needed for Multiple Comparisons. Epidemiology, 1, 43-46.

Rotter, J. B. (1966). Generalized expectancies for internal versus external control of reinforcement. Psychological Monographs (General & Applied), 80, 1-28.

Rummel, R. J. (1970). Applied Factor Analysis. Evanston, IL: Northwestern University Press.

Rumsey, D. J. (2002). Statistical literacy as a goal for introductory statistics courses. Journal of   Statistics Education, 10(3).

Russell, D. W. (2002). In search of underlying dimensions: The use (and abuse) of factor analysis in Personality and Social Psychology Bulletin. Personality and Social Psychology Bulletin 28, 1629-1646.

Sackett, D.L., Straus, S.E., Richardson, W.S., Rosenberg, W., & Haynes, R.B. (Eds.) (2000). Evidence-Based Medicine: How to Practice and Teach EBM. Edinburgh, Churchill Livingstone.

Sackett, D., Rosenberg, W., Gray, J., Haynes, R., Richardson, W. (1996). Evidence-Based Medicine: What it is and what it isn't. British Medical Journal, 312, 7023, 71-72.

Saldanha, L. A., & Thompson, P. W. (2001). Students' reasoning about sampling distributions and statistical inference. In R. Speiser & C. Maher (Eds.), Proceedings of The Twenty-Third Annual Meeting of the North American Chapter of the International Group for the Psychology of Mathematics Education (pp. 449-454), Snowbird, Utah. Columbus, Ohio.

Schafer, D. W., & Ramsey, F. L. (2003). Teaching the Craft of Data Analysis. Journal of Statistics Education, Volume 11, Number 1.

Schau, C. G., Dauphinee, T., & Del Vecchio, A. (1992). The development of the Survey of Attitudes Toward Statistics. Paper presented at the American Educational Research Association Conference (AERA), San Francisco, CA.

Schield, M. (2000). Statistical Literacy: Difficulties in Describing and Comparing Rates and Percentages. Paper presented at the Joint Statistical Meetings (American Statistical Association).

Schmitt, N. & Stults, D. M. (1985). Factors defined by negatively keyed items: The result of careless respondents? Applied Psychological Measurement, 9(4), 367-373.

Schmitt, N. (1996). Uses and abuses of coefficient Alpha. Psychological Assessment, 8, 350-353.

Schmitz, C.D., & Lucas, C.J. (1990). Seeking the right stuff: Attitudinal traits, personal style, and other affective variables as predictors of exemplary student teaching. Education, 110 (3), 270-282.

Sequeira, H., Hollins, S., & Howlin, P. (2003). Symptoms of psychological Disturbance associated with sexual abuse in people with intellectual disabilities. British Journal of Psychiatry, 183, 451-456

Shaffer, J. P. (1995). Multiple Hypothesis Testing. Ann. Rev. Psych. 46, 561-584.

Shepherd, B. H., Hartwick, J., & Warshaw, P. R. (1988). The theory of reasoned action: A meta-analysis of past research with recommendations for modifications and future research. Journal of Consumer Research, 15, 325–342.

Simms, L. J., Casillas, A., Clark, L .A., Watson, D., & Doebbeling, B. I. (2005). Psychometric evaluation of the Restructured Clinical Scales of the MMPI-2. Psychological Assessment, 17, 345-358.

Singer, J., & Willett, J. (1993). Lessons We Can Learn from Recent Research on Teaching: It's Not Just the Form, It's the Authenticity. Paper presented at the Joint Statistical Meetings (American Statistical Association), San Francisco, CA.

Skemp, R.R (1987). The psychology of learning mathematics. NJ Hillsdale: Lawrence Erlbaum Associates.

Skinner, B. F. (1974). About behaviorism. New York: Knopf.

Smith, M. B., Bruner, J.S., & White, R.W. (1956). Opinions and personality. New York: Wiley.

Snyder, M., & Kendzierski, D. (1982). Acting on one's attitudes: Procedures for linking attitude to behavior. Journal of Experimental Social Psychology, 18, 165–183.

Steinhorst, R. K., & Keeler, C. M. (1995). Developing Material for Introductory Statistics Courses from a Conceptual, Active Learning Viewpoint. Journal of Statistics Education v.3, n.3 (1995).

Stevens, J.P. (1992). Applied Multivariate Statistics for the Social Sciences (2nd edition). Hillsdale, NJ: Erlbaum.

Streiner, D. L. (1994). Figuring out factors: The use and misuse of factor analysis. Canadian Journal of Psychiatry, 39, 135-145.

Stroup, D. F., Goodman, R. A., Cordell, R., & Scheaffer. R. (2004). Teaching statistical principles using epidemiology: Measuring the health of populations. The American Statistician, 8, 77-84.

Tabachnick, B. G., & Fidell, L. S. (1996). Using multivariate statistics (3rd ed.). New York: Harper Collins.

Tabachnick, B. G., & Fidell, L. S. (2001). Using multivariate statistics (4th ed.). Boston: Allyn & Bacon.

Tapia, M., Marsh, G.E. (2004).  An Instrument to Measure Mathematics Attitudes. Academic Exchange Quarterly, Volume 8, Issue 2.

Taylor, S., & Todd, P.A. (1995). Understanding Information Technology Usage: A Test of  Competing Models. Information Systems Research, Vol. 6, No. 2, pp.144-176.

Thompson, A G. (1992). Teachers' Beliefs and Conceptions: A Synthesis of the Research. In D. A. Grouws (Ed.), Handbook of Research on Mathematics Teaching and Learning, (pp. 127-146). New York: Macmillan Publishing Company.

Thurstone, L., & Chave, E. (1929). The Measurement of Attitudes. Chicago: Univ. of Chicago Press.

Thurstone, L. L. (1947). Multiple factor analysis. Chicago: University of Chicago Press.

Thurstone, L.L. (1931). Measurement of social attitudes. Journal of Abnormal and Social Psychology, (26), 249-269.

Tiberghien, A., Jossem, E. L., & Barojas, J. (1997,1998) (General Editors). Connecting Research in Physics Education with Teacher Education An I.C.P.E. Book. The International Commission on Physics Education

Tinsley, H.E.A., & Tinsley, D.J. (1987). Uses of factor analysis in counseling psychology research. Journal of Counseling Psychology, 34, 414-424.

Tobin, K., Tippins, D. & Gallard, A.J. (1994). Research on instructional strategies for teaching science. In Gabel, D.L. (ed.), Handbook of research on science teaching and learning (pp. 45-93). New York: Macmillan.

Townsend, J.T. & Busemeyer, J.R. (1989). Approach-avoidance: Return to dynamic decision behavior. In C. Izawa (Ed.), Current Issues in Cognitive Processes: The Tulance Floweree Symposium on Cognition (pp. 107-133). Hillsdale, NJ: Lawrence Erlbaum Associates

Trafimow, D., & Sheeran, P. (1998). Some tests of the distinction between cognitive and affective beliefs. Journal of Experimental Social Psychology, 34, 378-397.

Trafimow, D., Sheeran, P., Conner, M., & Finlay, K. A. (2002). Evidence that perceived behavioral control is a multidimensional construct: Perceived control and perceived difficulty. British Journal of Social Psychology, 41, 101-121.

Triandis, H. C. (1977). Interpersonal behavior. Monterey, CA: Brooks/Cole.

Trigwell, K., & Prosser, M. (2004). Development and use of the Approaches to Teaching Inventory. Educational Psychology Review, 16(4), 409-424.

Trigwell, K. (1995). Increasing Faculty Understanding of Teaching. In W.A. Wright (Ed.) Successful Faculty Development Strategies. Anker Publishing Co, 76-100.

Troutman, A. P. (1991). Attitudes toward personal and school use of computers (ERIC Document # ED331480). Paper presented at the Annual Conference of the Eastern Educational Research Association, Boston, MA.

Tschannen-Moran, M., Uline, C., Hoy, A.W., & Mackley, T. (2000). Creating smarter schools through collaboration. Journal of Educational Administration, 38(3), 247–271.

Tucker, L. R., & MacCallum, R. C. (1997). Exploratory factor analysis. Unpublished manuscript, Ohio State University, Columbus.

Tversky, A., & Kahneman, D. (1974). Judgement Under Uncertainty: Heuristics and Biases. Science, 185, 1124-1131.

U.S. Senate (2004). Maria Cantwell Introduces Legislation to Address Critical Shortage of Healthcare Professionals.

van den Putte, B. (1993). On the theory of reasoned action. Unpublished dissertation, University of Amsterdam, the Netherlands.

Van Os, J., Altamura, A.C., Bobes, J., Cunningham-Owens, D., Gerlach, J., Hellewell, J.S., Kasper, S., Naber, D., Tarrier, N., & Robert, P. (2002). 2-COM an instrument to facilitate patient-professional communication in routine clinical practice. Acta Psychiatrica Scandinavica, 106: 446-452.

Venkatesh, V., & Davis, F.D. (2000). "A Theoretical Extension of the Technology Acceptance Model: Four Longitudinal Field Studies. Management Science, vol. 46, pp. 186-205.

Vere-Jones, D. (1995). The coming of age of statistical education. International Statistical Review, 63 (1), 3-23.

Verkoeijen, P., Imbos, T., van de Wiel, M., J., Berger, M. P. F., & Schmidt, H. G. (2002). Assessing Knowledge Structures in a Constructive Statistical Learning Environment. Journal of Statistics Education [Online], 10(2).

Verplanken, B., Hofstee, G., & Janssen, H.J.W. (1998). Accessibility of affective versus cognitive components of attitudes. European Journal of Social Psychology, 28, 23-36.

Visser, P. S., & Krosnick, J. A. (1998). The development of attitude strength over the life cycle: Surge and decline. Journal of Personality and Social Psychology, 75, 1389-1410.

Viswanathan, M. (1993). Measurement of individual differences in preference for numerical information. Journal of Applied Psychology, 78(5), 741–752.

Von Glasersfeld, E. (1987). Learning as a constructive activity. In C. Janvier (Ed.), Problems of representation in the teaching and learning of mathematics. Hillsdale, NJ: Lawrence Erlbaum Associates, 3-17.

Vygotsky, L.S. (1962). Thought and language. Cambridge, MA: MIT Press. (Original work published 1934)

Vygotsky, L.S. (1978). Mind in Society: The development of higher psychological processes. Cambridge, MA: Harvard University Press.

Wallace, D. S., Paulson, R. M., Lord, C. G., & Bond, C. F. (2005). Which behaviors do attitudes predict? Meta-analyzing the effects of social pressure and perceived difficulty. Review of General Psychology, 9, 214-227.

Wells, H.G. (1952). Scientific American, January 1952.

West, S.G., Finch, J.F. & Curran, P.J. (1995). Structural equation models with non-normal variables: Problems and remedies. In R. H. Hoyle (Ed.), Structural equation modeling: Concepts, issues and applications (pp. 56 – 75 ). Newbury Park, CA: Sage.

White, R.C. (1979). An investigation of allied health faculty attitudes toward faculty development concepts. J Allied Health. 1979 Nov;8(4): 197 - 211.

WHO (World Health Organization) (1978). Alma-Ata Declaration.

WHO (World Health Organization) (1986). Ottawa Charter for Health Promotion.

Widaman, K. F. (1985). Hierarchically nested covariance structure models for multitrait–multimethod data. Applied Psychological Measurement, 9, 1–26.

Widaman, K. F. (1993). Common factor analysis versus principal components analysis: Differential bias in representing model parameters. Multivariate Behavioral Research, 28: 263-311.

Wild, C., Triggs, C., & Pfannkuch M. (1997). Assessment on a Budget: Using Traditional Methods Imaginatively. In I. Gal & J. B. Garfield (Eds.), The Assessment Challenge in Statistics Education. Netherlands: IOS Press.

Wild, C. J., & Pfannkuch, M. (1999). Statistical thinking in empirical enquiry. International Statistical Review, 67, 223-265.

Winstead, M.S., (1996). The development of an instrument to measure two constructs of mathematics teachers attitudes toward teaching statistical concepts. UMI dissertation services, Ann Arbor.

Woolley, S. L., Benjamin, W. J., & Woolley, A. W. (2004). Construct validity of a self-report measure of teacher beliefs related to student-centered and traditional approaches to teaching and learning. Educational and Psychological Measurement, 64, 319-331.

Yamashita, J. (2004). Reading attitudes in L1 and L2, and their influence on L2 extensive reading. Reading in a Foreign Language, 16(1), 1-19.

Young, F. W. (1987) Multidimensional scaling: history, theory and applications, R. M. Hamer (Ed.). Hillsdale, NJ, Lawrence Erlbaum.

Yu, C. H. (2001). An introduction to computing and interpreting Cronbach Coefficient Alpha in SAS. Proceedings of 26th SAS User Group International Conference.

Yuen H.K., Ma, W.K. (2001), Teachers' attitudes toward computers: Development of a computer attitude scale for teachers, Curriculum and Instruction of IT Education for the 21st Century - A Joint Conference for Guangdong, Hong Kong and Macau areas, Department of Information and Applied Technology, Hong Kong Institute of Education, Hong Kong.

Yuen, H.K., Law, N. & Chan, H. (1999) Improving IT training for serving teachers through evaluation. In G. Cumming et al. (Eds.), Advanced Research in Computers and Communications in Education. Amsterdam: IOS Press. Volume 2, pp 441-448.

Yzer, M.C., Hennessy, M. & Fishbein, M. (2004). The usefulness of perceived difficulty for health research. Psychology, Health and Medicine, 9, 149-162.

Zimbardo, P.G. & Leippe, M.R. (1991). The psychology of attitude change and social influence. New York: McGraw-Hill.

**INITIAL MINI-SURVEY OF CONTENT AND MEASUREMENT SPECIALISTS (STATISTICS EDUCATION) TOWARD ESTABLISHING CONTENT VALIDITY**

Dear Colleagues:

I have embarked upon a research project through Touro University International, aimed at "DEVELOPING AND VALIDATING A SCALE TO MEASURE INSTRUCTORS' ATTITUDES TOWARD REFORM-BASED TEACHING OF INTRODUCTORY STATISTICS IN THE HEALTH AND BHEVAIORAL SCIENCES".

At this beginning stage, I am identifying salient and focused content areas for item development, and I will greatly appreciate your feedback to the attached instrument.

Your publications and/or other scholarly contributions have considerably influenced my conceptualization of this project, and the research team considers you suitably qualified to provide expert opinions.

This project has been approved by the IRB of Touro University International (www.tourou.edu – Tel: 714-816-0366). All information received will be treated confidentially, and your assistance as a group will be acknowledged.

We are behind schedule with this project, and cannot advance without your assistance. It is a 2-page instrument, and I am counting on your usual camaraderie, support and cooperation.

If you would prefer receiving this communication via regular mail (with an enclosed self-addressed stamped envelope for return mailing), please le me know.

I will be at USCOTS 2005 (Columbus, Ohio), and look forward to meeting with you.

Sincerely,

Rossi A. Hassad

Dear Colleagues:

I am seeking your help in establishing initial content validity as I embark upon developing the abovementioned instrument. I am using the tripartite conceptualization of attitude. Detailed below are the themes and subscales commonly used to measure this construct. Please indicate your perceived relevance of each on a scale of 1 to 5 (with higher numbers representing greater relevance). Also, please feel free to include and rate other content areas. In general, the attitude object will be concept-based teaching (with attention to content, pedagogy, assessment and integration of technology). Kindly perform the same rating for items in the second table which will be used to formulate a teaching practice inventory to classify teaching approach as concept-based versus non-concept based. Thanks, and I will keep you posted on the progress of this research.

| Cognitive – Beliefs about: | Not Relevant 1 | 2 | 3 | 4 | Very Relevant 5 |
|---|---|---|---|---|---|
| Value and utility | | | | | |
| Competence/Preparedness | | | | | |
| Difficulty | | | | | |
| Self-efficacy/Behavioral Control | | | | | |
| Importance | | | | | |
| Epistemology | | | | | |
| | | | | | |
| | | | | | |
| | | | | | |
| **Affective – Feelings:** | | | | | |
| Comfort | | | | | |
| Confidence | | | | | |
| Enjoyment | | | | | |
| Enthusiasm | | | | | |
| Optimism | | | | | |
| | | | | | |
| | | | | | |
| | | | | | |
| **Behavioral – Intentions:** | | | | | |
| Likelihood to act/perform | | | | | |
| Habitual behavior | | | | | |
| Consequences of action | | | | | |
| Normative beliefs | | | | | |
| | | | | | |

**Teaching Practice Inventory**

Participants will be asked to indicate the extent to which each of the following applies to their teaching of introductory statistics in the health and/or behavioral sciences. These items were developed based on the ASA/MAA guidelines and the GAISE project report.

| Practice | Always | Usually | Sometimes | Rarely | Never |
|---|---|---|---|---|---|
| **(1)** I emphasize rules and formulae as a basis for subsequent learning. | | | | | |
| **(2)** I promote statistics as a tool of research. | | | | | |
| **(3)** I assign drill and practice exercises for each topic. | | | | | |
| **(4)** I allow students to explore data in order to discover patterns and meanings. | | | | | |
| **(5)** I focus on developing calculation skills. | | | | | |
| **(6)** Group problem solving and discussion are core learning activities. | | | | | |
| **(7)** I assign homework from the textbook. | | | | | |
| **(8)** Students use a computer program for exploring and analyzing data. | | | | | |
| **(9)** The mathematical underpinning of each statistical test is emphasized. | | | | | |
| **(10)** I use real-life data for class demonstrations and assignments. | | | | | |
| **(11)** I require that students adhere to the material and methods in the textbook. | | | | | |
| **(12)** Assessment includes written and oral reports of data analysis. | | | | | |
| **(13)** I tell students that this is an introductory course, and they may not grasp all the concepts. | | | | | |
| **(14)** I integrate statistics with other subjects. | | | | | |

General comments:

……………………………………………………………………………………………………………
……………………………………………………………………………………………………………
……………………………………………………………………………………………………………
……………………………………………………………………………………………………………
……………………………………………………………………………………………………………
……………………………………………………………………………………………………………
……………………………………………………………………………………………………………
……………………………………………………………………………………………………………
……………………………………………………………………………………………………………
……………………………………………………………………………………………………………
……………………………………………………………………………………………………………
……………………………………………………………………………………………………………
……………………………………………………………………………………………………………
……………………………………………………………………………………………………………
……………………………………………………………………………………………………………
……………………………………………………………………………………………………………
……………………………………………………………………………………………………………
……………………………………………………………………………………………………………
……………………………………………………………………………………………………………
……………………………………………………………………………………………………………
……………………………………………………………………………………………………………
……………………………………………………………………………………………………………
……………………………………………………………………………………………………………
……………………………………………………………………………………………………………
……………………………………………………………………………………………………………
……………………………………………………………………………………………………………
……………………………………………………………………………………………………………
……………………………………………………………………………………………………………

Thanks for your support and cooperation.

Email: Rhassad@mercy.edu or Rhassad@tourou.edu

# <u>CONFIDENTIAL</u>

## FACULTY ATTITUDES TOWARD STATISTICS (FATS)

**DIRECTIONS:** The survey is designed to ascertain your attitudes toward the teaching of introductory statistics in the health and behavioral sciences using the concept-based approach. **<u>The concept-based approach</u>** is intended to promote statistical literacy by emphasizing concepts and their applications rather than calculations and formulas. It involves active learning strategies such as projects, group discussions, data collection, hands-on computer data analysis, critiquing research articles, report writing, oral presentations, and the use of real-life data. **<u>Statistical literacy</u>** is the ability to understand, critically evaluate, and use statistical information and data-based arguments.

The item scale has 5 possible responses ranging from *strongly disagree* through *undecided* to *strongly agree*. Please read each statement, and from the 5-point scale, clearly check one response that best represents your agreement with that statement.

**Email:**

*Please tell us what you think about introductory statistics and concept-based teaching.*

1. The concept-based approach to teaching introductory statistics is just a fad that will soon be forgotten.
☐ Strongly Disagree ☐ Disagree ☐ Undecided ☐ Agree ☐ Strongly Agree

2. Instructors should integrate introductory statistics with other subjects.
☐ Strongly Disagree ☐ Disagree ☐ Undecided ☐ Agree ☐ Strongly Agree

3. Statistics in any form is mathematics.
☐ Strongly Disagree ☐ Disagree ☐ Undecided ☐ Agree ☐ Strongly Agree

4. Statistical literacy is necessary for effective decision-making in everyday life.
☐ Strongly Disagree ☐ Disagree ☐ Undecided ☐ Agree ☐ Strongly Agree

5. Statistics by nature is a difficult subject.
☐ Strongly Disagree ☐ Disagree ☐ Undecided ☐ Agree ☐ Strongly Agree

6. The concept-based approach to teaching introductory statistics (rather than emphasizing calculations and formulas) makes students better prepared for work.
☐ Strongly Disagree ☐ Disagree ☐ Undecided ☐ Agree ☐ Strongly Agree

7. It is unreasonable to expect students to achieve statistical thinking and literacy from an introductory statistics course.

☐ Strongly Disagree ☐ Disagree ☐ Undecided ☐ Agree ☐ Strongly Agree

*In this section, we would like to learn about your perceptions regarding the effort involved in using the concept-based approach.*

8. The concept-based approach to teaching introductory statistics is straightforward.

☐ Strongly Disagree ☐ Disagree ☐ Undecided ☐ Agree ☐ Strongly Agree

9. Focusing on statistical literacy as a learning outcome of introductory statistics is a major shift from the way I was trained.

☐ Strongly Disagree ☐ Disagree ☐ Undecided ☐ Agree ☐ Strongly Agree

10. Teaching introductory statistics with emphasis on concepts and applications rather than calculations and formulas, can be time consuming.

☐ Strongly Disagree ☐ Disagree ☐ Undecided ☐ Agree ☐ Strongly Agree

11. I will adjust easily to teaching introductory statistics using the concept-based approach.

☐ Strongly Disagree ☐ Disagree ☐ Undecided ☐ Agree ☐ Strongly Agree

12. The preparation required to teach introductory statistics using the concept-based approach is burdensome.

☐ Strongly Disagree ☐ Disagree ☐ Undecided ☐ Agree ☐ Strongly Agree

13. Integrating hands-on computer analysis into the introductory statistics course is not a difficult task.

☐ Strongly Disagree ☐ Disagree ☐ Undecided ☐ Agree ☐ Strongly Agree

14. Using active learning strategies (such as projects, group discussions, oral and written presentations) in the introductory statistics course can make classroom management difficult.

☐ Strongly Disagree ☐ Disagree ☐ Undecided ☐ Agree ☐ Strongly Agree

*The items that follow relate to beliefs about the value of concept-based teaching.*

15. The concept-based approach to teaching introductory statistics (rather than emphasizing calculations and formulas) makes students better prepared for further studies.

☐ Strongly Disagree ☐ Disagree ☐ Undecided ☐ Agree ☐ Strongly Agree

16. Emphasizing concepts and applications in the introductory statistics course (rather than calculations and formulas) is a disservice to our students.

☐ Strongly Disagree ☐ Disagree ☐ Undecided ☐ Agree ☐ Strongly Agree

17. The concept-based approach to teaching introductory statistics is for low achievers only.

☐ Strongly Disagree ☐ Disagree ☐ Undecided ☐ Agree ☐ Strongly Agree

18. The concept-based approach to teaching introductory statistics enables students to understand research.

☐ Strongly Disagree ☐ Disagree ☐ Undecided ☐ Agree ☐ Strongly Agree

19. The concept-based approach to teaching introductory statistics is not theoretically sound.

☐ Strongly Disagree ☐ Disagree ☐ Undecided ☐ Agree ☐ Strongly Agree

*In this section, please tell us the extent to which you think you are prepared to use the concept-based approach.*

20. Concept-based teaching of introductory statistics may be problematic for me.

☐ Strongly Disagree ☐ Disagree ☐ Undecided ☐ Agree ☐ Strongly Agree

21. The benefits of concept-based teaching of introductory statistics are clear to me.

☐ Strongly Disagree ☐ Disagree ☐ Undecided ☐ Agree ☐ Strongly Agree

22. I do not understand how to organize my introductory statistics course to achieve statistical literacy.

☐ Strongly Disagree ☐ Disagree ☐ Undecided ☐ Agree ☐ Strongly Agree

23. I am engaged in the teaching of introductory statistics using the concept-based approach.

☐ Strongly Disagree ☐ Disagree ☐ Undecided ☐ Agree ☐ Strongly Agree

24. I will need training on how to integrate hands-on computer exercises into the introductory statistics course.

☐ Strongly Disagree ☐ Disagree ☐ Undecided ☐ Agree ☐ Strongly Agree

25. I am convinced that the concept-based approach to teaching introductory statistics enhances learning.

☐ Strongly Disagree ☐ Disagree ☐ Undecided ☐ Agree ☐ Strongly Agree

*Now, please let us know about personal and other issues which may influence your use of the concept-based approach.*

26. I may not use the concept-based approach to teach introductory statistics because of limited departmental resources.

☐ Strongly Disagree ☐ Disagree ☐ Undecided ☐ Agree ☐ Strongly Agree

27. It is important for me to network with instructors who are teaching introductory statistics using the concept-based approach.

☐ Strongly Disagree ☐ Disagree ☐ Undecided ☐ Agree ☐ Strongly Agree

28. I am hesitant to use computers in my introductory statistics class without the help of a teaching assistant.

☐ Strongly Disagree ☐ Disagree ☐ Undecided ☐ Agree ☐ Strongly Agree

29. I am concerned that using the concept-based approach to teach introductory statistics may result in me being poorly evaluated by my students.

☐ Strongly Disagree ☐ Disagree ☐ Undecided ☐ Agree ☐ Strongly Agree

30. It is mostly up to me whether or not I use the concept-based approach to teach introductory statistics.

☐ Strongly Disagree ☐ Disagree ☐ Undecided ☐ Agree ☐ Strongly Agree

31. Using active learning strategies (such as critiquing of research articles, group discussions, and hands-on computer analysis) in my introductory statistics course, may result in students asking me questions which I cannot answer.

☐ Strongly Disagree ☐ Disagree ☐ Undecided ☐ Agree ☐ Strongly Agree

***In this section, we want to know how you feel about the concept-based approach.***

32. I am not confident in my ability to successfully teach introductory statistics using the concept-based approach.

☐ Strongly Disagree ☐ Disagree ☐ Undecided ☐ Agree ☐ Strongly Agree

33. Teaching introductory statistics using the concept-based approach is likely to be a positive experience for me.

☐ Strongly Disagree ☐ Disagree ☐ Undecided ☐ Agree ☐ Strongly Agree

34. I am not comfortable using computer applications to teach introductory statistics.

☐ Strongly Disagree ☐ Disagree ☐ Undecided ☐ Agree ☐ Strongly Agree

35. Teaching introductory statistics with emphasis on concepts and their applications (rather than calculations and formulas) may be stressful for me.

☐ Strongly Disagree ☐ Disagree ☐ Undecided ☐ Agree ☐ Strongly Agree

36. Using computers to teach introductory statistics makes learning fun.

☐ Strongly Disagree  ☐ Disagree  ☐ Undecided  ☐ Agree  ☐ Strongly Agree

37. I am anxious about using active learning strategies (such as projects, group discussions, hands-on computer analysis, critiquing of research articles, oral and written presentations) for the teaching of introductory statistics.

☐ Strongly Disagree  ☐ Disagree  ☐ Undecided  ☐ Agree  ☐ Strongly Agree

38. I am interested in using the concept-based approach to teach introductory statistics.

☐ Strongly Disagree  ☐ Disagree  ☐ Undecided  ☐ Agree  ☐ Strongly Agree

*With reference to the concept-based approach, we would like to find out what you intend to do.*

39. I want to learn more about the concept-based approach to teaching introductory statistics.

☐ Strongly Disagree  ☐ Disagree  ☐ Undecided  ☐ Agree  ☐ Strongly Agree

40. Using the concept-based approach to teach introductory statistics is not a priority for me.

☐ Strongly Disagree  ☐ Disagree  ☐ Undecided  ☐ Agree  ☐ Strongly Agree

41. I plan on teaching introductory statistics according to the concept-based approach.

☐ Strongly Disagree  ☐ Disagree  ☐ Undecided  ☐ Agree  ☐ Strongly Agree

42. I will avoid using computers in my introductory statistics course.

☐ Strongly Disagree  ☐ Disagree  ☐ Undecided  ☐ Agree  ☐ Strongly Agree

43. I will incorporate active learning strategies (such as projects, hands-on data analysis, critiquing research articles, and report writing) into my introductory statistics course.

☐ Strongly Disagree  ☐ Disagree  ☐ Undecided  ☐ Agree  ☐ Strongly Agree

44. I will emphasize calculations and formulas in my introductory statistics course.

☐ Strongly Disagree  ☐ Disagree  ☐ Undecided  ☐ Agree  ☐ Strongly Agree

45. Statistics is mathematics, and that is how I intend to teach it.

☐ Strongly Disagree  ☐ Disagree  ☐ Undecided  ☐ Agree  ☐ Strongly Agree

*Now, please tell us the extent to which each of the following applies to your teaching of introductory statistics in the health and/or behavioral sciences.*

46. I emphasize rules and formulas as a basis for subsequent learning.
☐ Always ☐ Usually ☐ Sometimes ☐ Rarely ☐ Never

47. I integrate introductory statistics with other subjects.
☐ Always ☐ Usually ☐ Sometimes ☐ Rarely ☐ Never

48. Students use a computer program to explore and analyze data.
☐ Always ☐ Usually ☐ Sometimes ☐ Rarely ☐ Never

49. I assign homework primarily from the textbook.
☐ Always ☐ Usually ☐ Sometimes ☐ Rarely ☐ Never

50. Critiquing of research articles is a core learning activity.
☐ Always ☐ Usually ☐ Sometimes ☐ Rarely ☐ Never

51. The mathematical underpinning of each statistical test is emphasized.
☐ Always ☐ Usually ☐ Sometimes ☐ Rarely ☐ Never

52. I use real-life data for class demonstrations and assignments.
☐ Always ☐ Usually ☐ Sometimes ☐ Rarely ☐ Never

53. I require that students adhere to the procedures in the textbook.
☐ Always ☐ Usually ☐ Sometimes ☐ Rarely ☐ Never

54. Assessment includes written reports of data analysis.
☐ Always ☐ Usually ☐ Sometimes ☐ Rarely ☐ Never

55. I assign drill and practice exercises (mathematical calculations) for each topic.
☐ Always ☐ Usually ☐ Sometimes ☐ Rarely ☐ Never

*And finally, kindly tell us a bit more about yourself.*

56. **Gender:**
☐ Male
☐ Female

57. **Ethnicity:**
☐ Caucasian American
☐ African American
☐ Hispanic American

☐ Asian American
☐ Other

58. **Highest earned academic degree:**.
☐ Bachelor's
☐ Master's
☐ Doctoral
☐ Other (please specify) [                    ]

59. **Academic specialization/concentration:**
☐ Statistics
☐ Mathematics
☐ Psychology
☐ Sociology
☐ Health Sciences/Public Health
☐ Behavioral Sciences
☐ Epidemiology
☐ Social Sciences
☐ Education
☐ Business
☐ Engineering
☐ Other(please specify) [                    ]

60. **Age group:**
☐ 18 – 25
☐ 26- 30
☐ 31 – 40
☐ 41 – 50
☐ 51 – 60
☐ 60+

61. **How long have you been teaching statistics at the college/university level?**
[          ] **Years**

62. **I teach statistics in:**
☐ Health Sciences
☐ Social and Behavioral Sciences (including Psychology)
☐ Behavioral Sciences only
☐ Psychology only
☐ Other (please specify) [                    ]

63. **Faculty employment status:**
☐ Full-time
☐ Part-time

64. **Are you a member of any professional organization that addresses the teaching of statistics?**
☐ No

☐ Yes (Please Specify)

65. **General Comments:**

[Submit] [Reset]

# APPENDIX C

**LIST OF COUNTRIES REPRESENTED BY THE STUDY PARTICIPANTS**

1.  ARGENTIA
2.  AUSTRALIA
3.  BELIUM
4.  BRAZIL
5.  CANADA
6.  GERMANY
7.  GREECE
8.  INDIA
9.  IRELAND
10. ISRAEL
11. ITALY
12. MALAWI
13. MEXICO
14. NETHERLANDS
15. NEW ZEALAND
16. NORWAY
17. PORTUGAL
18. SAUDIA ARABIA
19. SLOVENIA
20. SOUTH AFRICA
21. SRI LANKA
22. TURKEY
23. UNITED KINGDOM (UK
24. UNITED STATES OF AMERICA (USA)

# APPENDIX D

## LIST OF INSTITUTIONS REPRESENTED BY THE STUDY PARTCIPANTS

**International**

Acadia University (Canada)
Aristotle University of Thessaloniki (Greece)
Cardiff University (UK)
City University Londodn (UK)
College of Medicine, University of Malawi
La Trobe University (Australia)
Maastricht University, The Netherlands
Macalester College (Canada)
McGill University (Canada)
McMaster University (Canada)
National Institute of Biology (Slovenia)
Open Universiteit Nederland (The Netherlands)
Rothamsted Research Centre (UK)
Simon Fraser University (Canada)
Tel Aviv University (Israel)
The Open University (UK)
The University of Auckland (New Zealand)
The University of Manchester (UK)
The University of Melbourbe (Australia)
The University of Sheffield (UK)
The University of Sydney (Australia)
Universidad Autónoma de Yucatán (Mexico)
Universidade do Algarve (Portugal)
Universidade Estadual Paulista (Brazil)
Universidade Federal da Bahia (Brazil)
Università "La Sapienza" (Italy)
Universitat Bremen (Germany)
Universität Witten/Herdecke (Germany)
University College Dublin (Ireland)
University Ghent (Belgium)
University of Aberdeen (UK)
University of Bergen (Norway)
University of Birmingham (UK)
University of Bradford (UK)
University of Cape Town (South Africa)
University of Crete, Heraklion (Greece)
University of East Anglia (UK)
University of Kelaniya (Sri Lanka)
University of London  (UK)
University of Melbourne (Australia)

University of Nottingham (UK)
University of Southern Queensland (Australia)
University of Sunderland (UK)
University of Ulster (UK)
University of Wales, Bangor (UK)
University of Western Ontario (Canada)
Vrije Universiteit Amsterdam

**USA**

Alabama State University
American University
Appalachian State University
Ashland University
Augsburg College, Minneapolis, Minnesota
Babson College
Barnard College
Boston College
Boston University
Brandeis University
Brigham Young University
Brown University
California Polytechnic State University
California State University
Case Western Reserve University
Columbia University
Cornell University
Duke University
East Carolina University
Eastern Kentucky University
Emory University
Georgia Institute of Technology
Georgia Southern University
Grand Valley State University
Harvard University
Hollins University
Humboldt State University
Hunter College
Iowa State University
Johns Hopkins University
Kansas State University
Kent State University
Lakeland College - Sheboygan, WI
Meredith College
Miami University
Missouri State University

New Jersey Institute of Technology
New York Medical College
Oakland University
Ohio State University
Oregon State University
Penn State University
Pomona College
Portland State University
Purchase College
Purdue University Indianapolis
Radford University
Rice University
Rutgers, The State University of New Jersey
Saginaw Valley State University
Seattle Pacific University
Shippensburg University
Southern Utah University - Cedar City, Utah
Sweet Briar College
Syracuse University
Towson University
University of Alabama
University of Chicago
University of Colorado
University of Connecticut
University of Georgia
University of Houston
University of Illinois
University of Iowa
University of Kansas Medical Center
University of Maine
University of Massachusetts
University of Michigan
University of Minnesota
University of Missouri
University of New Haven
University of North Carolina
University of North Carolina at Chapel Hill
University of Pennsylvania
University of Rochester
University of South Dakota
University of St. Thomas
University of Texas
University of Vermont
University of Wisconsin
Vanderbilt University
Wabash College, Crawfordsville, Indiana

Washington University in St. Louis
West Chester University of Pennsylvania
Western Illinois University
Yale University

# APPENDIX E

# Touro University International
## Institutional Review Board for the Protection of Human Subjects

## IRB REVIEW FORM

| | | |
|---|---|---|
| **PROJECT TITLE:** | **PROJECT INVESTIGATOR** | **PROJECT DATE:** |
| Development and Validation of an Attitudinal Scale ...... Rossi Hassad | | 4/2005 to 4/2006 |

**APPLICATION TYPE:** ☐ EXEMPT  **X**☐ **EXPEDITED REVIEW** ☐ FULL REVIEW

**APPLICATION STATUS:** ☐ APPROVED  **X**☐ **APPROVED WITH AMENDMENT**

☐ REQUIRES ADDITIONAL INFORMATION  ☐ NOT APPROVED

THE FOLLOWING ADDITIONAL INFORMATION/AMENDMENT IS REQUIRED BY THE IRB:

1. **Statement that data will be maintained in a secured location for 5 years;**
2. **Add IRB Chair contact information on the consent form;**
3. **Remove TUI name on the masthead of his consent form and instrument.**

**Afshin Afrookhteh          04/01/05**

_______________________________________________

IRB Chair                                    Date

# APPENDIX F

## INFORMED CONSENT

I hereby consent to my participation in the following research project:

**Title:** Development and Validation of a Scale to Measure Instructors' Attitudes toward Reform-Based Teaching of Introductory Statistics in the Health and Behavioral Sciences

**Department:** College of Health Sciences, Touro University International

**Address:** 5665 Plaza Dr., 3rd Floor, Cypress, CA 90630, U.S.A.

**Principal Investigator:** Rossi A. Hassad

**Telephone:** 800-375-9878 or 714-226-9840 **E-mail:** Rhassad@tourou.edu

Dr. Afshin Afrookhteh, Chair- Institutional Review Board
Touro University International Tel: (714) 226-9840, extension 2004

The primary objective of this study is to explore the attitudes and practices of instructors of introductory statistics in the health and behavioral sciences in the context of reform. A one-time self-administered questionnaire will be used to ascertain beliefs about, feelings toward, and intentions to adopt selected teaching methods. Sociodemographic and teaching practice data will also be obtained. This should allow for characterizing instructors' predisposition toward reform. Also, potential barriers to, and facilitators of effective teaching will be identified. All information obtained will be treated confidentially and in accordance with applicable laws. This study does not involve any experimentation. All data will be maintained in a secured location for 5 years.

As an incentive for participation, all participants will be given a chance to win one of three $100 (one hundred dollar) awards toward conference registration, journal subscription, continuing education courses or other professional development activities.

You are free to withdraw your consent and to discontinue participation in this study at any time, without prejudice. If you have questions regarding this study or your rights as a research participant, please contact Touro University International at the address above.

I fully understand the requirements of this study and the procedures described above. My questions and concerns have been satisfactorily addressed, and I hereby agree to participate in this study. I have been given a copy of this consent form.

**Full Name of Participant:**

I Agree to Participate      I DO NOT Agree to Participate